\documentclass{article}

\usepackage{graphics,graphicx}% Include figure files
\usepackage{dcolumn}% Align table columns on decimal point
\usepackage{bm}% bold math

\usepackage{amsmath}
\usepackage{amssymb,amsbsy}

\newcommand{\beq}{\begin{equation}}
\newcommand{\eeq}{\end{equation}}
\newcommand{\bea}{\begin{eqnarray}}
\newcommand{\eea}{\end{eqnarray}}
\newcommand{\bec}{\begin{center}}
\newcommand{\enc}{\end{center}}
\newcommand{\alp}{\alpha}
\newcommand{\om}{\omega}
\newcommand{\Om}{\Omega}
\newcommand{\eps}{\epsilon}
\newcommand{\ve}{\varepsilon}

\newcommand{\rmd}{{\rm d}}
\newcommand{\rmg}{{\rm g}}
\newcommand{\rmi}{{\rm i}}
\newcommand{\rmx}{{\rm x}}

\newcommand{\veck}{\bm{k}}

\newcommand{\vecp}{\bm{p}}
\newcommand{\vecf}{\bm{f}}
\newcommand{\vecr}{\bm{r}}

\newcommand{\vece}{\bm{e}}
\newcommand{\vecE}{\bm{E}}

\begin{document}
%%%%%%%%%%%%%%%%%%%%%%%%%%%%%%%%%%%%%%%%%%%%%%%%%%%%%%%%%%%%%%%%%%%%%%%%%%%%%%%

\title{\bf Quantum Zeno effect by general measurements}
% \title{\bf Quantum Zeno effect caused by general measurements}
% \title{\bf Quantum Zeno effect in general measurement processes}
\author{
Kazuki Koshino$^a$ and Akira Shimizu$^b$ \\ \\
$^a${\small\it CREST, Japan Science and Technology Corporation}\\
{\small\it c/o Department of Materials Engineering Science, 
%Graduate School of Engineering Science\\
Osaka University, Toyonaka, Osaka 560-8531, Japan}\\
{\small\it Email:} {\small\tt ikuzak@aria.mp.es.osaka-u.ac.jp}
\\ \\
$^b${\small\it Department of Basic Science, University of Tokyo,
% }\\ 
% {\small\it 
3-8-1 Komaba, Tokyo 153-8902, Japan}\\
% {\small\it and}\\
{\small\it and PRESTO, Japan Science and Technology Corporation,
% }\\ 
% {\small\it 
4-1-8 Honcho Kawaguchi, Saitama, Japan}\\
{\small\it Email:} {\small\tt shmz@ASone.c.u-tokyo.ac.jp}
}
\date{\small\it \today}
\maketitle
\begin{abstract}
It was predicted that frequently repeated measurements 
on an unstable quantum state may alter the decay rate of the state. 
This is called the quantum Zeno effect (QZE) 
or the anti-Zeno effect (AZE),
depending on whether the decay is suppressed or enhanced.
In conventional theories of the QZE and AZE,
effects of measurements are simply 
described by the projection postulate, 
assuming that each measurement is an instantaneous and ideal one.
However, real measurements are not 
instantaneous and ideal.
For the QZE and AZE by such general measurements,
interesting and surprising features 
have recently been revealed, 
which we review in this article.
The results are based on the quantum measurement theory, 
which is also reviewed briefly.
As a typical model, 
we consider a continuous measurement of the decay of an excited atom
by a photodetector that detects a photon emitted 
from the atom upon decay.
This measurement is an indirect negative-result one,
for which the curiosity of the QZE and AZE is emphasized.
It is shown that the form factor is renormalized 
as a backaction of the measurement,
through which the decay dynamics is modified.
In a special case of the flat response,
where the detector responds to every photon mode with an identical 
response time,
results of the conventional theories 
are reproduced qualitatively.
However, drastic differences emerge
in general cases where the detector 
responds only to limited photon modes.
For example, 
against predictions of the conventional theories,
the QZE or AZE may take place even for states
that exactly follow the exponential decay law.
We also discuss relation to the cavity quantum electrodynamics.
\end{abstract}

% \pagenumbering{roman}
\tableofcontents
% \newpage
% \pagenumbering{arabic}
% \setcounter{page}{1}

%%%%%%%%%%%%%%%%%%%%%%%%%%%%%%%%%%%%%%%%%%%%%%%%%%%%%%%%%%%%%%%%%%%%%%%%%%%%%%
\section{Introduction}
%%%%%%%%%%%%%%%%%%%%%%%%%%%%%%%%%%%%%%%%%%%%%%%%%%%%%%%%%%%%%%%%%%%%%%%%%%%%%%

The standard quantum theory assumes two principles for time 
evolution~\cite{vN,JJ};
the continuous {\em unitary evolution} 
in the absence of measurement, 
and the {\rm projection postulate} connecting the pre- and 
post-measurement states.
Using these principles, 
an interesting prediction was obtained by 
analyzing the gedanken experiment
in which  the initial state of a quantum system % the system 
is unstable
and one repeatedly checks whether the unstable state 
has decayed or not~\cite{BN,Kha,Misra}.  
Until the system is measured, 
the state vector undergoes the unitary evolution
according to the Schr\"{o}dinger equation.
It is then shown that 
at the very beginning of the decay
the survival probability $s(t)$ of 
the unstable state decreases very slowly
as a function of the elapsed time $t$
as $1- s(t) \propto t^2$,
whereas in the later time stage 
$s(t)$ decreases much faster, typically exponentially.
% with $t$.
The time scale $t_{\rm j}$ at which the crossover between 
these different behaviors takes place is called the 
{\em jump time}~\cite{jump}.  %shmz
On the other hand, the projection postulate tells us that
at every moment an observer confirms 
the survival of the system through the measurement,
the quantum state of the system is reset to the initial undecayed one.
Combining these two observations,
one is led to an interesting conclusion that
the decay rate is reduced 
if the intervals $\tau_{\rm i}$ of the repeated measurements is shorter than
$t_{\rm j}$. 
In the limit of $\tau_{\rm i}/t_{\rm j} \to +0$, in particular,   
the decay is completely suppressed, i.e., 
the system is frozen to the initial undecayed state.
This phenomenon is called the {\em quantum Zeno effect} 
(QZE)~\cite{POT,NNP,Panov,HW}.
It was also predicted later 
that for slightly longer $\tau_{\rm i}$ ($\sim t_{\rm j}$) 
the opposite effect, 
i.e., acceleration of decay, can occur in some quantum systems.
This is called the {\em quantum anti-Zeno effect} (AZE)
or inverse Zeno effect~\cite{Lane,KofPRA,Kau,FPreview}. % shmz
The QZE and AZE are sometimes called simply % as 
the {\em Zeno effect}.

Most importantly, 
the reasoning leading to these predictions is
independent of details of quantum systems, 
and thus the Zeno effect is 
expected to occur widely in quantum systems.
In particular, the complete suppression of the decay 
in the limit of $\tau_{\rm i}/t_{\rm j} \to +0$
is universal, common to all quantum systems.
However, the reasoning leading to such interesting % important 
and universal conclusions needs to be reexamined, because 
it assumes that each measurement is 
an {\em instantaneous ideal measurement}.
Here, the term `instantaneous' means that 
the response time $\tau_{\rm r}$ of the measuring apparatus 
is much shorter than 
other relevant time scales such as 
$\tau_{\rm i}$. % and $t_{\rm j}$.
The term `ideal' means that 
the post-measurement state is given by the projection postulate, 
which implies many conditions such as 
the measurement error is zero.
% which satisfies many conditions such as 
% (i) the response time $\tau_{\rm r}$ of a measuring apparatus
% is much shorter than $\tau_{\rm i}$, 
% (ii) the measurement error is zero,
% (iii) the post-measurement state is given by the projection postulate,
% and so on.
Unfortunately, these conditions are not strictly satisfied in % most 
real measurements. 
Therefore, the Zeno effect by such {\em general measurements} 
is interesting and to be explored.
% Therefore, the Zeno effects under such {\em general measurements} 
% are to be clarified.

To study % explore 
the time evolution of quantum systems 
under general measurements, 
one must apply the quantum measurement theory,
which has been developed for several decades % the last half century 
\cite{vN,LP,Glauber,Unruh,ozawa88,SF,FL,QNoise,simultaneousM,ozawa}.
The point of the quantum measurement theory is that 
one must apply the laws %principles 
of quantum theory to 
the joint quantum system composed of the 
target system S of interest and (a part of) 
the measuring apparatus A.
In other words, the `Heisenberg cut' separating 
the quantum system and the rest of the world should be 
located not between S and A but between S$+$A and the rest A$'$.
Although the boundary between A and A$'$ can be taken 
quite arbitrarily, 
one can obtain the same results if S$+$A is taken to be 
large enough, i.e., if the Heisenberg cut is properly located~\cite{vN}.
% shmz
%(see Ref. and Sec.~\ref{sec:HC}).
One can then calculate all relevant quantities, including the response time, 
measurement error, range of the measurement, 
and the post-measurement state, and so on. 
One can thus calculate the decay rate 
under general measurements as a function of these 
relevant quantities.
Furthermore, although the necessity of the projection postulate 
in the analysis of the Zeno effect
has been a controversial point for a long time~\cite{HW}, 
%(see, e.g., Ref.), 
the quantum measurement theory gives a clear answer:
As far as the Zeno effect is concerned
one can analyze it without using the projection postulate at all
if S$+$A is taken to be large enough. 
%(see Sec.~\ref{sec:Zeno-MMT}). %shmz

The purpose of the present article is to review 
results for the Zeno effect by general measurements.
% results for the Zeno effects under general measurements.
We show that 
the conclusions of the conventional theories, 
which assume instantaneous ideal measurements, 
are modified drastically depending on the natures of
real measurements.
In particular, 
some of common wisdoms deduced by the conventional theories
break in general measurement processes.
For example, 
the Zeno effect can take place even for systems with 
$t_{\rm j} \to +0$~\cite{KSPRL,KKPRL}, 
for which the conventional theories predicted
that the Zeno effect never occurs.

Note that in the original papers of the QZE 
% \cite{BN,Kha,Misra} % <--- Is it OK?
a truly decaying state was analyzed, 
for which $s(t)$ in the absence of measurements 
decreases monotonically. 
%with $t$ for $t \gg t_{\rm j}$.
However, the Zeno effect has also been discussed 
on other classes of states such as states for which 
$s(t)$ oscillates with $t$ 
(Rabi oscillation)~\cite{Cook,Itano}.
Furthermore, although 
the QZE was discussed as a result of measurements 
in the original papers, 
some works use the term QZE or AZE for % any 
changes of the decay rate              % that are 
induced by any 
external perturbations such as external noises~\cite{eiQZE1,eiQZE2,eiQZE3}.
The former 
may be called the Zeno effect in the narrow sense, 
whereas the latter may be called 
the Zeno effect in the broad sense.
Moreover, 
it is sometimes argued that 
the Zeno effect is curious or surprising only 
when the measurements are indirect and negative-result
% measurements
ones~\cite{HW}.
Although these different views concern merely the definition of 
the terms QZE and AZE, 
they have been the origins of certain confusion or 
controversy.
We will therefore notice the above points where it is needed.
% We therefore note the above points where it is needed.

The present article is organized as follows.
In Sec.~\ref{sec:fpus},
we review the {\it free} quantum dynamics
of unstable states, 
i.e., the dynamics 
while the system is not being measured,  
by solving the Schr\"{o}dinger equation.
After presenting a typical model of unstable systems,
we describe a simple technique to solve the model,
and explain characteristics of the survival probability of the unstable state.
By combining the results of Sec.~\ref{sec:fpus}
and the projection postulate, we review in Sec.~\ref{sec:ct}
the conclusions of the conventional theories of the Zeno effect, %shmz
which assumed instantaneous ideal measurements.
% for the Zeno effect.
The decay rate under repeated measurements is
presented as a function of the measurement intervals $\tau_{\rm i}$,
and the conditions for inducing the QZE or AZE by %under 
instantaneous ideal measurements are clarified.
The quantum measurement theory is briefly reviewed 
in Sec.~\ref{sec:mt}, in such a way that 
it provides for basic knowledge
that are required to understand 
not only the Zeno effect but also 
many other topics of quantum measurements.
After presenting the prescription for analyzing general measurements,  
we summarize relevant quantities 
such as the measurement error
and the range of the measurement,
as well as useful concepts, such as indirect measurements
and negative-result measurements.
We also explain why
% present an approximation which is
% often used in practical calculations,
% including those of Sec.~\ref{sec:rmt}, of the Zeno effect. \ASo
% Because of this approximation,
one can analyze the Zeno effect without using the projection postulate.
% We then give a simple explanation of the Zeno effect 
% by the quantum measurement theory. 
We then give a simple explanation of the Zeno effect 
using the quantum measurement theory. 
% \ASo
% (under the condition that S$+$A is taken to be large enough).
In Sec.~\ref{sec:rmt},
we analyze the Zeno effect
by the quantum measurement theory.
We employ a model which describes a continuous indirect negative-result
measurement of an unstable state. % of a target system.
It is shown that
the form factor is renormalized as an inevitable backaction of measurement, 
and this renormalization plays a crucial role in the Zeno effect.
We study the case of a continuous measurement with flat response 
in Sec.~\ref{sec:idealm},
and show that the results almost coincide with those 
obtained by the conventional theories, which assumed 
repeated instantaneous ideal measurements.
In contrast, we show in Secs.~\ref{sec:gim}, \ref{sec:eim} and \ref{sec:fm}
that dramatic differences emerge %dramatic
if the response is not flat.
In Sec.~\ref{sec:cqed}, 
relation between the Zeno effect and other phenomena, 
such as the motional narrowing, 
is discussed.
In particular, 
we discuss the close relationship between 
the cavity quantum electrodynamics (QED)
and the Zeno effect by %under 
a continuous indirect measurement.
Using the results of the cavity QED, 
we touch on the Zeno effect in case where the detectors are spatially 
separated from the target atom in Sec.~\ref{sec:atom-in-inhomo}.
In Sec.~\ref{sec:exper}, we introduce % discuss 
experimental studies on the Zeno effect on 
monotonically decaying %truly 
unstable states.
We also discuss how to {\em avoid} the Zeno effect 
in general experiments, %measurements, 
which are designed not to detect the Zeno effect but
to measure the free decay rate accurately.
Finally, the main points of this article are summarized 
in Sec.~\ref{sec:summary}.
Since Secs.~\ref{sec:mt} and \ref{sec:rmt} are rather long, 
guidelines are given at the beginnings of these sections,
which will help the readers who wish to read them faster. 

%%%%%%%%%%%%%%%%%%%%%%%%%%%%%%%%%%%%%%%%%%%%%%%%%%%%%%%%%%%%%%%%%%%%%%%%%%%%%%
\section{Fundamental properties of unstable quantum systems}
% \verb#{sec:fpus}#
\label{sec:fpus}
%%%%%%%%%%%%%%%%%%%%%%%%%%%%%%%%%%%%%%%%%%%%%%%%%%%%%%%%%%%%%%%%%%%%%%%%%%%%%%

%%%%%%%%%%%%%%%%%%%%%%%%%%%%%%%%%%%%%%%%%%%%%%%%%%%%%%%%%%%%%%%%%%%%%%%%%%%%%%
\subsection{A typical model of unstable quantum systems}
%%%%%%%%%%%%%%%%%%%%%%%%%%%%%%%%%%%%%%%%%%%%%%%%%%%%%%%%%%%%%%%%%%%%%%%%%%%%%%
In this section, 
we review the intrinsic dynamics
of unstable states in quantum systems,
which occurs while the system is not being measured.
There are many examples of unstable quantum systems:
excited atoms~\cite{RWA,Mandel}, 
unstable nuclei~\cite{Panov}, and so on.
The dynamics of these systems are characterized by irreversibility;
the initial unstable state decays with a finite lifetime,
and the system never returns to the initial state spontaneously.
Such an irreversible dynamics takes place
when the initial state is coupled to continua of states, 
whose energies extend over a wide energy range.
In the following, we employ an excited two-level atom 
with a finite radiative lifetime
as a typical example of unstable quantum systems,
but the main features of the dynamics are common to most unstable systems.

\begin{figure}%----------------------------------------------------------------
\bec
\includegraphics{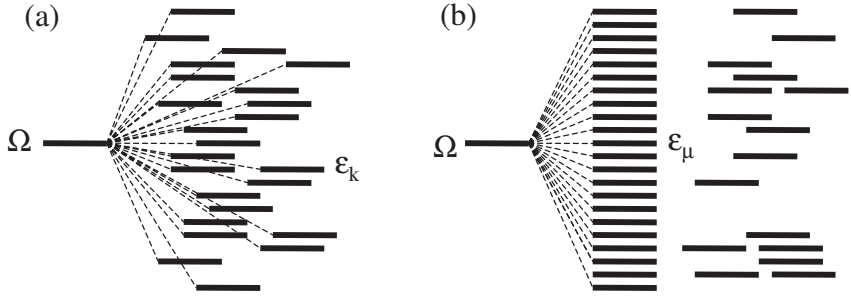}
\caption{\label{fig:enedi}
Schematic energy diagram of the Hamiltonian
in (a) the raw form [Eq.~(\ref{eq:H1})], 
and (b) after the interaction modes are extracted
[Eq~(\ref{eq:H1'})].
}% \verb#{fig:enedi}#
\enc
\end{figure}

The system is composed of a two-level atom and a photon field.
The eigenmodes of the photon field are labeled 
by the wavevector $\veck$ and the polarization $\lambda$.
For notational simplicity, 
we hereafter omit the label $\lambda$ and 
employ a single label $\veck$ % representing the wavevector
to discriminate photon eigenmodes.
We denote the atomic raising (lowering) operator 
by $\sigma_+$ ($\sigma_-$),
the creation (annihilation) operator of a photon
by $b_{\veck}^{\dagger}$ ($b_{\veck}$),
and the vacuum state (no atomic excitation and no photons) 
by $|0\rangle$.
At the initial moment ($t=0$),
the atom is in the excited state and there are no photons.
Taking $\hbar=c=1$, the Hamiltonian of this system is given by
\beq
\hat{H}_S =
\Om\sigma_+\sigma_- + \int \rmd\veck \ \left[(g_{\veck}\sigma_+ b_{\veck}+{\rm H.c.})
+\eps_{\veck}b_{\veck}^{\dagger}b_{\veck}\right],
% \verb#{eq:H1}#
\label{eq:H1}
\eeq
where $\Om$ is the atomic transition energy,
$\eps_{\veck}$ is the energy of the mode $\veck$,
and $g_{\veck}$ is the atom-photon coupling.
The schematic energy diagram is shown in Fig.~\ref{fig:enedi}(a).
Here, the dimension of $\veck$ is arbitrary,
and no specific forms of $\eps_{\veck}$ and $g_{\veck}$ are assumed.

Regarding the atom-photon interaction,
the rotating-wave approximation is employed~\cite{RWA,Mandel},
i.e., the counter-rotating terms such as $\sigma_- b_{\veck}$ 
and $\sigma_+ b_{\veck}^{\dagger}$ are neglected.
The effect of counter-rotating terms may be 
partly % shmz approximately 
incorporated 
% into $\hat{H}_1$ 
by renormalizing the atom-photon coupling 
$g_{\veck}$ for off-resonant photons.
Under the rotating-wave approximation, 
% we can confirm that 
the number of quanta, which is defined by
\beq
\hat{N}=\sigma_+ \sigma_- 
+ \int d\veck b_{\veck}^{\dagger} b_{\veck},
\eeq
is conserved in this system, i.e., $[\hat{H}_S, \hat{N}]=0$.
Because we have one quantum (atomic excitation) in the initial state,
the state vector evolves restrictedly in the one-quantum space,
which is spanned by the following states:
\bea
|{\rm x}\rangle &=& \sigma_+ |0\rangle, \\
|{\rm g},\veck\rangle &=& b_{\veck}^{\dagger} |0\rangle, 
\eea
where $|{\rm x}\rangle$ and $|{\rm g}\rangle$ in the left-hand-side 
represent the excited and ground states of the atom, respectively.
The initial state vector, $|{\mathrm i}\rangle$, is given by
$|{\mathrm i}\rangle=|{\mathrm x}\rangle$.

Throughout this article, we employ 
the Schr\"{o}dinger picture 
for describing temporal evolution.
The state vector evolves as
\beq
|\psi(t)\rangle = \exp(-\rmi \hat{H}_S t)|{\rm i}\rangle,
% \verb#{eq:Scheq}#
\label{eq:Scheq}
\eeq
which may be written as follows:
\beq
|\psi(t)\rangle = f(t)|\rmx\rangle 
+ \int d\veck f_{\veck}(t)|\rmg,\veck\rangle,
% \verb#{eq:psit}#
\label{eq:psit}
\eeq
where
\bea
f(t) &=& \langle\rmx|\exp(-\rmi\hat{H}_S t)|{\rm i}\rangle, 
% \verb#{eq:deff(t)}#
\label{eq:deff(t)}
\\
f_{\veck}(t) &=& \langle \rmg,\veck|\exp(-\rmi\hat{H}_S t)|{\rm i}\rangle. 
% \verb#{eq:deffk(t)}#
\label{eq:deffk(t)}
\eea
$f(t)$ is called the survival amplitude of the initial state,
and its square gives 
the survival probability $s(t)$;
\beq
s(t) = |f(t)|^2.
\eeq

%-----------------------------------------------------------------

%%%%%%%%%%%%%%%%%%%%%%%%%%%%%%%%%%%%%%%%%%%%%%%%%%%%%%%%%%%%%%%%%%%%%%%%%%%%%%
\subsection{Initial behavior of survival probability}
% \verb#{sec:ibsp}#
\label{sec:ibsp}
%%%%%%%%%%%%%%%%%%%%%%%%%%%%%%%%%%%%%%%%%%%%%%%%%%%%%%%%%%%%%%%%%%%%%%%%%%%%%%

In this subsection,
we briefly discuss short-time behaviors 
of the survival probability.
Expanding $\exp(-\rmi\hat{H}_S t)$ in Eq.~(\ref{eq:deff(t)}) 
in powers of $t$, $f(t)$ is given by
\beq
f(t)=\sum_{j=0}^{\infty}\frac{(-\rmi t)^j}{j!}\langle (\hat{H}_S)^j\rangle,
\eeq
where 
$\langle (\hat{H}_S)^j\rangle = 
\langle \rmi| (\hat{H}_S)^j| \rmi \rangle$.
\footnote{
%It should be remarked that 
This expansion is valid when $\langle (\hat{H}_S)^j\rangle$ is finite 
for any $j$.
Although this assumption seems to be satisfied
% $\langle (\hat{H}_S)^j\rangle$ is finite and this expansion is valid 
in real physical systems,
$\langle (\hat{H}_S)^j\rangle$ can diverge
if one takes a certain limit, such as
the $\Delta \to \infty$ limit % of the Lorentzian form factor.
taken in Sec.~\ref{sssec:exact_exp}.
Physically, such a limit should be understood as 
an abbreviated description of the case where 
$\Delta$ is larger than any other relevant energy scales.
% some mathematical models.
% (For example, see Sec.~\ref{sssec:exact_exp}.)
}
$s(t)$ is thus given by
\beq
s(t)=1-\langle (\Delta \hat{H}_S)^2 \rangle t^2 +{\cal O}(t^4),
% \verb#{eq:st2}#
\label{eq:st2}
\eeq
where
$\langle (\Delta \hat{H}_S)^2 \rangle
\equiv \langle \hat{H}_S^2 \rangle - \langle \hat{H}_S \rangle^2$ 
is a positive quantity.
It is easily confirmed that 
\beq
s(t)=|\langle\exp(-\rmi\hat{H}_S t)\rangle|^2 =
|\langle\exp(\rmi\hat{H}_S t)\rangle|^2 = s(-t),
\eeq
which implies that $s(t)$ is an even function of $t$
and therefore contains only even powers of $t$.
Equation (\ref{eq:st2}) states that 
$s(t)$ decreases quadratically in time at the beginning 
of decay, 
whereas in a later period $s(t)$ behaves differently 
depending on details of the system 
(see Sec.~\ref{sec:Lff} and Refs.~\cite{BN,Kha,Misra,NNP,Bern,exp_t2,Unn}).
The quadratic decrease becomes important 
when we consider the effects of frequently repeated measurements on the system
(see Sec.~\ref{sec:QZE}).
%as will be clarified in Section \ref{sec:QZE}.

%%%%%%%%%%%%%%%%%%%%%%%%%%%%%%%%%%%%%%%%%%%%%%%%%%%%%%%%%%%%%%%%%%%%%%%%%%%%%%
\subsection{Form factor}
% \verb#{sec:ff}#
\label{sec:ff}
%%%%%%%%%%%%%%%%%%%%%%%%%%%%%%%%%%%%%%%%%%%%%%%%%%%%%%%%%%%%%%%%%%%%%%%%%%%%%%

Before investigating the temporal evolution in the whole time region,
we transform the above Hamiltonian $\hat{H}_S$ into a simpler form,
where the atom is coupled to a single continuum
which is labeled one-dimensionally by the energy.

\subsubsection{Interaction mode and the form factor} % shmz
\label{sec:imff}

In order to explain the basic idea, 
we preliminarily consider a simplified case
where the atom interacts only with two photon modes $b_1$ and $b_2$
of the same energy $\mu$.
The Hamiltonian for this simplified system is given by
\begin{displaymath}
\hat{H}_{\rm sim}=\Om \sigma_+\sigma_-
+\mu b_1^{\dagger}b_1+\mu b_2^{\dagger}b_2
+\left( \gamma_1 \sigma_+ b_1+
\gamma_2 \sigma_+ b_2+{\rm H.c.} \right).
\end{displaymath}
By the following linear transformation, 
\begin{displaymath}
\left(\begin{matrix}
B_1 \cr B_2 \end{matrix}\right)=
\frac{1}{\sqrt{\gamma_1^2+\gamma_2^2}}
\left(\begin{matrix}
\gamma_1 & \gamma_2 \cr -\gamma_2 & \gamma_1
\end{matrix}\right)
\left(\begin{matrix}
b_1 \cr b_2 \end{matrix}\right),
\end{displaymath}
the Hamiltonian can be recast into the following form:
\begin{displaymath}
\hat{H}_{\rm sim}=
\Om \sigma_+\sigma_-
+\mu B_1^{\dagger}B_1
+\left[\sqrt{\gamma_1^2+\gamma_2^2} \ \sigma_+ B_1+{\rm H.c.}\right]
+\mu B_2^{\dagger}B_2.
\end{displaymath}
Here, only $B_1$ interacts with the atom 
with the renormalized coupling constant 
$\sqrt{\gamma_1^2+\gamma_2^2}$, 
while $B_2$ is decoupled from the atom.
We call $B_1$ the {\em interaction mode}~\cite{Toyo,Kaya}.

Now we return to the original Hamiltonian $\hat{H}_S$, Eq.~(\ref{eq:H1}).
% shmz>
We define the interaction mode at energy $\mu$ by
\beq
B_{\mu}= g_{\mu}^{-1}\int d\veck \
\delta(\eps_{\veck}-\mu) \ g_{\veck} \ b_{\veck},
% \verb#{eq:b_mu}#
\label{eq:b_mu}
\eeq
where $g_{\mu}$ is the {\em form factor} of the interaction, 
which is defined by\footnote{
The phase of $g_{\mu}$ can be taken arbitrary. 
For example, one can take it as $g_{\mu} \geq 0$.
}
\beq
|g_{\mu}|^2=\int d\veck |g_{\veck}|^2 \delta(\eps_{\veck}-\mu).
% \verb#{eq:g2}#
\label{eq:g2}
\eeq
% \bea
% B_{\mu}&=& g_{\mu}^{-1}\int d\veck 
% \delta(\eps_{\veck}-\mu)g_{\veck}b_{\veck},
% % \verb#{eq:b_mu}#
% \label{eq:b_mu}
% \\
% |g_{\mu}|^2&=&\int d\veck |g_{\veck}|^2 \delta(\eps_{\veck}-\mu),
% % \verb#{eq:g2}#
% \label{eq:g2}
% \eea
Here, $g_{\mu}$ has been determined so as to normalize $B_{\mu}$
as $[B_{\mu}, B_{\mu'}^{\dagger}]=\delta(\mu-\mu')$.
Using these quantities, the Hamiltonian Eq.~(\ref{eq:H1}) 
is transformed into the following form: 
\beq
\hat{H}_S = \Om\sigma_+\sigma_- + 
\int \rmd\mu \ \left[(g_{\mu}\sigma_+ B_{\mu}+{\rm H.c.})
+\mu B_{\mu}^{\dagger}B_{\mu}\right] + \hat{H}_{\rm rest},
% \verb#{eq:H1'}#
\label{eq:H1'}
\eeq
where $\hat{H}_{\rm rest}$ consists of modes that do not interact with the atom.
The schematic energy diagram after this transformation 
is shown in Fig.~\ref{fig:enedi}(b). Due to our initial condition, 
the existence of $\hat{H}_{\rm rest}$ does not affect the decay dynamics 
at all.

% koshino <
In Eq.~(\ref{eq:H1'}),
the atom is coupled to a single continuum $B_{\mu}$.
This type of Hamiltonian is called the Friedrichs model~\cite{Fri}.
In this model,
% >koshino 
the dynamics is determined solely by the functional 
form of $|g_{\mu}|^2$.
% which is called the {\it form factor} of interaction.
For example, if we apply the Fermi golden rule~\cite{FGR}, 
the radiative decay rate of the atom is given by 
\beq
\Gamma_{\rm FGR}=2\pi|g_{\Om}|^2
=2\pi\int d\veck |g_{\veck}|^2 \delta(\eps_{\veck}-\Om).
% \verb#{eq:FGR-0}#
\label{eq:FGR-0}
\eeq
In the following parts of this subsection,
% shmz<
concrete forms of the form factor are presented
for three realistic cases.

%%%%%%%%%%%%%%%%%%%%%%%%%%%%%%%%%%%%%%%%%%%%%%%%%%%%%%%%%%%%%%%%%%%%%%%%%%%%%%
\subsubsection{Free space}
% \verb#{sssec:freespace}#
\label{sssec:freespace}
%%%%%%%%%%%%%%%%%%%%%%%%%%%%%%%%%%%%%%%%%%%%%%%%%%%%%%%%%%%%%%%%%%%%%%%%%%%%%%

Firstly, we consider a case where an atom is placed 
in a free space~\cite{Mandel}.
The form factor is dependent on the dimension of the space.
Here, we discuss the three-dimensional case as an example.
Imposing the periodic boundary condition 
with the quantization length $L$, and 
reviving here the index $\lambda$ representing the photonic polarization,
the eigenmodes and eigenenergies are given by
\bea
\vecf_{\veck\lambda}(\vecr)&=&L^{-3/2}e^{\rmi\veck\cdot\vecr}
\bm{e}_{\veck\lambda},
\\
\eps_{\veck\lambda}&=&|\veck|,
\eea
where $\bm{e}_{\veck\lambda}$ is a unit vector 
in the direction of polarization,
which is normal to the wavevector $\veck$.
$\veck$ is discretized as $\veck=2\pi/L\times(n_x, n_y, n_z)$,
where $n_{x,y,z}=0, \pm1, \pm2, \cdots$.
The atom-photon coupling $g_{\veck\lambda}$ is given by
\beq
g_{\veck\lambda} = -\frac{e}{m}
\sqrt{\frac{2\pi}{\eps_{\veck\lambda}}}
\langle\rmx|\vecp\cdot\vecf_{\veck\lambda}(\vecr)|\rmg\rangle,
% \verb#{eq:apc1}#
\label{eq:apc1}
\eeq
where $m$, $e$, $\vecr$, and $\vecp$ are
the mass, charge, position, momentum of the electron in the atom.
If the $\vecr$-dependence of $\vecf_{\veck\lambda}(\vecr)$
within the atom is negligible (dipole approximation),
we obtain 
\beq
g_{\veck\lambda} = 
-\rmi\Om\sqrt{\frac{2\pi}{\eps_{\veck\lambda}}}
\bm{\mu}_{atom}\cdot \vecf_{\veck\lambda}(\vecr),
% \verb#{eq:apc2}#
\label{eq:apc2}
\eeq
where $\bm{\mu}_{atom}=e\langle\rmx|\vecr|\rmg\rangle$
is the transition dipole moment of the atom.

It should be noted that the atom is coupled 
only to photons within a finite energy range.
The lower-bound originates in the positiveness of the photonic energy.
The higher-cutoff $\om_c$ is introduced by the fact that
$\langle\rmx|\vecp\cdot\vecf_{\veck \lambda}(\vecr)|\rmg\rangle$
almost vanishes when $|\veck|$ is large,
due to rapid oscillation of $\vecf_{\veck \lambda}(\vecr)$.

\begin{figure}%----------------------------------------------------------------
\bec
\includegraphics{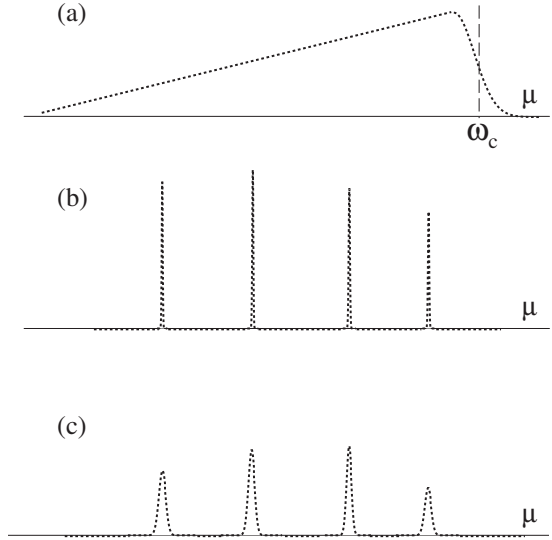}
\caption{\label{fig:ff3shu}
Schematic view of the form factors in  
(a) a three-dimensional free space,
(b) a perfect cavity, and
(c) a leaky cavity.
}% \verb#{fig:ff3shu}#
\enc
\end{figure}

Now we determine the form factor, using Eq.~(\ref{eq:apc2}).
The form factor is given, using the formula Eq.~(\ref{eq:g2}), by
\beq
|g_{\mu}|^2= \sum_{\veck,\lambda} 
|g_{\veck\lambda}|^2 \delta(\eps_{\veck}-\mu).
% \verb#{eq:g2_2}#
\label{eq:g2_2}
\eeq
Taking the $L \to \infty$ limit and replacing 
the summation over $\veck$ with the integral,
we obtain the following form factor:
\beq
|g_{\mu}|^2 = 
\begin{cases}
\frac{2\Om^2|\bm{\mu}_{atom}|^2 \mu}{3\pi} & (\mu \lesssim \om_c)
\\
0 & (\mu \gtrsim \om_c)
\end{cases}.
\eeq
Thus, the form factor has a continuous spectrum in a free space,
as shown in Fig.~\ref{fig:ff3shu}(a).

%-----------------------------------------------------------------

%%%%%%%%%%%%%%%%%%%%%%%%%%%%%%%%%%%%%%%%%%%%%%%%%%%%%%%%%%%%%%%%%%%%%%%%%%%%%%
\subsubsection{Perfect cavity}  % shmz
% \verb#{sssec:cc}#
\label{sssec:cc}
%%%%%%%%%%%%%%%%%%%%%%%%%%%%%%%%%%%%%%%%%%%%%%%%%%%%%%%%%%%%%%%%%%%%%%%%%%%%%%

Next, we discuss a case where the atom is placed in a perfect cavity,
whose eigenmodes do not suffer attenuation at all.
As a model of such a perfect cavity,
we consider a photon field bounded by perfect mirrors
placed at $x=0$ and $l_x$, $y=0$ and $l_y$, $z=0$ and $l_z$.
The eigenmodes and eigenenergies are given by
\bea
\vecf_{\veck\lambda}(\vecr)&=&
\sqrt{\frac{8}{l_x l_y l_z}}
% \sqrt{\frac{8}{L^3}}
\sin(k_x x)\sin(k_y y)\sin(k_z z)\bm{e}_{\veck\lambda},
\\
\eps_{\veck\lambda}&=&|\veck|,
\eea
where $\veck=(n_x\pi/l_x, n_y\pi/l_y, n_z\pi/l_z)$
with $n_{x,y,z}=1,2,\cdots$.
The atom-photon coupling constant and the form factor are
determined by Eqs.~(\ref{eq:apc1}) and (\ref{eq:g2_2}), respectively.

A distinct difference from the free-space case is that
the photonic modes are discretized.
In the present case,
the summation over $\veck$ cannot be replaced with the integral, 
and the form factor is composed 
of delta functions located at eigenenergies, %shmz
as shown in Fig.~\ref{fig:ff3shu}(b).
The energy separation becomes larger 
as the cavity lengths ($l_x, l_y, l_z$) are decreased.
By using a small cavity, one may realize a situation 
where the atom effectively interacts 
% shmz>
only with a single eigenmode of the cavity. 
Then, denoting the annihilation operator 
of that eigenmode by $a$,
the Hamiltonian of the whole system reads
\beq
\hat{H}_{\rm pc}=\Om \sigma_+\sigma_-  + (g \sigma_+ a + {\rm H.c.})
+ \om_0 a^{\dagger}a,
\label{eq:atom_cav}
\eeq
where $\om_0$ is the energy of the eigenmode,
and $g$ is the coupling constant between the atom and the eigenmode.

%%%%%%%%%%%%%%%%%%%%%%%%%%%%%%%%%%%%%%%%%%%%%%%%%%%%%%%%%%%%%%%%%%%%%%%%%%%%%%
\subsubsection{Leaky cavity}
% \verb#{ssec:LC}#
\label{ssec:LC}
%%%%%%%%%%%%%%%%%%%%%%%%%%%%%%%%%%%%%%%%%%%%%%%%%%%%%%%%%%%%%%%%%%%%%%%%%%%%%%

In usual optical cavities,
in order that one can input photons into the cavity 
from external photon modes,
one (or more) mirror composing the cavity should be weakly transmissive.
In this case, photons inside the cavity gradually escape
into external modes through the transmissive mirror,
i.e., the cavity is leaky.
Considering for simplicity a case 
where the atom effectively interacts with a single cavity mode,
the Hamiltonian for the whole system is given,
extending Eq.~(\ref{eq:atom_cav}), by
\beq
\hat{H}_{\rm lc} = \Om \sigma_+\sigma_- + (g \sigma_+ a + {\rm H.c.}) 
+ \om_0 a^{\dagger}a
+\int d\om \left[ 
\sqrt{\frac{\kappa}{2\pi}}(a^{\dagger}b_{\om}+b_{\om}^{\dagger}a)
+ \om b_{\om}^{\dagger}b_{\om} \right], 
\label{eq:atom_cav2}
\eeq
where $b_{\om}$ denotes the annihilation operator 
for the external photon mode with energy $\om$~\cite{Collett}.
The lifetime of the cavity mode is given by $\kappa^{-1}$,
and the Q-value of the cavity is given by $\om_0/\kappa$.

It is straightforward to derive the form factor 
from the above Hamiltonian,
by diagonalizing the interaction part of the Hamiltonian
between the cavity mode $a$ and the external modes $b_{\om}$.
To this end, 
hereafter denoting an infinitesimal positive constant by $\delta$,
we define the following operator~\cite{Fano},
\beq
B_{\mu} = \alp(\mu) a +
\int d\om \beta(\mu,\om) b_{\om},
\eeq
where
\bea
\alp(\mu) &=& \frac{(\kappa/2\pi)^{1/2}}
{\mu-\om_0+\rmi \kappa/2},
\\
\beta(\mu,\om) &=& \frac{\kappa/2\pi}
{(\mu-\om_0+\rmi \kappa/2)(\mu-\om+\rmi\delta)}
+\delta(\mu-\om).
\eea
Note that $B_{\mu}$ is orthonormalized as
$[B_{\mu}, B_{\mu'}^{\dagger}]=\delta(\mu-\mu')$.
The original operators, $a$ and $b_{\om}$,
are given in terms of $B_{\mu}$ by
\bea
a &=& \int d\mu \alp^{\ast}(\mu) B_{\mu},
\\
b_{\om} &=& \int d\mu \beta^{\ast}(\om,\mu) B_{\mu}.
\eea
Using $B_{\mu}$, Eq.~(\ref{eq:atom_cav2}) is rewritten as 
\beq
\hat{H}_{\rm lc} = \Om\sigma_+\sigma_- + 
\int d\mu  \left[(g\alp^{\ast}(\mu)\sigma_+B_{\mu}+H.c.)+
\mu B_{\mu}^{\dagger}B_{\mu} \right],
\eeq
in which the atom is coupled to 
a single continuum of $B_{\mu}$
and the energy diagram of Fig.~\ref{fig:enedi}(b) is realized.
Therefore, the form factor takes the following Lorentzian form:
\beq
|g_{\mu}|^2 =  |g\alp^{\ast}(\mu)|^2
= g^2\frac{\kappa/2\pi}{(\mu-\om_0)^2+(\kappa/2)^2},
\eeq
which satisfies the following sum rule:
\beq
\int d\mu |g_{\mu}|^2 = g^2,
\label{eq:Gsr}
\eeq
which holds for any $\kappa$.

To summarize,
when the cavity is perfect and has no leak,
the form factor % of the cavity mode 
is composed of delta functions,
as shown in Fig.~\ref{fig:ff3shu}(b).
Contrarily, when the cavity is leaky, 
% and the cavity mode has a finite lifetime $\kappa^{-1}$,
each delta function is broadened to be a Lorentzian,
keeping the sum rule of Eq.~(\ref{eq:Gsr}),
as shown in Fig.~\ref{fig:ff3shu}(c).

%%%%%%%%%%%%%%%%%%%%%%%%%%%%%%%%%%%%%%%%%%%%%%%%%%%%%%%%%%%%%%%%%%%%%%%%%%%%%%
\subsection{Perturbation theory}
% \verb#{sec:Pt}#
\label{sec:Pt}
%%%%%%%%%%%%%%%%%%%%%%%%%%%%%%%%%%%%%%%%%%%%%%%%%%%%%%%%%%%%%%%%%%%%%%%%%%%%%%

In the following part of Sec.~\ref{sec:fpus},
we investigate how the initial unstable state evolves
in time by the Schr\"{o}dinger equation, Eq.~(\ref{eq:Scheq}).
Here, we calculate the survival probability {\it etc} 
by an elementary perturbation theory.
For this purpose, 
we divide the Hamiltonian Eq.~(\ref{eq:H1})
into the diagonal and interaction parts as 
\bea
\hat{H}_S &=& \hat{H}_0+\hat{H}_1, \\
\hat{H}_0 &=& \Om\sigma_+\sigma_- 
+\int \rmd\veck \eps_{\veck}b_{\veck}^{\dagger}b_{\veck}, \\
\hat{H}_1 &=& 
\int \rmd\veck \ (g_{\veck}\sigma_+ b_{\veck}+{\rm H.c.}).
\eea
It is easy to derive the following perturbative expansion
for the evolution operator $\exp(-\rmi \hat{H}_S t)$:
\beq
\exp(-\rmi \hat{H}_S t)=\exp(-\rmi \hat{H}_0 t)
\left[\hat{1}+(-\rmi) \int_0^t dt'\hat{H}_1(t')
+(-\rmi)^2 \int_0^t dt' \int_0^{t'} dt'' \hat{H}_1(t')\hat{H}_1(t'')+\cdots
\right],
% \verb#{eq:pH}#
\label{eq:pH}
\eeq
where $\hat{H}_1(t)$ is the interaction representation of $\hat{H}_1$,
which is given by
\beq
\hat{H}_1(t)=e^{\rmi \hat{H}_0 t} \hat{H}_1 e^{-\rmi \hat{H}_0 t}
=\int \rmd\veck \ (g_{\veck}\sigma_+ b_{\veck}e^{-\rmi(\eps_{\veck}-\Om)t}
+{\rm H.c.}).
\eeq

Now we calculate the decay amplitude $f_{\veck}(t)$
defined in Eq.~(\ref{eq:deffk(t)})
within the lowest-order perturbation,
where the second-order and higher powers of $\hat{H}_1$
are neglected in Eq.~(\ref{eq:pH}).
Then, $f_{\veck}(t)$ is reduced to the following form:
\beq
f_{\veck}(t) \simeq -\rmi\langle 0| b_{\veck} 
e^{\rmi \hat{H}_0 t} \int_0^t dt' \hat{H}_1(t') 
\sigma_+ |0\rangle
=-\rmi e^{-\rmi (\Om+\eps_{\veck})t/2} \ g_{\veck}^{\ast} 
\ t \ {\rm sinc} [(\eps_{\veck}-\Om)t/2].
\eeq
The decay probability to the photon $\veck$ is given by
$|f_{\veck}(t)|^2$.
The survival probability is therefore given by
\bea
s(t)&=&1-\int d\veck |f_{\veck}(t)|^2 
= 1-t^2 \int d\veck |g_{\veck}|^2 {\rm sinc}^2 [(\eps_{\veck}-\Om)t/2]\\
&=&1-t^2\int d\mu |g_{\mu}|^2 {\rm sinc}^2 [(\mu-\Om)t/2].
% \verb#{eq:spert}#
\label{eq:spert}
\eea
In deriving the last equality,
Eq.~(\ref{eq:g2}) has been used.
% From the final expression of the survival probability,
% it is confirmed that the form factor $g_{\mu}$
% completely determines the survival probability.
Because Eq.~(\ref{eq:spert}) is based on the perturbation theory,
it is valid only for small $t$;
deviation from an exact result becomes significant
for large $t$, as we will see in Fig.~\ref{fig:s(t)}.
However, Eq.~(\ref{eq:spert}) serves as a convenient tool
as long as the short-time behavior is concerned,
such as discussion of the QZE and AZE.

Relation to the initial quadratic decay law, Eq.~(\ref{eq:st2}),
is easily observed.
At the very beginning of decay
(more strictly, when $t^{-1}$ is much larger than
the spectral width of $|g_{\mu}|^2$),
one can regard that 
${\rm sinc} [(\eps_{\veck}-\Om)t/2] \simeq 1$,
so the right-hand-side of Eq.~(\ref{eq:spert}) is approximated as
$s(t) \simeq 1-t^2\int d\mu |g_{\mu}|^2 = 
1-\langle(\Delta \hat{H}_S)^2\rangle t^2$.

%%%%%%%%%%%%%%%%%%%%%%%%%%%%%%%%%%%%%%%%%%%%%%%%%%%%%%%%%%%%%%%%%%%%%%%%%%%%%%
\subsection{Green function method}
% \verb#{sec:Gfm}#
\label{sec:Gfm}
%%%%%%%%%%%%%%%%%%%%%%%%%%%%%%%%%%%%%%%%%%%%%%%%%%%%%%%%%%%%%%%%%%%%%%%%%%%%%%

\begin{figure}%----------------------------------------------------------------
\bec
\includegraphics{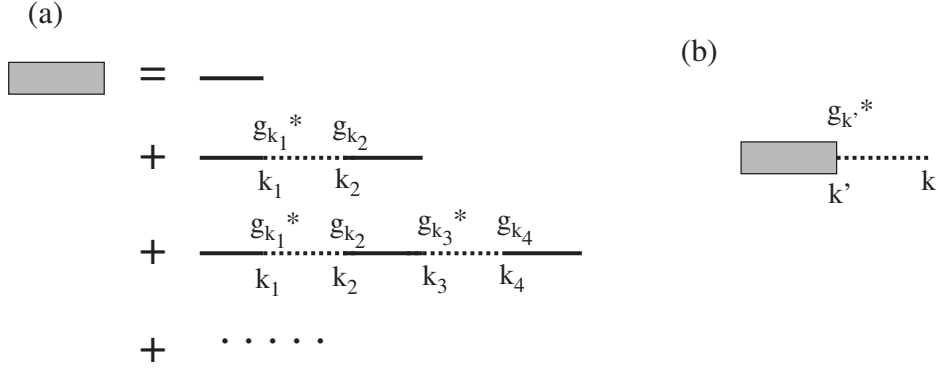}
\caption{\label{fig:Feyn}
(a) Feynman diagrams for the dressed atomic Green function.
The solid and dotted lines on the right hand side
represent the bare atom and photon Green functions
$A(\om)$ and $P(\om,\veck)$, respectively, 
while the bold line on the left hand side represents the dressed Green function.
(b) Feynman diagram representing the decay to photon $\veck$.}
\enc
\end{figure}%-----------------------------------------------------------------

In the previous subsection, 
the temporal evolution of unstable system
is investigated by the perturbation theory,
which is valid only for small $t$ in principle.
Here, we summarize the Green function method~\cite{FW,Mahan},
which is standardly used in calculating the temporal evolution 
of general quantum systems
and gives reliable results even for long $t$.

First, we define the {\it bare} atomic and photon Green functions
in the frequency representation. They are given,
in terms of the diagonal Hamiltonian $\hat{H}_0$, by
\bea
A(\om)&=&
\langle 0|\sigma_- \ 
\frac{1}{\om-\hat{H}_0 +\rmi\delta} \ 
\sigma_+|0\rangle
=\frac{1}{\om-\Om +\rmi\delta}, \\
P(\om, \veck, \veck')&=&\langle 0|b_{\veck'} \
\frac{1}{\om-\hat{H}_0 +\rmi\delta} \
b_{\veck}^{\dagger}|0\rangle
=\frac{\delta(\veck-\veck')}{\om-\eps_{\veck} +\rmi\delta}.
\eea
These bare Green functions are related,
through the Fourier transformation,
to the non-interacting dynamics of the atom and the photons.
For example,
\beq
\frac{\rmi}{2\pi}\int d\om e^{-\rmi\om t}A(\om)
=\langle 0|\sigma_- \ e^{-\rmi \hat{H}_0 t} \sigma_+|0\rangle =e^{-\rmi\Om t}.
\eeq
Similarly, we define the {\it dressed} atomic Green function by
\beq
\bar{A}(\om) = \langle 0|\sigma_- \ 
\frac{1}{\om - \hat{H}_S +\rmi\delta} \ 
\sigma_+|0\rangle.
\eeq
The dressed atomic Green function is related 
to the survival amplitude of the atom $f(t)$ by
\beq
\frac{\rmi}{2\pi}\int d\om e^{-\rmi\om t}\bar{A}(\om)=
\langle 0|\sigma_- \ e^{-\rmi \hat{H}_S t} \sigma_+|0\rangle
\equiv f(t).
\label{eq:f(t)byGreen}\eeq

It is known that the dressed atomic Green function
can be expanded in terms of the bare Green functions as follows,
\bea
\bar{A}(\om) &=& A(\om) + A(\om)\Sigma(\om)A(\om) 
+ A(\om)\Sigma(\om)A(\om)\Sigma(\om)A(\om) + \cdots
\nonumber
\\
&=& \frac{A(\om)}{1-A(\om)\Sigma(\om)},
% \verb#{eq:expansion}#
\label{eq:expansion}
\eea
where the self-energy $\Sigma(\om)$ is given by % shmz
\beq
\Sigma(\om)=\int \int d\veck_1 d\veck_2 
g_{\veck_1}^{\ast} g_{\veck_2} P(\om, \veck_1, \veck_2)
=\int d\veck \frac{|g_{\veck}|^2}{\om-\eps_{\veck}+\rmi\delta}. 
=\int d\mu \frac{|g_{\mu}|^2}{\om-\mu+\rmi\delta}. 
% \verb#{eq:slfene}#
\label{eq:slfene}
\eeq 
The Feynman diagrams corresponding to Eq.~(\ref{eq:expansion})
are drawn in Fig.~\ref{fig:Feyn}(a).
Equation (\ref{eq:g2}) is used 
in deriving the third equality of Eq.~(\ref{eq:slfene}).
We can reconfirm by Eqs.~(\ref{eq:f(t)byGreen})-(\ref{eq:slfene}) %shmz
that the decay dynamics of the atom 
is determined completely by the form factor $g_{\mu}$.

We also note that, besides the survival amplitude, 
the decay amplitude to a photon of a specific mode $\veck$
can be obtained using the dressed atomic Green function as
\beq
\langle 0| b_{\veck} \ 
\frac{1}{\om-\hat{H}_S +\rmi\delta} \ 
\sigma_+|0\rangle
=\int d\veck' g_{\veck'}^{\ast} \bar{A}(\om) P(\om, \veck', \veck)
=\frac{g_{\veck}^{\ast}\bar{A}(\om)}{\om-\eps_{\veck}+\rmi\delta}.
% \verb#{eq:a_to_p}#
\label{eq:a_to_p}
\eeq
The Feynman diagram corresponding to this amplitude
is drawn in Fig.~\ref{fig:Feyn}(b).

%%%%%%%%%%%%%%%%%%%%%%%%%%%%%%%%%%%%%%%%%%%%%%%%%%%%%%%%%%%%%%%%%%%%%%%%%%%%%%
\subsection{Decay dynamics under the Lorentzian form factor}
% \verb#{sec:Lff}#
\label{sec:Lff}
%%%%%%%%%%%%%%%%%%%%%%%%%%%%%%%%%%%%%%%%%%%%%%%%%%%%%%%%%%%%%%%%%%%%%%%%%%%%%%

\begin{figure}%----------------------------------------------------------------
\bec
\includegraphics{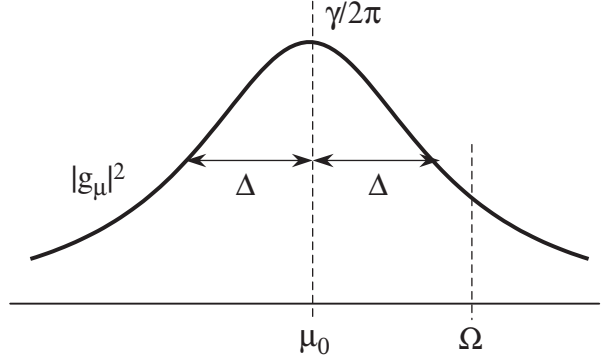}
\caption{\label{fig:Lor}
View of the Lorentzian form factor.
The atomic transition energy $\Om$ is arbitrary.}
% \verb#{fig:Lor}#
\enc
\end{figure}

The Green function method is applicable 
to any forms of $\eps_{\veck}$ and $g_{\veck}$.
In this section, we practically use the Green function method
to calculate the survival probability of an unstable state.
Here, we discuss the case where the form factor 
is given by a single Lorentzian as follows;
% mainly because most quantities can 
% be analytically calculated in this case.
% The Lorentzian form factor takes the following form:
\beq
|g_{\mu}|^2=\frac{\gamma}{2\pi}\frac{\Delta^2}{(\mu-\mu_0)^2+\Delta^2}.
% \verb#{eq:Lff}#
\label{eq:Lff}
\eeq
Here, $\mu_0$ and $\Delta$ denote the central energy 
and the spectral width of the form factor, respectively, 
% shmz>
and $\gamma/2\pi = |g_{\mu_0}|^2$ characterizes
the magnitude of the form factor (see Fig.~\ref{fig:Lor}).
% shmz<
% strength of the atom-photon coupling (see Fig.~\ref{fig:Lor}).
% This form factor has both lower- and higher-cutoff energies,
% which is required in general quantum unstable systems.

Such a Lorentzian form factor is realized, for example, 
by an atom placed in a leaky optical cavity,
as has been shown in Sec.~\ref{ssec:LC}. 
Although the form factors of general unstable systems 
are not necessarily approximated by Lorentzian,
it is expected that qualitative features of quantum dynamics 
are inferable by considering the case of a Lorentzian form factor,
as long as the form factor is single-peaked.
For example, the decay dynamics in a free space,
whose form factor is schematically shown in Fig.~\ref{fig:ff3shu}(a),
would be qualitatively reproducible by the Lorentzian form factor
if we take $\Om \ll \mu_0$.
%
% the asymmetry against the atomic transition energy
% may be incorporated into the model by adequate choice of $\Om-\mu_0$.

%-----------------------------------------------------------------

% shmz>
\subsubsection{Exact formulas}

\begin{figure}%----------------------------------------------------------------
\bec
\includegraphics{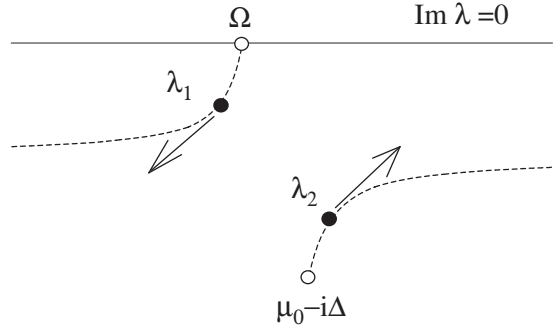}
\caption{\label{fig:lam12}
Solutions  
$\lambda_1$ and $\lambda_2$ of Eq.~(\ref{eq:lam_2ji})
are plotted in the complex $\lambda$-plane.
Dotted curves show the trajectories when $\gamma$ is changed,
and arrows indicate the directions 
into which $\lambda$'s move as $\gamma$ is increased.
}
% \verb#{fig:lam12}#
\enc
\end{figure}

In the case of the Lorentzian form factor, 
we can exactly calculate the self-energy $\Sigma(\om)$, 
the dressed Green function $\bar{A}(\om)$,
and the survival amplitude 
$f(t)=\langle{\rm i}|\exp(-\rmi{\cal H}t)|{\rm i}\rangle$
as follows:
% shmz<
\bea
\Sigma(\om) &=& \frac{\gamma \Delta}{2(\om-\mu_0+\rmi\Delta)},
\\
\bar{A}(\om) &=& \frac{\om-\mu_0+\rmi\Delta}
{(\om-\Om)(\om-\mu_0+\rmi\Delta)-\gamma\Delta/2},
\\
f(t)&=&\frac{\lambda_1-\mu_0+\rmi\Delta}{\lambda_1-\lambda_2}
\exp(-\rmi\lambda_1 t)+
\frac{\lambda_2-\mu_0+\rmi\Delta}{\lambda_2-\lambda_1}
\exp(-\rmi\lambda_2 t).
% \verb#{eq:f(t)}#
\label{eq:f(t)}
\eea
Here, $\lambda_1$ and $\lambda_2$ are the poles 
of the dressed Green function $\bar{A}(\om)$, and satisfy
\beq
(\om-\Om)(\om-\mu_0+\rmi\Delta)-\gamma\Delta/2
=(\om-\lambda_1)(\om-\lambda_2).
% \verb#{eq:lam_2ji}#
\label{eq:lam_2ji}
\eeq
Both of them lie in the lower half plane as shown in Fig.~\ref{fig:lam12},
and we choose them % $\lambda_1$ and $\lambda_2$ 
to satisfy $|Im(\lambda_1)|\leq |Im(\lambda_2)|$.
% shmz>
The survival probability $s(t)$ is exactly given by % squaring $|f(t)|$ as
\beq
s(t) = |f(t)|^2=
\left|
\frac{\lambda_1-\mu_0+\rmi\Delta}{\lambda_1-\lambda_2}
\exp(-\rmi\lambda_1 t)+
\frac{\lambda_2-\mu_0+\rmi\Delta}{\lambda_2-\lambda_1}
\exp(-\rmi\lambda_2 t)\right|^2.
% \verb#{eq:s(t)}#
\label{eq:s(t)}
\eeq
On the other hand, the Fermi golden rule, Eq.~(\ref{eq:FGR-0}), yields
the approximate decay rate as
\beq
\Gamma_{\rm FGR}=
\gamma\frac{\Delta^2}{\Delta^2+(\Om-\mu_0)^2}.
% \verb#{eq:FGR}#
\label{eq:FGR}
\eeq
We will compare $\Gamma_{\rm FGR}$ 
with the rigorous result Eq.~(\ref{eq:s(t)})
in the latter part of Sec.~\ref{sec:Lff} (Fig.~\ref{fig:decrate}).

%-----------------------------------------------------------------

%%%%%%%%%%%%%%%%%%%%%%%%%%%%%%%%%%%%%%%%%%%%%%%%%%%%%%%%%%%%%%%%%%%%%%%%%%%%%%
\subsubsection{Symmetric case}
\label{sec:simcase}
%%%%%%%%%%%%%%%%%%%%%%%%%%%%%%%%%%%%%%%%%%%%%%%%%%%%%%%%%%%%%%%%%%%%%%%%%%%%%%

\begin{figure}%----------------------------------------------------------------
\bec
\includegraphics{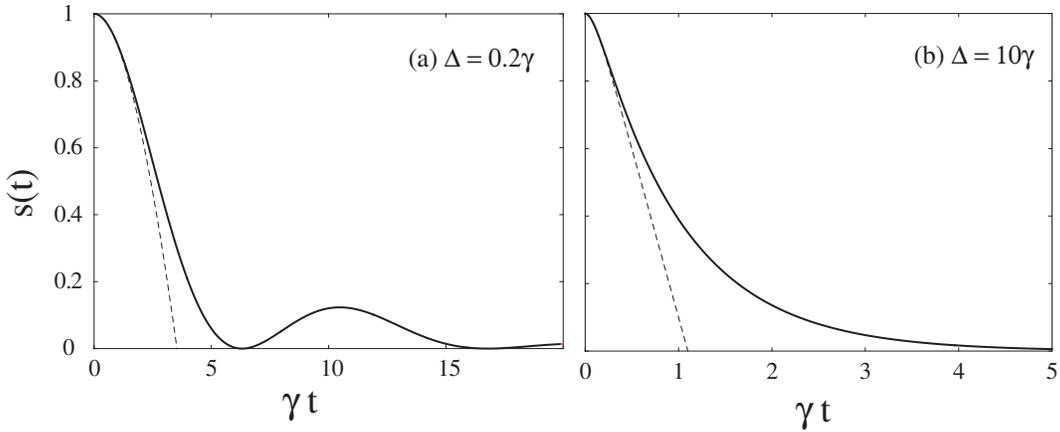}
\caption{\label{fig:s(t)}
Temporal evolution of the survival probability $s(t)$, for $\mu_0-\Om=0$.
$\Delta=0.2\gamma$ in (a), and $\Delta=10\gamma$ in (b).
The solid lines are drawn by the exact formula, Eq.~(\ref{eq:s(t)}).
The thin dotted lines are the results of perturbation:
combining Eqs.~(\ref{eq:spert}) and (\ref{eq:Lff}),
the survival probability is approximately given by
$s(t) \simeq 1-\gamma \Delta^2 |\mu_0-\Om+\rmi\Delta|^{-2} t
-\gamma \Delta {\rm Re} [(1-e^{\rmi(\mu_0-\Om+\rmi\Delta)t})
(\mu_0-\Om+\rmi\Delta)^{-2}]$.
}
% \verb#{fig:s(t)}#
\enc
\end{figure}%-----------------------------------------------------------------

When the form factor is symmetric about 
the atomic transition energy, i.e., $\mu_0=\Om$,
the above equations are particularly simplified;
\bea
\lambda_{1,2}&=&\Om-\rmi\Delta\frac{1\pm\sqrt{1-2\gamma/\Delta}}{2}, 
% \verb#{eq:lam12}#
\label{eq:lam12}
\\
f(t)&=&\frac{1+\sqrt{1-2\gamma/\Delta}}{2\sqrt{1-2\gamma/\Delta}}
\exp(-\rmi\lambda_1 t)-\frac{1-\sqrt{1-2\gamma/\Delta}}
{2\sqrt{1-2\gamma/\Delta}}\exp(-\rmi\lambda_2 t).
% \verb#{eq:f(t)2}#
\label{eq:f(t)2}
\eea
The temporal evolution of the survival probability $s(t)$
is plotted in Fig.~\ref{fig:s(t)}, 
for small and large $\Delta$.

First, we discuss a case of small $\Delta$ (satisfying $\Delta < 2\gamma$),
in which $Im(\lambda_1)=Im(\lambda_2)$
and $Re(\lambda_1)\neq Re(\lambda_2)$.
In this case, $s(t)$ shows a damped Rabi oscillation,
as shown in Fig.~\ref{fig:s(t)}(a), 
which implies that the emitted photon 
may be reabsorbed by the decayed atom.
This phenomenon, called the collapse and revival, 
has actually been observed in an atom 
in a high-Q cavity~\cite{c_and_r1,c_and_r2,c_and_r3}.
In the limit of $\Delta \to 0$ 
(and simultaneously $\gamma \to \infty$, keeping 
$\int d\mu |g_{\mu}|^2 = \gamma \Delta/2$ at a finite value),
the Rabi oscillation continues forever without damping.
Obviously, we cannot define the decay rate 
in the presence of the Rabi oscillation. % shmz

Next, we discuss a case of large $\Delta$.
In this case, $s(t)$ decreases monotonously as shown in Fig.~\ref{fig:s(t)}(b).
The radiative decay of an atom in free space belongs to this case.
In order to clarify the meanings of the parameters $\Delta$ and $\gamma$,
we focus on a case of $\Delta\gg\gamma$ in the following.
Then, $\lambda_1$ and $\lambda_2$ are approximated by
\bea
\lambda_1 & \simeq & \Om-\rmi\gamma/2, \\
\lambda_2 & \simeq & \Om-\rmi\Delta.
\eea
At the beginning of the decay, one can easily confirm, 
by expanding Eq.~(\ref{eq:f(t)2}) in powers of $t$,
that $s(t)$ decreases quadratically as 
\beq
s(t)=1-\gamma\Delta t^2/2+{\cal O}(t^4).
% \verb#{eq:quad}# 
\label{eq:quad}
\eeq
Noticing that $\langle(\Delta\hat{H}_S)^2 \rangle=\gamma\Delta/2$
in our example, 
Eq.~(\ref{eq:quad}) is in accordance with Eq.~(\ref{eq:st2}).
On the other hand, in the later stage of the decay 
($t \gtrsim \Delta^{-1}$),
the second term of Eq.~(\ref{eq:f(t)2}) becomes negligible 
and $s(t)$ follows the exponential decay law as
\beq
s(t)  \simeq  {\cal Z} \exp(-\Gamma(\infty)t),
% \verb#{eq:expdlaw}#
\label{eq:expdlaw}
\eeq
where
\bea
{\cal Z}&=&\left|\frac{\lambda_1-\mu_0+\rmi\Delta}
{\lambda_1-\lambda_2}\right|^2, 
% \verb#{eq:defZ}#
\label{eq:defZ}
\\
\Gamma(\infty) &=& -2 Im (\lambda_1).
% \verb#{eq:rigrate}#
\label{eq:rigrate}
\eea
The decay rate in the later stage of decay 
($t \gtrsim \Delta^{-1}$)
is rigorously given by Eq.~(\ref{eq:rigrate}).
We confirm that the rigorous rate $\Gamma(\infty)$
agrees to the lowest order of $\gamma/\Delta$ with 
the golden-rule decay rate $\Gamma_{\rm FGR}$,
which is given by Eq.~(\ref{eq:FGR}).
Thus, $\Gamma_{\rm FGR}$ serves as a good approximation 
of $\Gamma(\infty)$, as long as $\Delta \gg\gamma$.

In concluding this subsection, 
we summarize the results for the case of $\Delta\gg\gamma$,
which is satisfied in most unstable states of interest.
At the beginning of decay $s(t)$ decreases quadratically 
obeying Eq.~(\ref{eq:quad}),
and later follows the exponential decay law, Eq.~(\ref{eq:expdlaw}).
The transition between these two behaviors occurs at 
\beq
t \sim \Delta^{-1} \equiv t_{\rm j},
\label{eq:def_jt}
\eeq
which is called the {\em jump time}
\footnote{
Note that definition of the jump time is 
slightly different from the original one:
In Ref.~\cite{jump}, 
the jump time is defined as 
$t_{\rm j} \equiv \Gamma(\infty)/\langle(\Delta \hat{H}_S)^2\rangle$,
which is reduced here to 
$t_{\rm j} \simeq 2\Delta/[\Delta^2+(\Om-\mu_0)^2]$.
}~\cite{jump}.
The decay rate $\Gamma(\infty)$ in the later stage
is approximated well by the golden-rule decay rate $\Gamma_{\rm FGR}$.

% shmz>
%%%%%%%%%%%%%%%%%%%%%%%%%%%%%%%%%%%%%%%%%%%%%%%%%%%%%%%%%%%%%%%%%%%%%%%%%%%%%%
\subsubsection{Asymmetric case}  
%%%%%%%%%%%%%%%%%%%%%%%%%%%%%%%%%%%%%%%%%%%%%%%%%%%%%%%%%%%%%%%%%%%%%%%%%%%%%%

\begin{figure}%----------------------------------------------------------------
\bec
\includegraphics{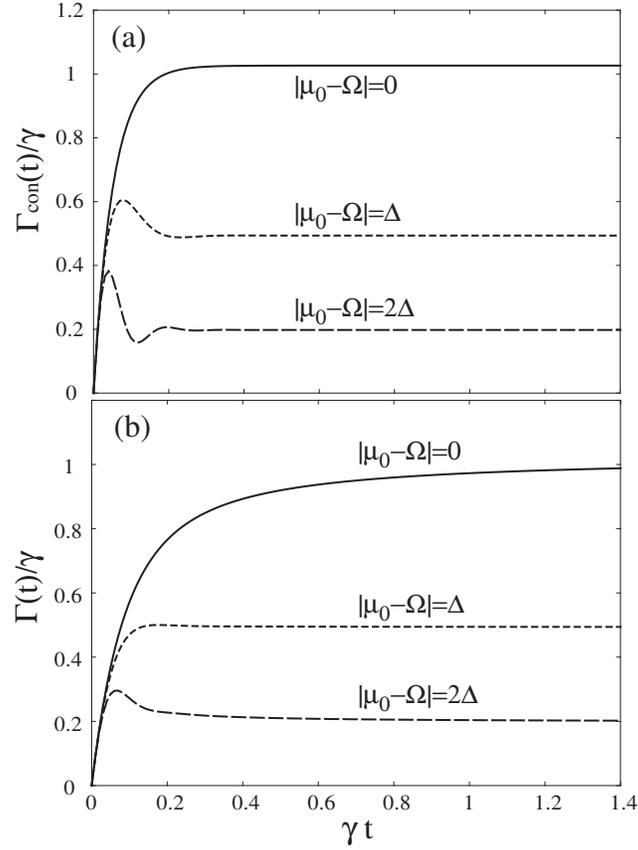}
\caption{\label{fig:decrate}
Temporal evolution of (a) $\Gamma_{\rm con}(t)$ 
and (b) $\Gamma(t)$.
We take $\Delta=20\gamma$ ($t_{\rm j} =0.05 \gamma^{-1}$), 
and $|\mu_0-\Om| = 0, \Delta, 2\Delta$.
The corresponding golden-rule decay rates, $\Gamma_{\rm FGR}=2\pi|g_{\Om}|^2$,
are $\gamma$, $0.5\gamma$, and $0.2\gamma$, respectively.
}
% \verb#{fig:decrate}#
\enc
\end{figure}%-----------------------------------------------------------------

Now we discuss the asymmetric case,
where $\mu_0$ is not necessarily equal to $\Om$.
The crossover between the damped Rabi oscillation 
and the monotonous decrease,
which was observed in Fig.~\ref{fig:s(t)},
is also observed in this case.
Here we focus on the latter situation assuming $\Delta \gg \gamma$,
which is usually satisfied in most of monotonically decaying systems.
% shmz<

Throughout this article,
much attentions are paid to the decay rate of an unstable quantum state.
The decay rate at time $t$ is conventionally defined by
\beq
\Gamma_{\rm con}(t) = 
-\left(\frac{ds}{dt}\right)/s = -\frac{d}{dt}\ln s(t).
\eeq
In Fig.~\ref{fig:decrate}(a), $\Gamma_{\rm con}(t)$ is plotted 
for three different values of $|\mu_0-\Om|$.
However, in the discussion of the Zeno effect, 
the following quantity is more significant:
\beq
\Gamma(t)=-\frac{\ln s(t)}{t}.
\eeq
In Sec.~\ref{sec:QZE}, it will be revealed that 
$\Gamma(\tau_{\rm i})$ gives the decay rate 
under repeated instantaneous ideal 
measurements with intervals $\tau_{\rm i}$.
$\Gamma(t)$ is plotted in Fig.~\ref{fig:decrate}(b). 
% with the same parameters as Fig.~\ref{fig:decrate}.
Comparing Figs.~\ref{fig:decrate}(a) and (b),
we find that the discrepancy between $\Gamma_{\rm con}(t)$ and $\Gamma(t)$ 
is significant in the early time stage, $t \lesssim t_{\rm j}$;
in particular, at the beginning of decay,
$\Gamma_{\rm con}(t)=2\Gamma(t)$.
However, the discrepancy becomes less significant as time evolves.

Focusing on $\Gamma(t)$, the following features 
are observed in common in three lines in Fig.~\ref{fig:decrate}(b):
Initially ($t \ll t_{\rm j}$), 
$s(t)$ is given by Eq.~(\ref{eq:quad}) regardless of $|\mu_0-\Om|$.
Therefore, $\Gamma(t)$ is approximately given by a linear function of $t$,
$\Gamma(t)=\gamma\Delta t$.
Sufficiently after the jump time ($t\gg t_{\rm j}$),
$\Gamma(t)$ approaches a constant value, $\Gamma(\infty)$,
which is given by Eq.~(\ref{eq:rigrate}).
As is observed in Fig.~\ref{fig:decrate},
$\Gamma(\infty)$ is in good agreement with 
the golden-rule decay rate, $\Gamma_{\rm FGR}$.

On the other hand, there is a remarkable qualitative difference
in the intermediate time region, $t\sim t_{\rm j}$.
For the case of $|\mu_0-\Om|=0$,
$\Gamma(t)$ is a monotonously increasing function of $t$
and $\Gamma(t) < \Gamma(\infty)$ for any $t$.
Contrarily, 
for the case of $|\mu_0-\Om|=2\Delta$,
$\Gamma(t)$ is not a monotonic function
and there exists a time region in which $\Gamma(t) > \Gamma(\infty)$.
This difference is crucial in determining whether
repeated measurements result in 
suppression of decay (QZE)
or enhancement of decay (AZE),
as will be discussed in Section \ref{sec:Z_AZ}.

%%%%%%%%%%%%%%%%%%%%%%%%%%%%%%%%%%%%%%%%%%%%%%%%%%%%%%%%%%%%%%%%%%%%%%%%%%%%%%
\section{Conventional theories of quantum Zeno and anti-Zeno effects}
% \verb#{sec:ct}#
\label{sec:ct}
%%%%%%%%%%%%%%%%%%%%%%%%%%%%%%%%%%%%%%%%%%%%%%%%%%%%%%%%%%%%%%%%%%%%%%%%%%%%%%

In this section, we summarize the main results of 
conventional theories
of the quantum Zeno and anti-Zeno effects,
where it is assumed that instantaneous and
ideal measurements to check the decay of the unstable state
are repeated frequently.

%%%%%%%%%%%%%%%%%%%%%%%%%%%%%%%%%%%%%%%%%%%%%%%%%%%%%%%%%%%%%%%%%%%%%%%%%%%%%%
\subsection{
% Description of quantum measurements by projection postulate
Ideal measurement on the target system
}
%%%%%%%%%%%%%%%%%%%%%%%%%%%%%%%%%%%%%%%%%%%%%%%%%%%%%%%%%%%%%%%%%%%%%%%%%%%%%%

In the previous section, we have reviewed 
the {\it free} unitary time evolution of an unstable quantum state, where
%in the case where
% by Sch\"{o}rodinger equation, 
% which is applicable while 
the system is not being measured. 
When one performs a measurement, on the other hand, 
the measurement accompanies considerable backaction 
on the measured system, according to quantum theory.
% shmz>
If the measurement is an ideal one, 
its influence on the quantum state of the measured system
is described as follows. 
% Here, we briefly summarize the conventional prescriptions
% for describing the influence of an ideal measurement
% on the quantum state of the measured system
%(see Sec.~\ref{sec:ideal_m} for more details).
%
Let us consider a situation in which 
one measures a physical quantity $Q$ of the system,
whose operator $\hat{Q}$ is assumed to have discrete eigenvalues.
The projection operator onto the subspace 
belonging to the eigenvalue $q$ of $\hat{Q}$ 
is denoted by $\hat{\cal P}(q)$,
and the state vector before the measurement 
is denoted by $|\psi \rangle$.
Then, the probability of obtaining $q$ as a measured value is given by
\beq
P(q)
= \| \hat{\cal P}(q)|\psi \rangle \|^2
=\langle \psi|\hat{\cal P}(q)|\psi \rangle,
% \verb#{eq:SPm}#
\label{eq:SPm}
\eeq
and the state vector just after the measurement is given by
\beq
|\psi_q \rangle = \frac{1}{\sqrt{P(q)}}\hat{\cal P}(q)|\psi\rangle,
% \verb#{eq:propos}#
\label{eq:propos}
\eeq
which is called the {\em projection postulate} of measurement~\cite{vN}.

Now we apply this prescription to a situation
in which an observer makes a measurement on the atom 
to check whether the atom has decayed or not.
In this case, the physical quantity to be measured is 
the number of excitations in the atom, $\sigma_+\sigma_-$.
The eigenvalues of the operator $\sigma_+\sigma_-$ are 1 and 0,
and the corresponding projection operators are given by
\bea
\hat{\cal P}(1) &=& |\rmx\rangle \langle\rmx| 
= \sigma_+|0\rangle \langle 0|\sigma_-,
% \verb#{eq:LP1}#
\label{eq:LP1}
\\
\hat{\cal P}(0) &=& \int d\veck|\rmg,\veck\rangle \langle \rmg,\veck| 
= \int d\veck b_{\veck}^{\dagger}|0\rangle \langle 0|b_{\veck}.
\eea 
If the state vector at $t=0$ is given by 
$|\rmi\rangle=|\rmx\rangle=\sigma_+|0\rangle$,
the state vector at time $t$ is given by Eq.~(\ref{eq:psit}).
The probability of observing the survival of the atom is given,
using Eqs.~(\ref{eq:SPm}) and (\ref{eq:LP1}), by
\beq
P(1)=|f(t)|^2=s(t),
% \verb#{eq:SP1}#
\label{eq:SP1}
\eeq
and the state vector just after this observation is given,
using Eqs.~(\ref{eq:propos}), (\ref{eq:LP1}), and (\ref{eq:SP1}), by
\beq
|\psi_1\rangle = \sigma_+|0\rangle = |\rmi\rangle,
\eeq
neglecting an irrelevant phase factor $f(t)/|f(t)|$.
Thus, when the survival of the atom is confirmed,
the state vector is reset to the initial one 
(the product of the atomic excited state and the photon vacuum)
as a backaction of the measurement. 

%%%%%%%%%%%%%%%%%%%%%%%%%%%%%%%%%%%%%%%%%%%%%%%%%%%%%%%%%%%%%%%%%%%%%%%%%%%%%%
\subsection{Decay rate under repeated 
% \ASi instantaneous ideal \ASo 
measurements}
% \verb#{sec:DRR}#
\label{sec:DRR}
%%%%%%%%%%%%%%%%%%%%%%%%%%%%%%%%%%%%%%%%%%%%%%%%%%%%%%%%%%%%%%%%%%%%%%%%%%%%%%

In the preceding subsection,
we have summarized the influence 
of a single ideal measurement on the atomic state.
We now investigate, using the projection postulate,
how the decay dynamics is affected
by repeated measurements to check the decay of the atom,
assuming that each measurement is instantaneous and ideal.

Suppose that instantaneous ideal measurements are 
performed periodically at
% To be more concrete, we discuss 
% how the decay dynamics is affected
% by a series of instantaneous ideal measurements at 
$t=j\tau_{\rm i}$ ($j=1,2,\cdots$),
where $\tau_{\rm i}$ is the intervals between measurements.
We hereafter denote the survival probability 
just after the $n$-th measurement by $S(t=n\tau_{\rm i})$.
This probability is identical to the probability 
of confirming survival of the atom in all measurements
($j=1, 2, \cdots, n$),
because, once the atom has decayed and emitted a photon, 
the revival probability %of reabsorption of the emitted photon 
is negligibly small in monotonically decaying systems.
Noticing that (i) if the atom is in the excited state at $t=0$,
the survival probability at $t=\tau_{\rm i}$ is given by $s(\tau_{\rm i})$,
and that 
(ii) the state is reset to the atomic excited state
after every confirmation of survival, we obtain 
$S(t=n\tau_{\rm i})$ simply as 
\beq
S(t=n\tau_{\rm i}) = [s(\tau_{\rm i})]^n = [s(\tau_{\rm i})]^{t/\tau_{\rm i}}.
\label{eq:S(t)}\eeq
Therefore, the decay rate $\Gamma$ under such 
repeated measurements 
is given, as a function of the measurement intervals $\tau_{\rm i}$, by
\beq
\Gamma(\tau_{\rm i}) = -\tau_{\rm i}^{-1} \ln s(\tau_{\rm i}).
% \verb#{eq:LGam}#
\label{eq:LGam}
\eeq
This equation clearly demonstrates that 
the decay rate depends 
not only on the original unitary dynamics of the system
[which determines $s(t)$]
but also on the measurement intervals $\tau_{\rm i}$.

\begin{figure}%----------------------------------------------------------------
\bec
\includegraphics{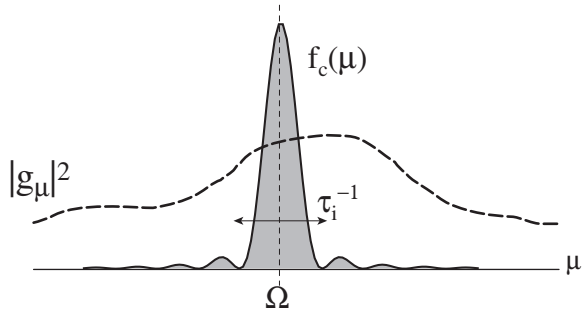}
\caption{\label{fig:drr}
Calculation of the decay rate $\Gamma(\tau_{\rmi})$
under repeated measurement.
$\Gamma(\tau_{\rmi})$ is given by integrating the form factor $|g_{\mu}|^2$
with a weight function $f_{\rm c}(\mu)$.
}
% \verb#{fig:drr}#
\enc
\end{figure}%-----------------------------------------------------------------

Throughout this article, our main concern is focused on the 
case % where measurements are frequently repeated, namely, the cases 
of short $\tau_{\rmi}$.
As we have observed in Sec.~\ref{sec:Pt},
the initial behavior of the survival probability $s(t)$ 
can be well evaluated by Eq.~(\ref{eq:spert}).
Using Eq.~(\ref{eq:spert}), 
Eq.~(\ref{eq:LGam}) is recast into the following form~\cite{KofNature}:
\bea
\Gamma(\tau_{\rm i}) &=& \int d\mu |g_{\mu}|^2 \times f_{\rm c}(\mu),
% \verb#{eq:DRR}#
\label{eq:DRR}
\\
f_{\rm c}(\mu) &=&
\tau_{\rmi} \ {\rm sinc}^2 \left[\frac{\tau_{\rmi}(\mu-\Om)}{2}\right].
% \verb#{eq:wei1}#
\label{eq:wei1}
\eea
Namely, the decay rate under repeated 
instantaneous ideal measurements
is given by integrating the form factor $|g_{\mu}|^2$
with a weight function $f_{\rm c}(\mu)$,
as illustrated in Fig.~\ref{fig:drr}.
The weight function $f_{\rm c}(\mu)$ has the following properties:
(i) $f_{\rm c}(\mu)$ is a positive function
centered at the atomic transition energy $\Om$
with a spectral width $\sim \tau_{\rmi}^{-1}$,
and (ii) $f_{\rm c}(\mu)$ is normalized as 
$\int d\mu f_{\rm c}(\mu) = 2\pi$.

As a reference, we first consider a situation where
the unstable state is not measured, namely,
$\tau_{\rmi} \to \infty$.
In this limit, 
the weight function is reduced to a delta function 
as $f_{\rm c}(\mu) \to 2\pi\delta(\mu-\Om)$,
so $\Gamma \to 2\pi |g_{\Om}|^2$.
This is nothing but the Fermi golden rule 
for an unobserved system, Eq.~(\ref{eq:FGR-0}).

Generally, $\Gamma(\tau_{\rmi})$ depends on 
the measurement intervals $\tau_{\rmi}$
through the width of the weight function $f_{\rm c}(\mu)$.
However, there exists a notable exception:
a system whose form factor is a constant function as
$|g_{\mu}|^2=\gamma/2\pi$.
It is known that such a system follows 
an exact exponential decay law, $s(t)=e^{-\gamma t}$, 
as can be seen by taking 
the $\Delta \to \infty$ limit of the results of Sec.~\ref{sec:Lff}.
For such a system,
the measurement-modified decay rate, Eq.~(\ref{eq:DRR}),
is always reduced to the {\it free} decay rate $\gamma$,
irrespective of $\tau_{\rmi}$.
Thus, % if we apply the projection postulate to the atomic state,
we are led to a well-known conclusion that 
the decay rate of exactly exponentially decaying systems 
are not affected by repeated measurements at all.
Note, however,  that we have assumed that each measurement is
instantaneous and ideal.
For general measurements, the above conclusion is {\it not} necessarily 
true, as will be shown in Sec.~\ref{sec:rmt}.

%%%%%%%%%%%%%%%%%%%%%%%%%%%%%%%%%%%%%%%%%%%%%%%%%%%%%%%%%%%%%%%%%%%%%%%%%%%%%%
\subsection{Quantum Zeno effect}
% \verb#{sec:QZE}#
\label{sec:QZE}
%%%%%%%%%%%%%%%%%%%%%%%%%%%%%%%%%%%%%%%%%%%%%%%%%%%%%%%%%%%%%%%%%%%%%%%%%%%%%%

In the preceding subsection,
it is observed that the decay rate $\Gamma$ is 
generally modified by repeated measurements, 
and $\Gamma$ is dependent on the measurement intervals $\tau_{\rm i}$.
A particularly interesting phenomenon is expected 
when $\tau_{\rm i}$ is extremely short: 
For very small $t$, 
the behavior of $s(t)$ is described by the quadratic decay law,
Eq.~(\ref{eq:st2}).
Combining Eqs.~(\ref{eq:st2}) and (\ref{eq:LGam}),
the decay rate under very frequent measurements is given by
\beq
\Gamma(\tau_{\rm i})=\langle(\Delta \hat{H}_S)^2 \rangle \tau_{\rm i},
% \verb#{eq:prop_taui}#
\label{eq:prop_taui}
\eeq
which states that the decay rate is proportional 
to the measurement intervals $\tau_{\rm i}$.\footnote{
Eq.~(\ref{eq:prop_taui}) can also be obtained 
by Eqs.~(\ref{eq:DRR}) and (\ref{eq:wei1});
When $\tau_{\rmi}$ is very short
[$\tau_{\rmi}^{-1} \gg$ (spectral width of $|g_{\mu}|^2$)],
$f_{\rm c}(\mu)\simeq \tau_{\rmi}$ and
$\Gamma(\tau_{\rmi}) \simeq \tau_{\rmi}\times \int d\mu |g_{\mu}|^2
=\tau_{\rmi}\langle(\Delta \hat{H}_S)^2 \rangle$.
}
Thus, as one measures 
the system more frequently 
(i.e., as $\tau_{\rm i}$ is made shorter),
the decay of the system is more suppressed.
In the limit of infinitely frequent measurements ($\tau_{\rm i} \to 0$),
the decay of the system is perfectly inhibited.
This phenomenon is called the {\it quantum Zeno effect}~\cite{Misra} 
(or several other names~\cite{Wol,Kraus,Joos}),
which is hereafter abbreviated as the QZE.

Note that the above argument 
does not invoke
any system-dependent features.
Thus, the QZE is a universal phenomenon,
which is expected in general quantum systems
if instantaneous ideal measurements are possible for the unstable 
states of interest.

%%%%%%%%%%%%%%%%%%%%%%%%%%%%%%%%%%%%%%%%%%%%%%%%%%%%%%%%%%%%%%%%%%%%%%%%%%%%%%
\subsection{Quantum anti-Zeno effect}
% \verb#{sec:Z_AZ}#
\label{sec:Z_AZ}
%%%%%%%%%%%%%%%%%%%%%%%%%%%%%%%%%%%%%%%%%%%%%%%%%%%%%%%%%%%%%%%%%%%%%%%%%%%%%%

\begin{figure}%----------------------------------------------------------------
\bec
\includegraphics{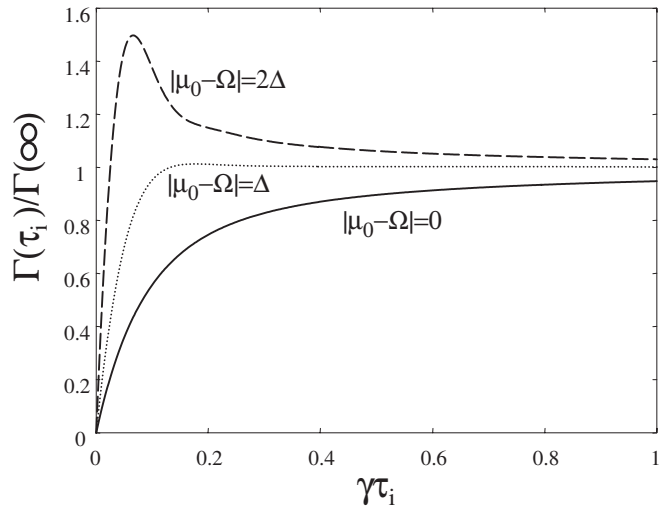}
\caption{\label{fig:norm_dr}
The decay rate $\Gamma(\tau_{\rm i})$, 
normalized by the free %natural 
decay rate $\Gamma(\infty)$, 
under repeated instantaneous ideal measurements with
intervals $\tau_{\rm i}$,
for the case of the Lorentzian form factor, Eq.~(\ref{eq:Lff}). 
The parameters are chosen as follows:
$\Delta=20\gamma$ (which gives $t_{\rm j} \sim 0.05 \gamma^{-1}$) 
and $|\Om-\mu_0|=0, \Delta, 2\Delta$.
In the case of $|\Om-\mu_0|=2\Delta$ and $\tau_{\rm i} \sim t_{\rm j}$,
it is observed that the decay is accelerated
from the unobserved case (the anti-Zeno effect).
The decay rate is maximized when $\tau_{\rm i} \simeq 0.066 \gamma^{-1}$.
}
% \verb#{fig:norm_dr}#
\enc
\end{figure}%-----------------------------------------------------------------

In the preceding subsection,
the QZE is derived by combining the initial quadratic decrease,
Eq.~(\ref{eq:st2}), 
and the measurement-modified decay rate, Eq.~(\ref{eq:LGam}).
However, Eq.~(\ref{eq:st2}) holds only for extremely small $t$,
and hence 
Eq.~(\ref{eq:prop_taui}) 
is not applicable for longer measurement intervals $\tau_{\rm i}$.
In this subsection, we study the case of longer $\tau_{\rm i}$.
For simplicity, 
we focus on an unstable state with the Lorentzian form factor,
for which the rigorous form of $s(t)$ is known for any $t$,
as discussed in Sec.~\ref{sec:Lff}.

Combining Eqs.~(\ref{eq:s(t)}) and (\ref{eq:LGam}),
we can obtain the decay rate for 
general values of $\tau_{\rm i}$.
Since the decay rate under repeated instantaneous ideal measurements 
is given by Eq.~(\ref{eq:LGam}), 
$\Gamma(\tau_{\rm i})$ has already been plotted in Fig.~\ref{fig:decrate}(b),
if one regards the horizontal axis 
as the measurement intervals $\tau_{\rm i}$ (in units of $\gamma^{-1}$).
In order to emphasize the effect of measurements, we plot 
the normalized decay rate 
$\Gamma(\tau_{\rm i})/\Gamma(\infty)$
in Fig.~\ref{fig:norm_dr}, where $\Gamma(\infty)$
is the free decay rate.
The three curves correspond to three different values 
($0, \Delta$, and $2\Delta$) of the energy discrepancy
between the atomic transition energy $\Om$ and 
the central energy of the form factor $\mu_0$.
It should be recalled that
the jump time is related to the width of the form factor
and is roughly evaluated as $t_{\rm j} \sim \Delta^{-1}$. 
[Since a multiplicative factor of order unity is unimportant, 
we have defined $t_{\rm j}$ as $t_{\rm j} \equiv \Delta^{-1}$ in 
% Sec.~\ref{sec:simcase}.
Eq.~(\ref{eq:def_jt})]

The following points are observed in common:
(i) When the measurement intervals $\tau_{\rm i}$ is long 
($\tau_{\rm i} \gg t_{\rm j}$),
the decay rate is almost unaffected by measurement,
i.e., $\Gamma(\tau_{\rm i})\simeq\Gamma(\infty)$.
(ii) A large deviation from the unobserved decay rate
is observed when $\tau_{\rm i}$ 
is short enough to satisfy
\beq
\tau_{\rm i} \lesssim t_{\rm j}.
% \verb#{eq:condi}#
\label{eq:condi}
\eeq
This condition is in accordance with our expectation,
because $s(t)$ significantly deviates from the exponential law
only for $t \lesssim t_{\rm j}$.
(iii) When $\tau_{\rm i}$ is extremely short ($\tau_{\rm i} \ll t_{\rm j}$),
$\Gamma(\tau_{\rm i})$ is proportional to $\tau_{\rm i}$, 
in accordance with Eq.~(\ref{eq:prop_taui}).
%the usual QZE is observed.

Qualitative difference is observed in the intermediate region, 
$\tau_{\rm i} \sim t_{\rm j}$. In the case of $\Om-\mu_0=0$,
$\Gamma(\tau_{\rm i})$ is always smaller than 
the free decay rate $\Gamma(\infty)$.
The decay is more suppressed as the measurements become more frequent.
Thus, in this case, the argument of Sec.~\ref{sec:QZE}
can be smoothly 
extended to a larger $\tau_{\rm i}$ region without qualitative change. 
Contrarily, in the case of $\Om-\mu_0=2\Delta$,
$\Gamma(\tau_{\rm i})$ is not a monotonous function of $\tau_{\rm i}$,
and $\Gamma(\tau_{\rm i})$ may become larger than the free decay rate
for $\tau_{\rm i}\sim t_{\rm j}$.
Thus, the decay is {\em accelerated} by successive measurements.
This opposite effect is called the 
{\it quantum anti-Zeno effect}~\cite{KofPRA,Kau,March,KofNature,Anton},
which is hereafter abbreviated as the AZE,
or the {\it inverse-Zeno effect}~\cite{FPreview,FPPRL,Pano}.
The QZE and AZE are sometimes called simply 
the {\em Zeno effect}.

%%%%%%%%%%%%%%%%%%%%%%%%%%%%%%%%%%%%%
\subsection{QZE--AZE phase diagram}
% \subsection{Transition between Zeno and anti-Zeno effects}
%%%%%%%%%%%%%%%%%%%%%%%%%%%%%%%%%%%%%

It should be remarked that, whereas the QZE may be observed 
for any unstable quantum system if $\tau_{\rm i}$ is sufficiently small,
the AZE does not necessarily takes place;
a counterexample is the case of $|\Om-\mu_0|=0$,
where decay is always suppressed for any $\tau_{\rm i}$
(see Fig.~\ref{fig:norm_dr}).
From this respect, quantum unstable states can be 
classified into the following two types:
(a) The QZE is always observed for any value of $\tau_{\rm i}$, and 
(b) the QZE is observed for $\tau_{\rm i}<\tau^{\ast}$,
whereas the AZE is observed for $\tau_{\rm i}>\tau^{\ast}$,
i.e., the QZE-AZE transition takes place
at $\tau_{\rm i}=\tau^{\ast}$.
% when the measurement intervals is $\tau^{\ast}$.

\begin{figure}%----------------------------------------------------------------
\bec
\includegraphics{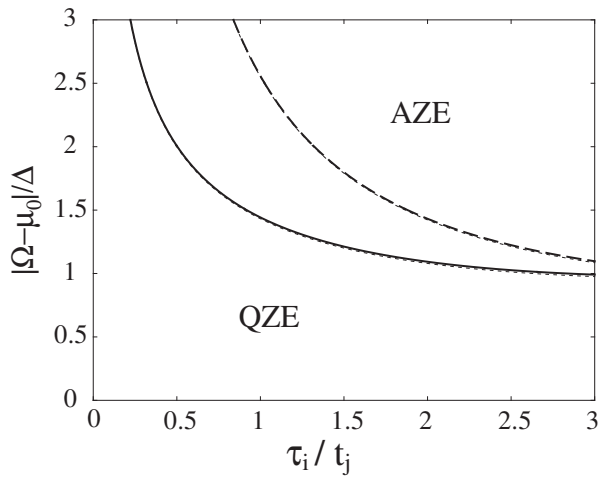}
\caption{\label{fig:pdpp}
The phase diagram of the QZE and AZE 
under repeated instantaneous ideal measurements,
for the case of the Lorentzian form factor [Eq.~(\ref{eq:Lff})]. 
The solid line divides the QZE region and the AZE region.
The broken line shows the optimum measurement intervals $\tau_{\rm i}$, 
at which the decay rate is maximized, for each value of $|\Om-\mu_0|$.
Although $\Delta = 40 \gamma$ in this Figure, 
the results are insensitive to the value of $\Delta$.
In fact, the thin dotted lines, for which $\Delta=20 \gamma$, 
almost overlap the corresponding lines for $\Delta = 40 \gamma$.
% In this Figure, $\Delta$ is chosen at $40 \gamma$,
% but the results are insensitive to $\Delta$
% (see thin dotted lines, where $\Delta=20 \gamma$).
}
% \verb#{fig:pdpp}#
\enc
\end{figure}

By analyzing Eqs.~(\ref{eq:s(t)}) and (\ref{eq:LGam}),
a phase diagram discriminating the QZE and AZE 
is generated in Fig.~\ref{fig:pdpp}
for the case of the Lorentzian form factor, Eq.~(\ref{eq:Lff}), 
% for the case of the Lorentzian form factor, 
as a function of $|\Om-\mu_0|$ and $\tau_{\rm i}$.
The phase boundary (solid line) is drawn 
by solving the following equation:
\beq
\Gamma(\tau^{\ast})=\Gamma(\infty).
% \verb#{eq:GGeq}#
\label{eq:GGeq}
\eeq
If Eq.~(\ref{eq:GGeq}) has a solution 
in $0<\tau^{\ast}<\infty$,
the unstable system belongs to type (b).
In case of the Lorentzian form factor,
Fig.~\ref{fig:pdpp} indicates that 
the system belongs to type (a) if $|\Om-\mu_0|<\Delta$,
and to type (b) if $|\Om-\mu_0|>\Delta$.

%-----------------------------------------------------------------

In order to judge whether Eq.~(\ref{eq:GGeq}) has a solution or not,
it is useful to remember the fact that,
$s(t)$ generally follows the exponential decay law
in the later stage of decay
as $s(t)\simeq {\cal Z} \exp(-\Gamma(\infty)t)$,
where ${\cal Z}$ is a positive constant.
For example, in the case of Lorentzian form factor, 
$s(t)$ behaves as Eq.~(\ref{eq:expdlaw})
in the later stage of decay.
Therefore, the asymptotic form of $\Gamma(t)$ is given by
$\Gamma(t)=\Gamma(\infty)-\ln{\cal Z}/t$.
We thus find that, as $t \to \infty$, 
$\Gamma(t)$ approaches to $\Gamma(\infty)$ 
from below when ${\cal Z}>1$,
and from above when ${\cal Z}<1$.
Combining this fact and the fact that $\Gamma(t)\to 0$ as $t \to 0$,
one is intuitively led to the following criterion:
a system belongs to type (a) when ${\cal Z}>1$,
and to type (b) when ${\cal Z}<1$~\cite{FPreview,FPPRL}.

Now we check the validity of this criterion
for the case of Lorentzian form factor, as an example.
In this case, 
${\cal Z}$ is given by Eq.~(\ref{eq:defZ}),
where $\lambda_1$ and $\lambda_2$ are given by Eq.~(\ref{eq:lam_2ji}).
When $\gamma$ is small,
$\lambda_1$ and $\lambda_2$ are approximated as
\bea
\lambda_1 &=& \Om+\frac{\gamma\Delta}{2}\frac{1}{\Om-\mu_0+\rmi\Delta},
\\
\lambda_2 &=& \mu_0-\rmi\Delta-\frac{\gamma\Delta}{2}\frac{1}{\Om-\mu_0+\rmi\Delta}.
\eea
${\cal Z}$ is therefore approximately given by
\beq
{\cal Z}\simeq \left|1-\frac{\gamma\Delta}{2(\Om-\mu_0+\rmi\Delta)^2}\right|^2
\simeq 1-\gamma\Delta\frac{(\Om-\mu_0)^2-\Delta^2}{|\Om-\mu_0+\rmi\Delta|^4}.
\eeq
Thus, the condition of type (a), i.e., ${\cal Z}>1$, 
is reduced to $|\Om-\mu_0|<\Delta$.
This is in agreement with numerical results (see Fig.~\ref{fig:pdpp}),
which indicates the validity of the criterion.

To summarize this subsection, there are two types of unstable 
systems.
In type (a), only the QZE is induced by repeated measurements.
The decay is more suppressed as the measurements become more frequent.
In type (b), whereas the QZE is induced 
when the measurements are very frequent,
the opposite effect -- the AZE -- takes place 
when the measurements are less frequent.
The decay rate depends on $\tau_{\rm i}$ in a non-monotonous way.
The former (latter) type of systems satisfy ${\cal Z}>1$ (${\cal Z}<1$),
where ${\cal Z}$ is the prefactor of the exponential decay law
in the later stage of decay.

These conclusions have been drawn 
under the assumptions that 
each measurement is instantaneous and ideal.
However, as we will discuss in Sec.~\ref{sec:IMasLC},
such measurements 
are unrealistic and in some sense unphysical.
Therefore, we must explore the Zeno effect in 
realistic measurement processes.
For this purpose, 
we should apply the quantum measurement theory, 
which is briefly summarized in the next section.
% The reader who is familiar with the quantum measurement theory
% can skip to Sec.~\ref{sec:Zeno-MMT}, and then to Sec.~\ref{sec:rmt}.

%%%%%%%%%%%%%%%%%%%%%%%%%%%%%%%%%%%%%%%%%%%%%%%%%%%%%%%%%%%%%%%%%%%%%%%%%%%%%%
\section{Quantum measurement theory}
\label{sec:mt}
%%%%%%%%%%%%%%%%%%%%%%%%%%%%%%%%%%%%%%%%%%%%%%%%%%%%%%%%%%%%%%%%%%%%%%%%%%%%%%

Since quantum systems exhibit probabilistic natures,  
one has to perform many 
runs of measurements in an experiment. % on a quantum system.
In ordinary experiments, 
one resets the system before each run 
in order to prepare the same quantum state $|\psi \rangle$ for all runs.
Or, alternatively, one prepares many equivalent systems in 
the same state $|\psi \rangle$, and performs the same
measurement independently for each system.
In either case,
one does not need to know the {\em post-measurement state}
$|\psi' \rangle$
(i.e., the state after the measurement) in order to 
predict or analyze the results of the experiment.

However, one can perform another measurement (of either the 
same observable or a different observable) on 
$|\psi' \rangle$ before he resets the system.
That is, one can perform two subsequent measurements 
in each run,
one for the pre-measurement state $|\psi \rangle$ and the other
for the post-measurement state $|\psi' \rangle$.
Or, alternatively, if one prepares many equivalent systems in 
the same state $|\psi \rangle$, he can perform the two subsequent 
measurements for each system.
In order to predict or analyze the results of such subsequent 
measurements, 
one needs to know $|\psi' \rangle$,
i.e., the state after the first measurement.
To calculate  $|\psi' \rangle$,
one must use something like the so-called projection postulate
(Sec.~\ref{sec:ideal_m}).
By many studies in the last several decades, it has been revealed 
{\em both theoretically and experimentally}
that a naive application of the 
projection postulate gives wrong results that do {\em not} agree with 
experiments on subsequent measurements.
To resolve this discrepancy, the quantum measurement theory
has been developed 
\cite{vN,LP,Glauber,Unruh,ozawa88,SF,FL,QNoise,simultaneousM,ozawa},
which will be briefly explained in this section.
Although Landau and Lifshitz \cite{bookLL} were pessimistic 
about the possibility of the calculations of the post-measurement states,
it has been revealed that the calculations {\em are} possible in many cases.
Furthermore, most importantly,  
the results of the calculations have been {\em confirmed by many 
experiments} (mostly on quantum optics; see, e.g., Refs.\
\cite{QNoise} and \cite{Mandel}).
This demonstrates the power of the quantum measurement theory.

% The main results of this section will be summarized in 
% Sec.~\ref{sec:summary}.

To explain all the points which may be questioned in studying 
the Zeno effect, we describe all the basic things of 
the quantum measurement theory in this section.
As a result, this section provides for basic knowledge
that are required to understand 
not only the Zeno effect but also 
many other topics of quantum measurements. % other than the Zeno effects, 
Actually, the full powers of the quantum measurement theory 
are manifest in studying the other topics, such as those 
in Refs.~\cite{Glauber,Unruh,ozawa88,SF,FL,simultaneousM,ozawa,
IHY,PRA91,QND1,QND2,reversible,AY},
whereas 
in studying the Zeno effect one can make great simplifications
for the reasons explained in Sec.~\ref{sec:Zeno-MMT}
if the underlying logic and approximations
% such as that explained in Sec.~\ref{sec:ua}, 
are taken for granted.
Hence, if the reader is interested only in the Zeno effect
and wish to read this section faster, 
we suggest the reader to read only Secs.~\ref{sec:ideal_m}, \ref{sec:ev_S+A},
\ref{sec:HC}, \ref{sec:vNmixture}, \ref{sec:additional}, 
% \ref{sec:probe}, \ref{sec:ua} 
and \ref{sec:Zeno-MMT}.\footnote{
The readers who are quite familiar with the quantum measurement theory
can skip directly to Sec.~\ref{sec:Zeno-MMT}, and then to Sec.~\ref{sec:rmt}.
}
When a question arises on the underlying logic upon reading later sections, 
the reader can go back to the rest of this section.

\subsection{Ideal measurement}
\label{sec:ideal_m}

Consider measurement of an observable $Q$
of a quantum system S.
The observable
$Q$ can be either the position, momentum, spin, or any 
other observable.
However, for simplicity, we assume throughout this section that 
the operator $\hat Q$ representing $Q$
has discrete eigenvalues.
The case of continuous eigenvalues can be described in a similar manner, 
which, however, requires some technical cares concerning the 
mathematical treatment of continuous eigenvalues.

In an early stage of the development of quantum theory, 
measurement of $Q$
is formulated simply as follows.
The probability $P^{\rm ideal}_R(r)$ of getting a value $r$ of the 
readout observable $R$ of a measuring apparatus
is given by
\begin{eqnarray}
P^{\rm ideal}_R(r)=
\begin{cases}
P(q)
& \mbox{for $r =q$,  an eigenvalue of $\hat Q$},
\\
0 & \mbox{otherwise},
\end{cases}
\label{P(r)ideal}\end{eqnarray}
where
\begin{equation}
P(q)
\equiv 
\left\| \hat {\cal P}(q) | \psi \rangle \right\|^2
\label{P(q)}\end{equation}
is the probability given by the Born rule.
Here, $\hat {\cal P}(q)$ denotes the projection operator
onto the subspace belonging to the eigenvalue $q$ of $\hat Q$, 
and $| \psi \rangle$ is the {\em pre-measurement state}, i.e., 
the state vector of S just before the measurement.
When an eigenvalue $q$ is obtained as readout $r$ of this measurement, 
the {\em post-measurement state} $| \psi^{\rm ideal}_q  \rangle$, i.e., 
the state vector of S just after the measurement, 
is given by
\begin{equation}
| \psi^{\rm ideal}_q \rangle = 
{1 \over \sqrt{P(q)}} \hat {\cal P}(q) | \psi \rangle,
\label{psi(r)ideal}\end{equation}
where the prefactor $1/\sqrt{P(q)}$ is simply a normalization factor.
In other words, the density operator of S just after the measurement
is given by
\begin{equation}
\hat \rho^{\rm ideal}_q  = 
| \psi^{\rm ideal}_q \rangle \langle \psi^{\rm ideal}_q |
=
{1 \over P(q)}
\hat {\cal P}(q) | \psi \rangle \langle \psi |\hat {\cal P}(q).
\label{rho(r)ideal}\end{equation}
This postulate is often called 
the {\em projection postulate} after the work of von Neumann \cite{vN}, 
although he considered not $\hat \rho^{\rm ideal}_q$ but the following mixture;
\begin{equation}
\hat \rho^{\rm ideal}_{\rm vN} 
\equiv \sum_q P(q) \hat \rho^{\rm ideal}_q
= \sum_q \hat {\cal P}(q) | \psi \rangle \langle \psi |\hat {\cal P}(q),
\label{rho_vN-ideal}\end{equation}
which we call the {\em von Neumann mixture}.
Its physical meaning will be described in Sec.~\ref{sec:vNmixture}.

A measurement process satisfying Eqs.\ (\ref{P(r)ideal}) and 
(\ref{psi(r)ideal}) is called an {\em ideal measurement}
or a {\em projective measurement}\footnote{
It is sometimes called a {\em first-kind measurement}.
However, this term is also used for a general measurement
in which 
the post-measurement state is in the subspace 
that is spanned by the eigenvectors belonging to 
eigenvalues which are close to the readout $r$.
If $[\hat Q, \hat H_{\rm S}]=0$, in particular, 
a first-kind measurement in this sense is called 
a {\em quantum non-demolition measurement} \cite{SF,QND1,QND2}.
A measurement which is not of the first kind is said to be
of the second kind.
In this article, however, we do not use these terms 
in order to avoid possible confusions.
}.
The conventional theories of the Zeno effect in Sec.~\ref{sec:ct} 
assumed that measurements are ideal (and instantaneous).
However, real measurement processes
do not satisfy Eqs.\ (\ref{P(r)ideal}) and (\ref{psi(r)ideal}) strictly, 
and thus are called 
{\em general measurements} or {\em imperfect measurements}
or {\em non-ideal measurements}.
For example, a real measuring apparatus has a non-vanishing 
error, % $\delta q_{\rm err}$. This means that 
and thus Eqs.\ (\ref{P(r)ideal}) and (\ref{psi(r)ideal})
are satisfied only approximately.
In order to analyze such general measurements, 
one has to use the quantum measurement theory.

%%%%%%%%%%%%%%%%%%%%%%%%%%%%%%%%%%%%%%%%%%%%%%%%%%%%%%%%%%%%%%%%%%%%%%%%%%%%%%
\subsection{Time evolution of the system and apparatus}
%%%%%%%%%%%%%%%%%%%%%%%%%%%%%%%%%%%%%%%%%%%%%%%%%%%%%%%%%%%%%%%%%%%%%%%%%%%%%%
\label{sec:ev_S+A}

\begin{figure}[htbp]
\begin{center}
\includegraphics[width=0.4\linewidth]{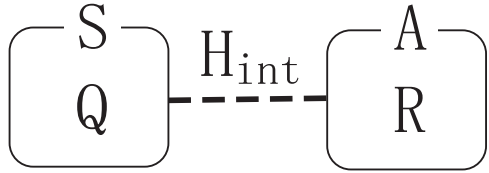}
\end{center}
\caption{
A schematic diagram of a general measurement of 
an observable $Q$ of a quantum system S
using a measuring apparatus A, whose readout observable is $R$.
}
\label{fig:S+A}
\end{figure}

\begin{figure}[htbp]
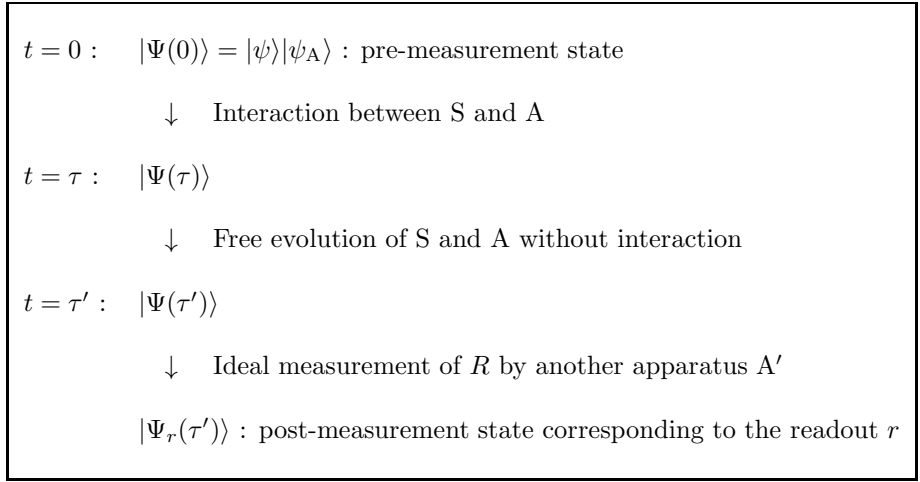

\begin{center}
\begin{tabular}[htbp]{|ll|}
\hline
& \\
$t=0$ : 
& 
$|\Psi(0) \rangle = | \psi \rangle | \psi_{\rm A} \rangle$ :
pre-measurement state 
\\ & \\
& 
\quad $\downarrow$ \quad Interaction between S and A
\\ & \\
$t=\tau$ : 
& 
$|\Psi(\tau) \rangle$ 
\\ & \\
& 
\quad $\downarrow$ \quad Free evolution of S and A without interaction
\\ & \\
$t=\tau'$ : 
& 
$|\Psi(\tau') \rangle$
\\ & \\
& 
\quad $\downarrow$ \quad Ideal measurement of $R$ by another apparatus A$'$
\\ & \\
& 
$|\Psi_r(\tau') \rangle$ : 
post-measurement state corresponding to the readout $r$
\\ & \\
\hline
\end{tabular}
\end{center}
\caption{
The time evolution of the joint quantum system % S$+$A
which is shown in Fig.\ \ref{fig:S+A}.
The unitary evolution during $0 \leq t \leq \tau$, 
by which correlations between the observable $Q$ 
and the readout observable $R$ is established, is called
the {\em unitary part} of the measurement.
See the text for details.
}
\label{fig:ev}
\end{figure}

The starting point of the quantum measurement theory is 
the key observation that 
not only the system S to be observed but also 
the measuring apparatus A should obey the laws of quantum theory.
Therefore, one must analyze 
the time evolution of the joint quantum system S$+$A using 
the laws of quantum theory, as schematically shown in 
Figs.\ \ref{fig:S+A} and \ref{fig:ev} 
\cite{vN,Glauber,Unruh,ozawa88,SF,FL,QNoise,IHY, PRA91, NQE}.
In this case, A is sometimes called a {\em probe quantum system}.

If S and A can be 
described by the Hilbert spaces 
${\bf H}_{\rm S}$ and ${\bf H}_{\rm A}$, respectively, 
then the joint system S$+$A can be 
described by the product space 
${\bf H}_{\rm S+A} \equiv
{\bf H}_{\rm S} \otimes {\bf H}_{\rm A}$.
If the Hamiltonians of S and A are 
$\hat H_{\rm S}$ (which operates on ${\bf H}_{\rm S}$) and 
$\hat H_{\rm A}$ (on ${\bf H}_{\rm A}$), respectively, 
the Hamiltonian of S$+$A is given by 
\begin{equation}
\hat H_{\rm S+A} = 
\hat H_{\rm S} \otimes \hat 1 + \hat 1 \otimes \hat H_{\rm A}
+ \hat H_{\rm int},
\label{eq:H_S+A}\end{equation}
which is simply written as
$
\hat H_{\rm S+A} = 
\hat H_{\rm S} + \hat H_{\rm A} + \hat H_{\rm int}
$.

Assuming that any correlations are erased during the preparation 
processes of the measurement, we can take the pre-measurement state
(i.e., the state just before the measurement at $t=0$) of S$+$A as
a simple product state,
\begin{equation}
|\Psi(0) \rangle = | \psi \rangle | \psi_{\rm A} \rangle 
\quad (\in {\bf H}_{\rm S+A}),
\end{equation}
where
$| \psi \rangle$ ($\in {\bf H}_{\rm S}$) 
and 
$| \psi_{\rm A} \rangle$ ($\in {\bf H}_{\rm A}$) 
denote the pre-measurement states of S and A, respectively.
Here, we assume for simplicity that the pre-measurement states
are pure states.
The generalization to a mixed state is straightforward, and one will 
then find that the main conclusions and ideas that will be explained in the 
following are not changed at all.

Let $|q, l \rangle$'s ($\in {\bf H}_{\rm S}$) be 
orthonormalized eigenvectors of $\hat Q$, 
the operator representing the observable $Q$ to be measured.
Here, $q$ is an eigenvalue of $\hat Q$ 
and $l$ denotes a set of quantum numbers labeling 
degenerate eigenvectors.
Since $|q, l \rangle$'s form a complete set of ${\bf H}_{\rm S}$, 
we can expand the pre-measurement state of S as
\begin{equation}
| \psi \rangle = \sum_{q,l} \psi(q,l) |q, l \rangle.
\label{eq:psi-pre}\end{equation}
The readout is an observable of $A$, because it will be 
observed by another apparatus or an observer (see Sec.~\ref{sec:vNC}
for details).
It is denoted by $R$, %(see Fig.\ \ref{fig:S+A})
and the operator (on ${\bf H}_{\rm A}$) 
representing it by $\hat R$.
Let $|r, m \rangle$'s ($\in {\bf H}_{\rm A}$) be 
orthonormalized eigenvectors of $\hat R$, 
where $r$ is an eigenvalue of $\hat R$ 
and $m$ denotes a set of quantum numbers labeling 
degenerate eigenvectors.
We can expand the pre-measurement state of A as
\begin{equation}
| \psi_{\rm A} \rangle = \sum_{m} \psi_{\rm A}(m) |r_0,m \rangle,
\end{equation}
where $r_0$ is the pre-measurement value of the readout.
Note that every operator of S commutes with every operator of A.
For example, 
$ %\begin{equation}
[\hat Q, \hat R]=[\hat Q, \hat H_{\rm A}]=
[\hat H_{\rm S}, \hat R]=[\hat H_{\rm S}, \hat H_{\rm A}]=0
$, %\end{equation}
which will be used in the following calculations.

If S$+$A can be regarded as an isolated system 
in the time interval $0 \leq t \leq \tau$
during which the interaction between S and A takes place,
its state vector evolves into
\begin{eqnarray}
|\Psi(\tau) \rangle 
&=&
e^{-i \hat H_{\rm S+A} \tau} | \psi \rangle | \psi_{\rm A} \rangle
\label{Psi(tau)-0}\\
&=&
\sum_{q,l} \sum_{m} \psi(q,l) \psi_{\rm A}(m) 
e^{-i \hat H_{\rm S+A} \tau} |q,l \rangle |r_0, m \rangle.
\label{Psi(tau)}\end{eqnarray}
Let us express 
the factor in the last line 
% $e^{-i \hat H_{\rm S+A} \tau} |q,l \rangle |r_0,m \rangle$
as the superposition of 
$|q',l' \rangle |r',m' \rangle$'s as
\begin{equation}
e^{-i \hat H_{\rm S+A} \tau} |q,l \rangle |r_0,m \rangle
=
\sum_{q',l'} \sum_{r',m'} 
u_{q,l,m}^{q',l',r',m'}
|q',l' \rangle |r',m' \rangle,
\label{eq:evol.general}\end{equation}
where the coefficient $u_{q,l,m}^{q',l',r',m'}$
is a function of 
$\hat H_{\rm S+A}$, $\tau$ and $r_0$, 
all of which can be tuned by tuning the experimental setup.\footnote{
Note that one can tune not only 
$\hat H_{\rm S+A}$ and $\tau$ but also $r_0$.
The significance of this fact is discussed in Refs.~\cite{SF,FL}.
}
Then, Eq.~(\ref{Psi(tau)}) can be expressed as
\begin{equation}
|\Psi(\tau) \rangle 
=
\sum_{q,l} \left[ \psi(q,l)
\sum_{q',l'} \left( |q',l' \rangle 
\sum_{m, r',m'} 
\psi_{\rm A}(m) u_{q,l,m}^{q',l',r',m'} |r',m' \rangle
\right) \right].
\label{Psi(tau)-2}\end{equation}
As we will show shortly, 
the results for a general measurement reduce to 
those for an ideal one if
$u_{q,l,m}^{q',l',r',m'}$ takes the following form;\footnote{
A more general form of Eq.\ (\ref{eq:cond-IM}) is
\begin{equation}
u_{q,l,m}^{q',l',r',m'} = 
u_{q,m}^{m'}
\delta_{q',q} \delta_{l',l} \delta_{r', f(q)}.
\end{equation}
where $f$ is an invertible function of $q$.
By relabeling $r$ appropriately, 
we can reduce this to Eq.\ (\ref{eq:cond-IM}).
}
\begin{equation}
u_{q,l,m}^{q',l',r',m'} = 
u_{q,m}^{m'}
\delta_{q',q} \delta_{l',l} \delta_{r',q}.
\label{eq:cond-IM}\end{equation}
In this case, 
Eq.~(\ref{Psi(tau)-2}) reduces to 
\begin{equation}
|\Psi(\tau) \rangle 
=
\sum_{q,l} \left[
\psi(q,l) |q,l \rangle 
 \left( \sum_{m, m'} \psi_{\rm A}(m) u_{q,m}^{m'} 
|q,m' \rangle
\right) \right],
\label{Psi(tau)-IM}\end{equation}
which clearly shows that S and A get entangled\footnote{
That is, this state is not generally a simple product of two vectors,
one is in ${\bf H}_{\rm S}$ and the other is in ${\bf H}_{\rm A}$.
},
by the interaction $\hat H_{\rm int}$ in $\hat H_{\rm S+A}$,
in such a way that $Q$ and $R$ are strongly correlated.
This is the main part of the measurement process, 
which we call the {\em unitary part}, 
because it is 
described as a unitary evolution due to the Schr\"odinger equation.
Since  $Q$ and $R$ are correlated, 
one can get information about $Q$ by measuring $R$, 
as shown below.
For general measurements, the correlation between $Q$ and $R$ may be weaker
than that in Eq.~(\ref{Psi(tau)-IM}).
However, non-vanishing correlation {\em should} be established in order to 
get non-vanishing information.
Since non-vanishing information should be obtained by any measurement
(see Sec.~\ref{sec:info}), 
$\hat H_{\rm int}$ should be such an interaction that 
creates non-vanishing correlation between $Q$ and $R$.

For $t > \tau$, for which the interaction is over 
(or ineffective\footnote{
For example, when the wavefunction of S takes a wavepacket form
the interaction becomes ineffective after the wavepacket passes 
through the apparatus.
}), 
the state vector further evolves as
\begin{eqnarray}
|\Psi(t) \rangle 
&=&
e^{-i (\hat H_{\rm S}+\hat H_{\rm A}) (t-\tau)} |\Psi(\tau) \rangle
\end{eqnarray}
until the readout $R$ is measured by another apparatus or an observer A$'$
at $t=\tau'$ ($\geq \tau$).
If this measurement of $R$ at $t=\tau'$ can be regarded as an 
instantaneous ideal 
measurement of $R$ ({\em not} of $Q$), then 
the probability $P_R(r)$ of getting a readout $r$ 
(which is an eigenvalue of $\hat R$) is given by
\begin{eqnarray}
P_R(r)
&=& 
\left\| \hat {\cal P}_R(r) |\Psi(\tau') \rangle \right\|^2
\label{P(r)general}\\
&=& 
\left\|  
\hat {\cal P}_R(r) 
e^{-i (\hat H_{\rm S}+\hat H_{\rm A}) (\tau'-\tau)} |\Psi(\tau) \rangle
\right\|^2.
\label{P(r)general2}\end{eqnarray}
Here, $\hat {\cal P}_R(r)$ denotes the projection operator
onto the subspace belonging to the eigenvalue $r$ of $\hat 1 \otimes \hat R$;
\begin{equation}
\hat {\cal P}_R(r) \equiv \hat 1 \otimes \sum_m |r,m \rangle \langle r,m |.
\end{equation}
When an eigenvalue $r$ is thus obtained as the readout, 
the post-measurement state $| \Psi_r(\tau') \rangle$ of S$+$A
is given by
\begin{eqnarray}
| \Psi_r(\tau') \rangle 
&=& 
{1 \over \sqrt{P_R(r)}}
\hat {\cal P}_R(r) |\Psi(\tau') \rangle
\label{Psi(r)general}\\
&=& 
{1 \over \sqrt{P_R(r)}}
\hat {\cal P}_R(r) 
e^{-i (\hat H_{\rm S}+\hat H_{\rm A}) (\tau'-\tau)} |\Psi(\tau) \rangle.
\label{Psi(r)general2}\end{eqnarray}
If we denote the trace operation over ${\bf H}_{\rm A}$ by
${\rm Tr}_{\rm A}$, the post-measurement state of S 
is represented by the reduced density operator, 
\begin{equation}
\hat \rho_r(\tau') 
=
{\rm Tr}_{\rm A} 
\left( | \Psi_r(\tau') \rangle \langle \Psi_r(\tau') | \right),
\label{rho(r)general}\end{equation}
because the expectation value $\langle X \rangle_r$
of any observable $X$ of S is given by
\begin{equation}
\langle X \rangle_r = 
\langle \Psi_r(\tau') | \hat X | \Psi_r(\tau') \rangle 
= {\rm Tr}[\hat \rho_r(\tau') \hat X].
\end{equation}
Since the entanglement of S and A is 
not generally dissolved in $| \Psi_r(\tau') \rangle$, 
$\rho_r(\tau')$ generally becomes a mixed state. 

Equations (\ref{P(r)general}) and (\ref{rho(r)general})
% (or (\ref{rho(r)general-t}))
for a general measurement of $Q$ should be compared with 
Eqs.~(\ref{P(r)ideal}) and (\ref{rho(r)ideal})
for an ideal measurement.
The general equations reduce to the ideal ones if
Eq.\ (\ref{eq:cond-IM}) is satisfied.
In fact, we have in this case 
\begin{equation}
\hat {\cal P}_R(r) |\Psi(\tau) \rangle
=
\begin{cases}
\displaystyle
\left( \sum_{l} \psi(q,l) |q,l \rangle \right)
\left( \sum_{m, m'} \psi_{\rm A}(m) u_{q,m}^{m'} 
|q,m' \rangle \right)
& \mbox{for $r =q$, an eigenvalue of $\hat Q$},
\\
0 & \mbox{otherwise}.
\label{PRPSI-ideal}
\end{cases}
\end{equation}
Since we can take $\tau'=\tau$ as will be discussed in 
Sec.~\ref{sec:tau'}, 
and noting 
that $\sum_{l} \psi(q,l) |q,l \rangle = {\cal P}(q) |\psi \rangle$,
we find that 
Eqs.\ (\ref{P(r)general}) and (\ref{rho(r)general}) reduce 
in this case to 
Eqs.\ (\ref{P(r)ideal}) and (\ref{rho(r)ideal}), respectively.
Therefore, in order to realize an ideal measurement, one must construct an 
experimental setup by which Eq.\ (\ref{eq:cond-IM}) is satisfied.
Implications of this condition will be discussed in subsection 
\ref{sec:IMasLC}.

%%%%%%%%%%%%%%%%%%%%%%%%%%%%%%%%%%%%%%%%%%%%%%%%%%%%%%%%%%%%%%%%%%%%%%%%%%%%%%
\subsection{von Neumann chain}
%%%%%%%%%%%%%%%%%%%%%%%%%%%%%%%%%%%%%%%%%%%%%%%%%%%%%%%%%%%%%%%%%%%%%%%%%%%%%%
\label{sec:vNC}

\begin{figure}[htbp]
\begin{center}
\includegraphics[width=0.95\linewidth]{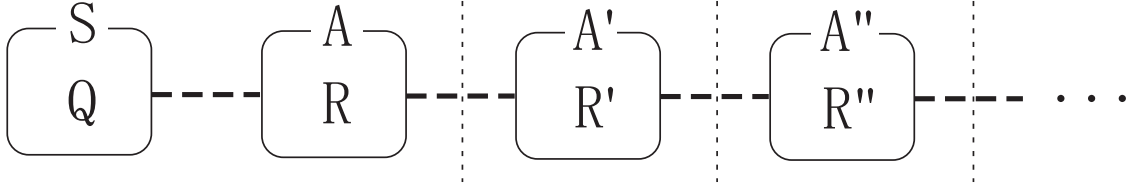}
\end{center}
\caption{
The {\em von Neumann chain}.
To measure an observable $Q$ of a quantum system S, 
an apparatus A is coupled to  S and the information on $Q$ 
is transferred to an observable $R$ of A.
To measure $R$, another apparatus A$'$ is coupled to A
and the information on $R$ 
is transferred to an observable $R'$ of A$'$, and so on.
The vertical dotted lines indicate possible locations
of the {\em Heisenberg cut} (see Sec.~\ref{sec:HC}),
inside which the laws of quantum theory are applied whereas 
the outside is just taken as a device for measuring $R$ (or $R'$, or $R''$,
$\cdots$, depending on the location of the Heisenberg cut).
}
\label{fig:vNC}
\end{figure}

In the above argument, the observable $Q$ 
of the quantum system S is measured by 
the apparatus A, and the readout observable $R$ of A is measured by another 
apparatus or an observer A$'$ \cite{vN}.
Such a sequence, as shown in Fig.\ \ref{fig:vNC}, 
is sometimes called the {\em von Neumann chain}.
We here describe its basic notions.

\subsubsection{When measurement is completed?}
\label{sec:tau'}

The measurement process described above is completed at $t=\tau'$.
We now show, however, that 
one can also say that
the measurement is completed at $t=\tau$.

To show this, let us calculate the state after $\tau'$.
For $t > \tau'$, S and A evolve freely, hence the 
state vector of S$+$A at $t$ ($> \tau'$) is given by
\begin{eqnarray}
| \Psi_r(t) \rangle 
&=& 
e^{-i (\hat H_{\rm S}+\hat H_{\rm A}) (t-\tau')} |\Psi_r(\tau') \rangle
\label{Psi(rt)general}\\
&=& 
{1 \over \sqrt{P_R(r)}}
e^{-i (\hat H_{\rm S}+\hat H_{\rm A}) (t-\tau')} 
\hat {\cal P}_R(r) 
e^{-i (\hat H_{\rm S}+\hat H_{\rm A}) (\tau'-\tau)} |\Psi(\tau) \rangle.
\label{Psi(rt)general2}\end{eqnarray}
For the apparatus A to work well, the readout $R$ should be stable
for $t \geq \tau$.
That is, 
\begin{equation}
P_R(r) =
\mbox{independent of $\tau'$},
\label{P(r)=const}\end{equation}
to a good approximation.
This is satisfied if 
\begin{equation}
[\hat R, \hat H_{\rm A}] = 0,
\label{[R,H]=0}\end{equation}
because this implies $[\hat {\cal P}_R(r), \hat H_{\rm A}]=0$, 
and Eq.\ (\ref{P(r)general2}) then reduces to 
\begin{equation}
P_R(r)
= 
\left\|  
e^{-i (\hat H_{\rm S}+\hat H_{\rm A}) (\tau'-\tau)} 
\hat {\cal P}_R(r) |\Psi(\tau) \rangle
\right\|^2
=
\left\|  
\hat {\cal P}_R(r) |\Psi(\tau) \rangle
\right\|^2
=
\mbox{independent of $\tau'$}.
\label{P(r)=const-2}\end{equation}
Although Eq.\ (\ref{[R,H]=0}) is not a necessary condition
but a sufficient condition for Eq.\ (\ref{P(r)=const}),\footnote{
Condition (\ref{[R,H]=0}) implies Eq.\ (\ref{P(r)=const-2}) for 
{\em every} vector $|\Psi(\tau) \rangle$ in ${\bf H}_{\rm S+A}$.
However, it is sufficient for Eq.\ (\ref{P(r)=const}) 
that Eq.\ (\ref{P(r)=const-2}) is satisfied only 
for $|\Psi(\tau) \rangle$ given by Eq.\ (\ref{Psi(tau)}).
}
we henceforth assume Eq.\ (\ref{[R,H]=0}) for simplicity.\footnote{
It is worth mentioning that Eq.\ (\ref{[R,H]=0}) is a natural 
assumption if $R$ and $H_{\rm A}$ are macroscopic variables,
because then they must be additive observables \cite{SM02,MSS}
and the volume $V_{\rm A}$ of A is quite large, and thus
Eq.~(\ref{[R,H]=0}) is {\em always} satisfied to a good approximation 
in the sense that \cite{MSS} 
\begin{equation}
\left[ {\hat R \over V_{\rm A}} , {\hat H_{\rm A} \over V_{\rm A}}
\right] = O\left( {1 \over V_{\rm A}} \right).
\label{[R/V,H/V]=O(1/V)}\end{equation}
% where $V_{\rm A}$ is the volume of A \cite{MSS}.
}
Then, Eq.\ (\ref{Psi(rt)general2}) reduces to 
\begin{equation}
| \Psi_r(t) \rangle 
= 
{1 \over \sqrt{P_R(r)}}
e^{-i (\hat H_{\rm S}+\hat H_{\rm A}) (t-\tau)} 
\hat {\cal P}_R(r) |\Psi(\tau) \rangle,
\end{equation}
which is independent of $\tau'$.
Therefore, 
\begin{equation}
\hat \rho_r(t) =
{\rm Tr}_{\rm A} \left( | \Psi_r(t) \rangle  \langle \Psi_r(t)| \right)
\end{equation}
is also independent of $\tau'$.
We thus find that 
the state of S$+$A, as well as  
the state of S, for $t > \tau'$ 
(i.e., after the measurement is completed)
are independent of $\tau'$,
the instance at which 
$R$ is read by another apparatus or an observer.

Because of this reasonable property, 
one can take $\tau'$ as an arbitrary time after $\tau$. 
In particular, we can take $\tau'=\tau$, from which 
{\em we can say that the measurement is completed at $t=\tau$}.

%%%%%%%%%%%%%%%%%%%%%%%%%%%%%%%%%%%%%%%%%%%%%%%%%%%%%%%%%%%%%%%%%%%%%%%%%%%%%%
\subsubsection{Where is the Heisenberg cut?}
%%%%%%%%%%%%%%%%%%%%%%%%%%%%%%%%%%%%%%%%%%%%%%%%%%%%%%%%%%%%%%%%%%%%%%%%%%%%%%
\label{sec:HC}

In the above argument, apparatus 
A has been treated as a quantum system that evolves according to 
the Schr\"odinger equation. On the other hand, 
another apparatus or an observer A$'$ which measures R of A 
has been treated as a device that performs the measurement of R.
This means that we have assumed in the von Neumann chain 
a {\em hypothetical} boundary between A and A$'$, 
inside which the laws of quantum theory are applied, whereas 
the outside is just taken as such a device.
Such a boundary is called the {\em Heisenberg cut}.
Since this boundary is hypothetical and artificial, 
it must be able to be moved to a large extent freely, 
without causing any observable effects, for quantum theory 
to be consistent. von Neumann called this property 
the {\em psychophysical parallelism} \cite{vN}.
He showed that quantum theory indeed has this property
in the following sense.\footnote{
Although he showed this for the von Neumann mixture $\hat \rho_{\rm vN}$,
we here show it more generally for $\hat \rho_{r}$. 
}

Suppose that the Heisenberg cut is moved to the boundary between 
A$'$ and A$''$ of Fig.\ \ref{fig:vNC}, 
and that both the measurement of $R$ by A$'$ (at $t=\tau' > \tau$)
and that of $R'$ by A$''$ (at $t=\tau'' > \tau'$) are ideal
(and instantaneous, for simplicity).
By calculating the time evolution of the enlarged joint system S$+$A$+$A$'$
in a manner similar to the calculations of Secs.~\ref{sec:ev_S+A}
and \ref{sec:tau'},
one can calculate the probability distribution $P'_{R'}(r')$ of 
the readout $r'$ as well as 
the reduced density operator $\hat \rho'_{r'}(t)$ of S for $t>\tau''$.
From such calculations, one can show that 
$P'_{R'}(r')$ and $\hat \rho'_{r'}(t)$ coincide with 
$P_{R}(r)$ and $\hat \rho_{r}(t)$, which have been obtained 
above as Eqs.~(\ref{P(r)general}) and (\ref{rho(r)general}), 
respectively. That is, 
\begin{equation}
P'_{R'}(\cdot) = P_{R}(\cdot),
\end{equation}
\begin{equation}
\hat \rho'_{r'}(t) = \hat \rho_{r}(t)
\ \mbox{ for every pair of $r'$ and $r$ such that $r'=r$}.
\end{equation}
This shows that the Heisenberg cut can be located either between 
A and A$'$ or between A$'$ and A$''$,
without causing any observable effects.
In contrast, 
the Heisenberg cut {\em cannot}
be moved to the boundary between S and A 
for general measurements, because it would then give
Eqs.~(\ref{P(r)ideal}) and (\ref{rho(r)ideal}), 
which do not agree with the correct equations 
(\ref{P(r)general}) and (\ref{rho(r)general}).

Therefore, we conclude that 
{\em the Heisenberg cut can be located at any place 
at which the interaction process can be regarded as 
the unitary part} (in the terminology of Sec.~\ref{sec:ev_S+A}) 
{\em of an ideal measurement}.

% \subsubsection{Mixed state of all post-measurement states}
\subsubsection{Average over all possible values of the readout}
\label{sec:vNmixture}

Suppose that the Heisenberg cut is located between A and A$'$.
The post-measurement state corresponding to each readout $r$ is given by 
\begin{equation}
\hat \rho_r(\tau) 
=
{\rm Tr}_{\rm A} 
\left( | \Psi_r(\tau) \rangle \langle \Psi_r(\tau) | \right),
\label{rho(r)general2}\end{equation}
where we have taken $\tau'=\tau$ as discussed in Sec.~\ref{sec:tau'}. 
The expectation value $\langle X \rangle_r$
of an observable $X$ of S is calculated, for each readout $r$, as
\begin{equation}
\langle X \rangle_r = {\rm Tr}[\hat \rho_r(\tau) \hat X].
\end{equation}

In some cases, 
the mixture of $\hat \rho_r(\tau)$'s over all possible values of $r$,
\begin{equation}
\hat \rho_{\rm vN}(\tau) 
\equiv \sum_r P_R(r) \hat \rho_r(\tau),
\label{rho_vN}\end{equation}
is also used as the post-measurement state.
As in the case of an ideal measurement (Sec.~\ref{sec:ideal_m}), 
we call it the {\em von Neumann mixture}.
This density operator is useful when one discusses 
{\em average} properties of the post-measurement states.
That is, when one is interested in the average 
$\langle X \rangle_{\rm vN}$
of $\langle X \rangle_r$ over all possible values of $r$,
it can be calculated as
\begin{equation}
\langle X \rangle_{\rm vN}
=
\sum_r P_R(r) 
\langle X \rangle_r
=
\sum_r P_R(r) {\rm Tr}[\hat \rho_r(\tau) \hat X]
=
{\rm Tr}[\hat \rho_{\rm vN}(\tau) \hat X].
\label{<X>}\end{equation}
In this case, $\hat \rho_{\rm vN}(\tau)$ is equivalent to
the set of $\{ P_R(r), \hat \rho_r(\tau) \}$.
Notice, however, that 
$\hat \rho_{\rm vN}(\tau)$ has less information 
in more general cases, such as the case where
one is interested in properties of the post-measurement state 
corresponding to {\em each} value of $r$.
In fact, 
the decomposition of $\hat \rho_{\rm vN}(\tau)$
into the form of the right-hand side of
Eq.\ (\ref{rho_vN}) is not unique, and hence
one {\em cannot} get the set of $\{ P_R(r), \hat \rho_r(\tau) \}$ uniquely 
from $\hat \rho_{\rm vN}(\tau)$.

Equation (\ref{<X>}) can be simplified if we
note that 
$[\hat X, \hat {\cal P}_R(r)]=0$ for all $r$
(because $X$ is an observable of S whereas
$R$ is an observables of A).
Using this and 
$\hat {\cal P}_R(r) \hat {\cal P}_R(r) = \hat {\cal P}_R(r)$
and 
$\sum_r \hat {\cal P}_R(r) = \hat 1$,
we can rewrite Eq.~(\ref{<X>}) as
\begin{equation}
\langle X \rangle_{\rm vN}
=
\sum_r 
\langle \Psi(\tau) | \hat {\cal P}_R(r) 
\hat X
\hat {\cal P}_R(r) |\Psi(\tau) \rangle
=
\sum_r 
\langle \Psi(\tau) | \hat {\cal P}_R(r) 
\hat X
|\Psi(\tau) \rangle
=
\langle \Psi(\tau) | \hat X |\Psi(\tau) \rangle.
\label{<X>-commute}\end{equation}
% where we have used 
% $\hat {\cal P}_R(r) \hat {\cal P}_R(r) = \hat {\cal P}_R(r)$
% and 
% $\sum_r \hat {\cal P}_R(r) = 1$.
This shows that 
one can calculate $\langle X \rangle_{\rm vN}$ from 
$|\Psi(\tau) \rangle$, which is the final state of
the unitary part (in the terminology of Sec.~\ref{sec:ev_S+A}). 
Therefore, 
{\em when one is interested only in $\langle X \rangle_{\rm vN}$,
it is sufficient to calculate the unitary part of the measurement process},
and one can forget about the ideal measurement of $R$ by A$'$,
for which we have used the projection postulate.
% That is, 
% one does not have to use the projection postulate at all
% to calculate $\langle X \rangle_{\rm vN}$,
% which is the average of $\langle X \rangle_r$
% over possible values of the readout $r$.
Note, however, that this is not generally 
the case for repeated measurements, 
as will be discussed in Sec.~\ref{sec:CvsD}.

\subsection{Prescription for analyzing general measurements}
\label{sec:prescription}

% \subsubsection{Prescription}

From discussions in Secs.~\ref{sec:ev_S+A} and \ref{sec:vNC}, 
we can deduce the prescription for analyzing general measurements
as follows:
\begin{enumerate}
\item 
Write down the von Neumann chain S, A$_1$, A$_2$, $\cdots$.

\item
Find a place at which the interaction process can be 
regarded as the unitary part of an ideal measurement.
Locate the Heisenberg cut there.
Although two or more such places may be found, 
you can choose any of them.
However, to simplify calculations, it is better to 
choose the one that is closest to S.

\item
If the Heisenberg cut thus located lies between A$_k$ and A$_{k+1}$, 
apply the laws of quantum theory to 
the joint system S$+$A$_1+ \cdots$A$_k$, 
taking A$_{k+1}$ as 
a device that performs an ideal measurement of the readout observable R$_k$
of A$_k$.
If the interaction in the joint system is effective 
during the time interval $0 \leq t \leq \tau$, 
one can say that the measurement is performed during this interval.
% as discussed in Sec.~\ref{sec:tau'}.

\item
Evaluate the probability distribution of the readout $r_k$ of R$_k$
and the post-measurement state, 
in the same way as we have done in Sec.~\ref{sec:ev_S+A}.
\end{enumerate}

We can regard the subsystem A$_1+ \cdots$A$_k$
of the joint system S$+$A$_1+ \cdots$A$_k$ as system A
of Sec.~\ref{sec:ev_S+A}, and A$_{k+1}$ as A$'$.
We can therefore apply the formulation of Sec.~\ref{sec:ev_S+A}
to general cases.
We will thus use the equations and notations of Sec.~\ref{sec:ev_S+A}
in the following discussions.

%%%%%%%%%%%%%%%%%%%%%%%%%%%%%%%%%%%%%%%%%%%%%%%%%%%%%%%%%%%%%%%%%%%%%%%%%%%%%%
\subsection{Properties of general measurements}
%%%%%%%%%%%%%%%%%%%%%%%%%%%%%%%%%%%%%%%%%%%%%%%%%%%%%%%%%%%%%%%%%%%%%%%%%%%%%%
\label{sec:prop-gm}

%%%%%%%%%%%%%%%%%%%%%%%%%%%%%%%%%%%%%%%%%%%%%%%%%%%%%%%%%%%%%%%%%%%%%%%%%%%%%%
\subsubsection{Response time}
%%%%%%%%%%%%%%%%%%%%%%%%%%%%%%%%%%%%%%%%%%%%%%%%%%%%%%%%%%%%%%%%%%%%%%%%%%%%%%
\label{sec:rt}

In an early stage of the development of quantum theory, 
it was sometimes argued that the measurement should be made 
instantaneously.
Such a measurement is called an {\em instantaneous measurement}.
However, as we will discuss in Sec.~\ref{sec:IMasLC}, 
any physical measurement takes a finite time. 
This finite time $\tau$ has been defined in Sec.~\ref{sec:ev_S+A}
as the time after which the interaction 
between S and A becomes ineffective.
Therefore, if the Hamiltonian $\hat H_{\rm S+A}$ 
and the pre-measurement state $|\Psi(0) \rangle$ are known, 
one can evaluate $\tau$ by solving the Schr\"odinger equation.
This $\tau$ is usually called the {\em response time} of 
the apparatus.\footnote{
See Sec.~\ref{sec:CvsD} for the response time of continuous
measurements.
}

To be more precise, $\tau$ should be called the 
{\em lower limit of the response time},
because {\em practical} response times of real experiments usually become 
longer for many practical reasons.
In the model of Sec.~\ref{sec:rmt}, for example, 
$\tau$ (which will be denoted by $\tau_{\rm r}$ there)
is the time required for generating an elementary excitation 
in the detector.
Such a microscopic excitation should be magnified to obtain 
a macroscopic signal.
Due to possible delays in the magnification and 
the signal transmission processes,
the practical response time will become longer
in real experiments. 

However, in discussing fundamental physics, 
the limiting value is more significant
than practical values,\footnote{
For example, suppose that
the measurement of the readout $R$ by A$'$ is performed
not at $t=\tau$ but at a later time $t=\tau' > \tau$.
Then the total response time
of the experiment becomes longer.
However, 
we have shown in Sec.~\ref{sec:tau'}
that the value of $\tau'$ is irrelevant.
} 
which depend strongly  
on detailed experimental conditions.
For this reason, we simply call $\tau$ the {\em response time} in this
article. 
For the same reason, 
we shall drop in the following subsections the words 
`lower limit of' or `upper limit of' 
from the terms such as the 
{\em lower limit of the measurement error}, 
the {\em upper limit of the range of measurement},
the {\em upper limit of the amount of information obtained by measurement},
and the {\em lower limit of the backaction of measurement}.

It is worth stressing that 
if one makes $\tau$ shorter without increasing 
the strength of $\hat H_{\rm int}$
then the measurement error % $\delta q_{\rm err}$ (see below) 
would be increased.
Therefore, there is a tradeoff between 
(the reduction of) the response time and (that of) the measurement error.
This and related tradeoffs, as well as their deep implications, 
were discussed in Ref.~\cite{FL}.

%%%%%%%%%%%%%%%%%%%%%%%%%%%%%%%%%%%%%%%%%%%%%%%%%%%%%%%%%%%%%%%%%%%%%%%%%%%%%%
\subsubsection{Measurement error}
%%%%%%%%%%%%%%%%%%%%%%%%%%%%%%%%%%%%%%%%%%%%%%%%%%%%%%%%%%%%%%%%%%%%%%%%%%%%%%
\label{sec:me}

For general measurements, the probability distribution 
$P_R(r)$ of the readout of measuring apparatus is 
different from that for an ideal measurement, $P^{\rm ideal}_R(r)$.
This means that a general measurement has a non-vanishing
measurement error.

For example, consider a special case where
$| \psi \rangle$ is an eigenstate of $\hat Q$;
\begin{equation}
| \psi \rangle = \sum_l \varphi_q(l) |q,l \rangle \equiv | \varphi_q \rangle,
\label{eq:sum-psi(l)}\end{equation}
where $\varphi_q(l)$'s are arbitrary coefficients satisfying 
$\sum_l |\varphi_q(l)|^2=1$.
In this case, $P^{\rm ideal}_R(r) = \delta_{r,q}$ from 
Eqs.\ (\ref{P(r)ideal}) and (\ref{P(q)}), whereas
$P_R(r)$ (for $\tau'=\tau$) is evaluated from 
Eqs.\ (\ref{Psi(tau)}), (\ref{eq:evol.general}) and
(\ref{P(r)general}) as
\begin{eqnarray}
P_R(r)
&=&
\left\| \hat {\cal P}_R(r) 
e^{-i \hat H_{\rm S+A} \tau} | \varphi_q \rangle | \psi_{\rm A} \rangle
\right\|^2
\nonumber\\
&=&
\left\| 
\sum_l \varphi_q(l) \sum_{m} \psi_{\rm A}(m) \sum_{q',l',m'} 
u_{q,l,m}(q',l',r,m') |q',l' \rangle |r,m' \rangle
\right\|^2.
\end{eqnarray}
It is then clear that 
$P_R(r) \neq P^{\rm ideal}_R(r)$ in general, except when condition 
(\ref{eq:cond-IM}) is satisfied.

Since the predictions of quantum theory are of probabilistic nature, 
the definition of the measurement error is not so trivial, 
as will be discussed shortly.
In principle, however, 
the measurement error should be quantified by
an appropriate measure of the difference between 
$P_R(r)$ and $P^{\rm ideal}_R(r)$. 
For example, it may be quantified by the 
{\em Kullback-Leider distance} or {\em relative entropy}
\cite{InfoTh};
\begin{equation}
D(P^{\rm ideal}_R \| P_R)
\equiv
\sum_r P^{\rm ideal}_R(r) \log {P^{\rm ideal}_R(r) \over P_R(r)}.
\label{eq:D}\end{equation}
One can also use $D(P_R \| P^{\rm ideal}_R)$, which 
is not equal to $D(P^{\rm ideal}_R \| P_R)$ in general.
Or, one can use other measures which are used in 
probability theory and/or information theory \cite{InfoTh}.

However, it is customary, and sometimes convenient,
to quantify the measurement error in a different way
using a few parameters.
One of such parameters is
the difference between 
the expectation values of the two probability distributions,
\begin{eqnarray}
\delta r_{\rm bias} 
&\equiv& 
\langle R \rangle - \langle R \rangle^{\rm ideal}
\nonumber\\
&\equiv& 
\sum_r r P_R(r) - \sum_r r P^{\rm ideal}_R(r) 
\nonumber\\
&=& 
\langle \Psi(\tau) | \hat R | \Psi(\tau) \rangle
- 
\langle \psi | \hat Q | \psi \rangle.
\end{eqnarray}
When $\delta r_{\rm bias} =0$, the measurement 
is said to be {\em unbiased}.
In certain cases, one can easily 
calibrate (i.e., relabel) $r$ in such a way that
the unbiased condition is satisfied.

Another parameter used to quantify the measurement error
is related to the standard deviation.
When the pre-measurement state is an eigenstate of $\hat Q$,
$| \varphi_q \rangle$, 
the readout $r$ of an ideal measurement always agrees with $q$,
showing no fluctuation.
Hence, its standard deviation
\begin{equation}
\delta r^{\rm ideal}_{\rm sd}
\equiv
[\langle (\Delta R)^2 \rangle^{\rm ideal}]^{1/2}
\equiv 
\big[ 
\sum_r (r-\langle R \rangle^{\rm ideal})^2 P^{\rm ideal}_R(r) 
\big]^{1/2}
\label{eq:sd-eigen-ideal}\end{equation}
vanishes.
On the other hand,  for the same state $| \varphi_q \rangle$,
the standard deviation of the 
readout of a general measurement
\begin{equation}
\delta r_{\rm sd}
\equiv
[\langle (\Delta R)^2 \rangle]^{1/2}
\equiv 
\big[ 
\sum_r (r-\langle R \rangle)^2 P_R(r) 
\big]^{1/2}
\label{eq:sd-eigen-general}\end{equation}
is finite. 
Therefore, a set of 
$\delta r_{\rm bias}$ and $\delta r_{\rm sd}$ may be used 
to quantify the measurement error
when the pre-measurement state is an eigenstate of $\hat Q$.
However, for a general pre-measurement state 
$| \psi \rangle$, 
the readout fluctuates % from run to run 
even for an ideal measurement, i.e., 
$\delta r^{\rm ideal}_{\rm sd} \geq 0$.
As a result, there exist various ways of 
quantifying the measurement error by (something like) the standard deviation.
For example, many works on quantum nondemolition measurement 
\cite{Unruh,QND1,QND2}
quantified it by a set of $\delta r_{\rm bias}$ and 
the increase of the variance \cite{SF,FL,IHY,PRA91,NQE},
\begin{equation}
(\delta r_{\rm sd})^2
-
(\delta r^{\rm ideal}_{\rm sd})^2.
\label{me-delta-variance}\end{equation}
Although this quantification is convenient for many applications, 
one of its disadvantages is that its vanishment (along with  
$\delta r_{\rm bias}=0$) 
does not guarantee $P_R(r)=P^{\rm ideal}_R(r)$.
Another important work \cite{ozawa} quantified the measurement error by
\begin{equation}
\langle \Psi_{\rm H} | \left( \hat R_{\rm H}(\tau)-\hat Q_{\rm H}(0) \right)^2
| \Psi_{\rm H} \rangle,
\label{eq:ozawa}\end{equation}
where $\hat R_{\rm H}, \hat Q_{\rm H}, | \Psi_{\rm H} \rangle$ 
are $\hat R, \hat Q, | \Psi \rangle$ in the Heisenberg picture, respectively, 
i.e., $| \Psi_{\rm H} \rangle = | \Psi(0) \rangle$ and so on.
Although this quantity has good mathematical properties, 
its physical meaning is not clear enough.
For example, suppose that we are given two pieces of 
apparatus A and A$^{\rm ideal}$
which perform general and ideal measurements, respectively.
By performing two experiments, 
one using A and the other using A$^{\rm ideal}$, 
we can measure all of 
$\delta r_{\rm bias}$, $\delta r_{\rm sd}$, 
$\delta r^{\rm ideal}_{\rm sd}$ and $D(P^{\rm ideal}_R \| P_R)$,
for any states.
However, it is impossible to measure the quantity of 
Eq.\ (\ref{eq:ozawa}) using A and A$^{\rm ideal}$
for general states.

In the following, we do not specify 
the detailed quantification of the measurement error 
$\delta q_{\rm err}$ except when it is needed\footnote{
If the reader feels uneasy about this, 
you can assume for example 
that $\delta q_{\rm err}$ of an apparatus is defined only for
eigenstates of $\hat Q$,
and that $\delta q_{\rm err}$ is a set of 
$\delta r_{\rm bias}$ and $\delta r_{\rm sd}$.
}.
However, we simply say that $\delta q_{\rm err}=0$ 
when $P_R(r)=P^{\rm ideal}_R(r)$.

%%%%%%%%%%%%%%%%%%%%%%%%%%%%%%%%%%%%%%%%%%%%%%%%%%%%%%%%%%%%%%%%%%%%%%%%%%%%%%
\subsubsection{Range of measurement}
%%%%%%%%%%%%%%%%%%%%%%%%%%%%%%%%%%%%%%%%%%%%%%%%%%%%%%%%%%%%%%%%%%%%%%%%%%%%%%
\label{sec:range}

Let us denote the eigenvalue spectrum of $\hat Q$ by ${\cal Q}$, 
and the number of eigenvalues by $| {\cal Q} |$.
For example, when $\hat Q$ is the $z$ component of the 
spin of a spin-$S$ system,
${\cal Q} = \{ -S \hbar, \cdots, (S-1) \hbar, S \hbar \}$ 
and $| {\cal Q} | = 2S+1$.

Consider the case where 
the pre-measurement state is an eigenstate $| \varphi_q \rangle$ of $\hat Q$.
Then, $\delta q_{\rm err}$ can be taken as a set of 
$\delta r_{\rm bias}$ and $\delta r_{\rm sd}$.
Note that both 
$\delta r_{\rm bias}$ and $\delta r_{\rm sd}$ are generally functions
of $q$, i.e., $\delta q_{\rm err}$ varies 
in ${\cal Q}$.
Let $\delta q^*_{\rm err}$ be the upper limit of the measurement error
allowable for the purpose of the experiment.
For example, 
if $Q$ is a component of a spin and if
one wants to distinguish different spin states,
then $\delta q^*_{\rm err}$ should be less than $\hbar/2$, say 
$\delta q^*_{\rm err} = \hbar/4$. 
In this case, $\delta q^*_{\rm err}$ is of the same order of magnitude
as the minimum spacing $\Delta q_{\rm min}$ 
between the eigenvalues of $\hat Q$.
On the other hand, $\delta q^*_{\rm err} \gg \Delta q_{\rm min}$
in many optical experiments on condensed-matter physics using 
photodetectors, in which $Q$ is the photon number and thus 
$\Delta q_{\rm min}=1$.

If $\delta q_{\rm err} \leq \delta q^*_{\rm err}$ 
in a region ${\cal Q}_{\rm range}$ in ${\cal Q}$, 
we say that the {\em range} of the measurement 
(or of the measuring apparatus) is ${\cal Q}_{\rm range}$ \cite{SF,FL}.
For an ideal measurement,
$\delta q_{\rm err}=0$ everywhere in ${\cal Q}$, 
and hence ${\cal Q}_{\rm range} = {\cal Q}$.
In this case, we say that the range of the measurement 
covers the whole spectrum of $\hat Q$.
This is not necessarily the case for general measurements.
For example, a photon counter 
cannot count the photon number correctly if 
the number is too large.

Although the importance of the range of the measurement has not been 
stressed in many theoretical works, it often plays crucial roles
as stressed in Refs.~\cite{SF,FL}, and 
as will be explained in Secs.~\ref{sec:info}, \ref{sec:gim} and \ref{sec:fm}.

%%%%%%%%%%%%%%%%%%%%%%%%%%%%%%%%%%%%%%%%%%%%%%%%%%%%%%%%%%%%%%%%%%%%%%%%%%%%%%
\subsubsection{Information obtained by measurement}
%%%%%%%%%%%%%%%%%%%%%%%%%%%%%%%%%%%%%%%%%%%%%%%%%%%%%%%%%%%%%%%%%%%%%%%%%%%%%%
\label{sec:info}

We have seen that for general measurements 
the measurement error $\delta q_{\rm err}$ 
may be nonzero and the range ${\cal Q}_{\rm range}$ of the measurement 
may be narrower than the spectrum ${\cal Q}$ of the observable $\hat Q$
to be measured.
This implies that the amount $I$ of information that is 
obtained by the measurement is smaller for 
a general measurement than for an ideal measurement \cite{SF,FL, reversible}.

To see this, 
consider again the case where 
the pre-measurement state is an eigenstate $| \varphi_q \rangle$
of $\hat Q$,  for which $\delta q_{\rm err}$ is specified by %a set of 
$\delta r_{\rm bias}$ and $\delta r_{\rm sd}$.
We assume that $\delta r_{\rm bias} = 0$ for simplicity,
so that $\delta q_{\rm err} = \delta r_{\rm sd}$.
Let $J$ be the number of different eigenstates 
that can be distinguished from each other by this measurement.
 As will be illustrated shortly, $J$ depends on 
$\delta q_{\rm err}$ and ${\cal Q}_{\rm range}$. 
We may define $I$ by
\begin{equation}
I \equiv \log_2 J.
\end{equation}
Although more elaborate definition of $I$ would be possible, 
this simple definition will suffice the present discussion.

For an ideal measurement, 
$\delta q_{\rm err} =0$
and $J = | {\cal Q} |$.
Therefore, $I$ takes the maximum value,
\begin{equation}
I^{\rm ideal} = \log_2 | {\cal Q} |.
\label{Iideal}\end{equation}
For a general measurement, however, 
$I \leq I^{\rm ideal}$ in general.
For example, 
when $\delta q_{\rm err}$ is smaller 
than the minimum spacing $\Delta q_{\rm min}$
between the eigenvalues of $\hat Q$,
we have $J \simeq | {\cal Q}_{\rm range} |$, hence
\begin{equation}
I \simeq \log_2 | {\cal Q}_{\rm range} | \leq I^{\rm ideal}.
\end{equation}
When $\delta q_{\rm err} \gtrsim \Delta q_{\rm min}$, on the other hand, 
one cannot distinguish between
$| \varphi_q \rangle$ and 
$| \varphi_{q'} \rangle$ with certainty if $| q-q'| < \delta q_{\rm err}$.
Therefore, $J < | {\cal Q}_{\rm range} |$ and $I$ becomes even smaller.

It should be stressed 
that {\em an interaction process between S and A can be called 
a measurement process only when $I$ is large enough} 
(at least $I \gtrsim 1$),\footnote{
This is common to both quantum and classical physics.
}
because measurement of $Q$ is a process by which an observer gets 
information about $Q$.
For example, 
as the temperature of A is increased
$I$ is generally decreased because of the thermal noise, 
until $I \simeq 0$ at a high temperature\footnote{
This may be seen simply as follows:
Since $R$ can change through the interaction with S, 
the change of $R$ is not forbidden by a boundary condition 
which could be imposed on A.
Then, according to the fluctuation-dissipation theorem \cite{KTH},
$R$ fluctuates at a finite temperature $T$, and 
the magnitude of the fluctuation is proportional to $T$
(apart from possible $T$ dependence of the response function).
Therefore, with increasing $T$, 
$\delta r_{\rm sd}$ increases, and thus $\delta q_{\rm err}$ increases, 
and consequently $I$ decreases, approaching zero at the high-temperature
limit.}.
In such a case, 
the interaction process between S and A is not 
a measurement process 
because an observer cannot get any information about $Q$.
Hence, it should be called a {\em non-informative disturbance}
of S by A.
Another, rather trivial, example of non-informative disturbances
is the case where $\hat H_{\rm int}$ is such an interaction 
that does not generate the correlation between $Q$ and $R$.
It is obvious that such an interaction is possible.

The distinction between measurement and 
a non-informative disturbance is crucial  when 
discussing many problems about measurement, such as the 
quantum nondemolition measurement \cite{SF,FL} and the reversible
measurement \cite{reversible}.
For example, the state before the interaction with A
can be physically recovered
only for a non-informative disturbance \cite{reversible}.
In discussions of the Zeno effect, however, 
the distinction was sometimes disregarded in the literature.
That is, there are two ways of defining the Zeno effect:
one is as an effect of measurements, 
which may be called the Zeno effect {\em in the narrow sense}, 
while the other,
which may be called the Zeno effect {\em in the broad sense},
 is as an effect of any kinds of disturbances 
including non-informative disturbances.
In the latter sense, it was concluded for example that 
the Zeno effect would become stronger as the temperature of A is increased 
\cite{high-T}.
Furthermore, the well-known shortening of lifetimes  
of quasi-particles with increasing the temperature of solids could be 
called the AZE.
However, these are {\em not} the Zeno effect in the narrow sense because
one cannot get information from a high-temperature apparatus.
It should be noticed that 
universal conclusions, which are independent of details of models,
can be drawn only for the Zeno effect in the narrow sense
(see Sec.~\ref{sec:Zeno-MMT}).

%%%%%%%%%%%%%%%%%%%%%%%%%%%%%%%%%%%%%%%%%%%%%%%%%%%%%%%%%%%%%%%%%%%%%%%%%%%%%%
\subsubsection{Backaction of measurement}
%%%%%%%%%%%%%%%%%%%%%%%%%%%%%%%%%%%%%%%%%%%%%%%%%%%%%%%%%%%%%%%%%%%%%%%%%%%%%%
\label{sec:ba}

If the measurement were not made (i.e., if 
$\hat H_{\rm int}=0$), the state of S at $t=\tau$ would be given by
\begin{equation}
\hat \rho^{\rm free}
=
e^{-i \hat H_{\rm S} \tau} | \psi \rangle 
\langle \psi | e^{i \hat H_{\rm S} \tau}.
\end{equation}
When defining the backaction, however, 
$\tau$ in this expression is often taken $0$
in order to exclude the effect of the trivial change induced 
by $\hat H_{\rm S}$.
% $\tau$ for $\hat \rho^{\rm free}$ is often taken $0$
We will not specify which is used for $\hat \rho^{\rm free}$, except
when the specification is needed.

If the measurement has been made, 
the post-measurement state corresponding to each readout $r$ 
is given by Eq.\ (\ref{rho(r)general2}).
When quantifying the backaction, however, it is customary to take 
the von Neumann mixture $\hat \rho_{\rm vN}(\tau)$, 
Eq.\ (\ref{rho_vN}),
as the post-measurement state.
We call the difference between  
$\hat \rho_{\rm vN}(\tau)$ (or $\hat \rho_r(\tau)$) and $\hat \rho^{\rm free}$ 
the {\em backaction} of the measurement.
Its magnitude
should be quantified by a measure
of the difference between 
the two density operators.
For example, it may be quantified by the 
{\em quantum relative entropy} \cite{Helst,NC};
\begin{equation}
D(\hat \rho^{\rm free} \| \hat \rho_{\rm vN}(\tau))
\equiv
{\rm Tr} \left[ \hat \rho^{\rm free} \left(
\log_2 \hat \rho^{\rm free} - \log_2 \hat \rho_{\rm vN}(\tau)
\right) \right].
\label{eq:Dquantum}\end{equation}
One can also use $D( \hat \rho_{\rm vN}(\tau) \| \hat \rho^{\rm free})$, which 
is not equal to $D(\hat \rho^{\rm free} \| \hat \rho_{\rm vN}(\tau))$
 in general.
Or, one can use other measures which are used 
in quantum information theory \cite{Helst,NC}.

However, it is customary, and sometimes convenient,
to quantify the backaction in the following way. 
If an ideal measurement of an observable $X$ of S is
performed for the post-measurement state $\hat \rho_{\rm vN}(\tau)$,
its probability distribution will be
\begin{equation}
{\rm Tr} \left[ \hat \rho_{\rm vN}(\tau) \hat {\cal P}_X(x) \right]
\equiv 
P_X^{\rm vN}(x),
\end{equation}
where 
$\hat {\cal P}_X(x)$ denotes the projection operator
onto the subspace belonging to an eigenvalue $x$ of $\hat X$.
For $\hat \rho^{\rm free}$, 
on the other hand, the probability distribution would be
\begin{equation}
{\rm Tr} \left[ \hat \rho^{\rm free} \hat {\cal P}_X(x) \right]
\equiv 
P^{\rm free}_X(x).
\end{equation}
The backaction is sometimes quantified 
by the difference between 
$P_X^{\rm vN}(x)$
and $P^{\rm free}_X(x)$ 
of properly chosen observables, such as $Q$ and/or its 
canonical conjugate $P$.\footnote{
When $Q$ is a position coordinate, for example, 
$P$ is the conjugate momentum.
% When $Q$ is a component of a spin, $P$ may be taken as  
% another component of the spin.
}
In particular, the difference between
$P_Q^{\rm vN}(q)$ 
and  $P^{\rm free}_Q(q)$ is called the {\em backaction 
on the measured observable},
whereas the difference between 
$P_P^{\rm vN}(p)$
and $P^{\rm free}_P(p)$ 
may be called the {\em backaction on the conjugate observable}
\cite{Unruh,QND1,QND2}.

The difference between 
$P_X^{\rm vN}(x)$ and $P^{\rm free}_X(x)$ 
can be quantified, for example, by
the relative entropies. 
However, they are 
sometimes quantified more simply by 
the differences between the averages, 
$\langle X \rangle_{\rm vN}$ and $\langle X \rangle^{\rm free}$, 
and the variances, 
$\langle  (\Delta X)^2 \rangle_{\rm vN}$ 
and $\langle (\Delta X)^2 \rangle^{\rm free}$, 
of 
$P_X^{\rm vN}(x)$
and $P^{\rm free}_X(x)$;
\begin{eqnarray}
\delta \langle X \rangle
&\equiv&
\langle X \rangle_{\rm vN} - \langle X \rangle^{\rm free},
\\
\delta \langle  (\Delta X)^2 \rangle
&\equiv&
\langle  (\Delta X)^2 \rangle_{\rm vN} 
- \langle (\Delta X)^2 \rangle^{\rm free}.
\label{ba-delta-variance}
\end{eqnarray}
In this quantification, 
the backaction on the measured observable is represented by the set of 
$\delta \langle Q \rangle$ and $\delta \langle (\Delta Q)^2 \rangle$,
whereas the backaction on the conjugate observable
by the set of
$\delta \langle P \rangle$ and
$\delta \langle (\Delta P)^2 \rangle$.
% It should be reminded, however, that 
% $\delta \langle (\Delta Q)^2 \rangle$
% and $\delta \langle (\Delta P)^2 \rangle$ are not necessarily 
% positive in general.
%
Heisenberg used $\delta \langle (\Delta P)^2 \rangle$ 
in his famous gedanken experiment on the uncertainty principle.
It may thus be tempting to think that 
the measurement error and the backaction would be related simply by
Heisenberg's  uncertainty relation.
However, {\em this is false}, as we will explain in Sec.~\ref{sec:UCR}.

In the following, we do not specify 
the detailed quantification of the backaction
except when it is needed.

%%%%%%%%%%%%%%%%%%%%%%%%%%%%%%%%%%%%%%%%%%%%%%%%%%%%%%%%%%%%%%%%%%%%%%%%%%%%%%
\subsubsection{Instantaneous measurement and ideal measurement 
as limiting cases}
%%%%%%%%%%%%%%%%%%%%%%%%%%%%%%%%%%%%%%%%%%%%%%%%%%%%%%%%%%%%%%%%%%%%%%%%%%%%%%
\label{sec:IMasLC}

It is sometimes assumed that the response time $\tau \to +0$.
In order to get a non-vanishing information $I$, 
however, 
such an {\em instantaneous measurement} is possible only
in the limit of infinite coupling constant $\xi$ of $\hat H_{\rm int}$.
In fact, if $\xi$ is finite we have
\begin{equation}
\lim_{\tau \to +0} |\Psi(\tau) \rangle 
=
\lim_{\tau \to +0} e^{-i \hat H_{\rm S+A} \tau} 
| \psi \rangle | \psi_{\rm A} \rangle
=
| \psi \rangle | \psi_{\rm A} \rangle,
\end{equation}
which clearly shows that one cannot get any information about S by
measuring $R$ of A.
Since the coupling constant of any physical interaction is finite, 
an instantaneous measurement is, in its exact definition, an unphysical limit.
It becomes physical only in the sense that
$\tau$ is shorter than any other relevant time scales.

On the other hand, 
an ideal measurement can be regarded as the following limit of 
a general measurement;
$\delta q_{\rm err} \to 0$,
and 
${\cal Q}_{\rm range} \to {\cal Q}$,
and 
$I \to \log_2 |{\cal Q}|$,
and 
the backaction 
$\to D(\hat \rho^{\rm free} \| \hat \rho^{\rm ideal}_{\rm vN})$.
These conditions are satisfied if Eq.\ (\ref{eq:cond-IM})
is satisfied
for every $q,l,m, q',l',r', m'$.
Therefore, to realize an ideal measurement, 
one must construct an experimental setup whose
$\hat H_{\rm S+A}, \tau$ and $r_0$ satisfy this condition.
This is generally very hard and somewhat unrealistic, 
particularly when the size of A is small \cite{Wig,AY}.
Moreover, a fundamental tradeoff among 
the measurement error, 
range, 
and backaction has been suggested for measurements of 
a certain classes of physical quantities,\footnote{
Strictly speaking, such quantities 
cannot be called observables, although they {\em can} be measured,
because the word ``observable''
should be defined as a quantity for which 
an ideal measurement is possible, at least in principle,
to an arbitrarily good approximation.
} 
such as the photon number \cite{FL}.
Furthermore, it is sometimes assumed that $\tau \to +0$ for ideal
measurements,
although such a limit is unphysical as mentioned above.
To avoid confusion, we call such an ideal measurement as
an {\em instantaneous ideal measurement}.

Since most of real measurements do not satisfy 
these limiting conditions, 
it is important to explore properties of general measurements.

%%%%%%%%%%%%%%%%%%%%%%%%%%%%%%%%%%%%%%%%%%%%%%%%%%%%%%%%%%%%%%%%%%%%%%%%%%%%%%
\subsection{Various types of measurements}
%%%%%%%%%%%%%%%%%%%%%%%%%%%%%%%%%%%%%%%%%%%%%%%%%%%%%%%%%%%%%%%%%%%%%%%%%%%%%%
\label{sec:additional}

In discussing the Zeno effect, more characterizations of measurements are used,
which are explained in this subsection.

%%%%%%%%%%%%%%%%%%%%%%%%%%%%%%%%%%%%%%%%%%%%%%%%%%%%%%%%%%%%%%%%%%%%%%%%%%%%%%
\subsubsection{Direct versus indirect measurements}
%%%%%%%%%%%%%%%%%%%%%%%%%%%%%%%%%%%%%%%%%%%%%%%%%%%%%%%%%%%%%%%%%%%%%%%%%%%%%%
\label{sec:DvsI}

Suppose that S can be decomposed into two parts, S$_0$ and S$'$.
This does not necessarily mean that S$_0$ and S$'$ are spatially 
separated.
They can be, for example, 
different sets of variables such as
different quantized fields. 
Let $Q$ and $Q'$ be observables of S$_0$ and S$'$, respectively,
and assume that they are correlated strongly,
where $Q$ is the observable to be measured.
For example, 
$Q$ may be the electron energy in an excited atom, 
by measurement of which one can detect the decay of the atom, 
and $Q'$ the energy of photons emitted from the atom:
They are strongly correlated with each other because of 
the energy conservation.

Because of the strong correlation, 
the information about $Q$ can be obtained 
through either an interaction between S$_0$ and A
or another interaction between S$'$ and A.
In the former case, the measurement is called a
{\em direct measurement} because apparatus A 
interacts directly with S$_0$ which includes $Q$, 
whereas in the latter case it is called an 
{\em indirect measurement}
because A does not interact directly with S$_0$.\footnote{
It might be tempting to regard S$'$ in a indirect measurement 
as a part of a measuring apparatus.
However, this is not recommended because they are different in 
the following point:
S$'$ always couples to S$_0$, 
whereas the apparatus couples to S$_0$ (or S$'$) 
only during a measuring process.
}
When discussing decay of an unstable state, for example, 
$Q'$ may be regarded as a {\em decay product}, which 
is produced by the decay.
In such a case, an indirect measurement is a measurement 
of a decay product(s).
Note that in indirect measurements
properties of the measurement of $Q'$
become important.
For example, the range ${\cal Q}'_{\rm range}$ 
of the measurement of $Q'$ plays crucial roles 
in Secs.~\ref{sec:eim} and \ref{sec:fm}.

It is often criticized that the Zeno effect
by direct measurements is not the `genuine' Zeno effect \cite{HW},
because the appearance of change of 
$Q$ is not very surprising if an apparatus
acts directly on S$_0$.
It seems therefore that 
theories and experiments on the Zeno effect 
by indirect measurements are to be explored more intensively.

%%%%%%%%%%%%%%%%%%%%%%%%%%%%%%%%%%%%%%%%%%%%%%%%%%%%%%%%%%%%%%%%%%%%%%%%%%%%%%
\subsubsection{Positive- versus negative-result measurements}
%%%%%%%%%%%%%%%%%%%%%%%%%%%%%%%%%%%%%%%%%%%%%%%%%%%%%%%%%%%%%%%%%%%%%%%%%%%%%%
\label{sec:PvsN}

Consider an excited atom, which will emit a photon when it 
decays to the ground state.
If one monitors the decay by a photodetector that 
detects a photon emitted from the atom,
the photodetector reports no signal if the decay does not
occur.
One can confirm that the atom does not decay
by the fact that nothing happens.
Such a measurement, in which one can get information 
even when a measuring apparatus reports no signal, 
is called a {\em negative-result measurement}.
In terms of the formulation of Sec.~\ref{sec:ev_S+A}, 
this means that one can get information even when $r=r_0$.
On the other hand, 
when the spin of an electron is measured by 
the Stern-Gerlach apparatus, the apparatus reports 
either $r=+\hbar/2$ or $-\hbar/2$, whereas
$r_0$ takes another value, say $r_0 = 0$. 
Such a measurement, in which $r$ after the measurement 
is always different from $r_0$, 
is called a {\em positive-result measurement}.

The Zeno effect looks more interesting when it is 
induced by 
negative-result measurements than by positive-result measurements,
because seemingly nothing happens in the former case \cite{HW}.
In Sec.~\ref{sec:rmt}, we will analyze the Zeno effect induced by
indirect negative-result measurements.

%%%%%%%%%%%%%%%%%%%%%%%%%%%%%%%%%%%%%%%%%%%%%%%%%%%%%%%%%%%%%%%%%%%%%%%%%%%%%%
\subsubsection{Repeated instantaneous measurements 
versus continuous measurement}
%%%%%%%%%%%%%%%%%%%%%%%%%%%%%%%%%%%%%%%%%%%%%%%%%%%%%%%%%%%%%%%%%%%%%%%%%%%%%%
\label{sec:CvsD}

\begin{figure}[htbp]
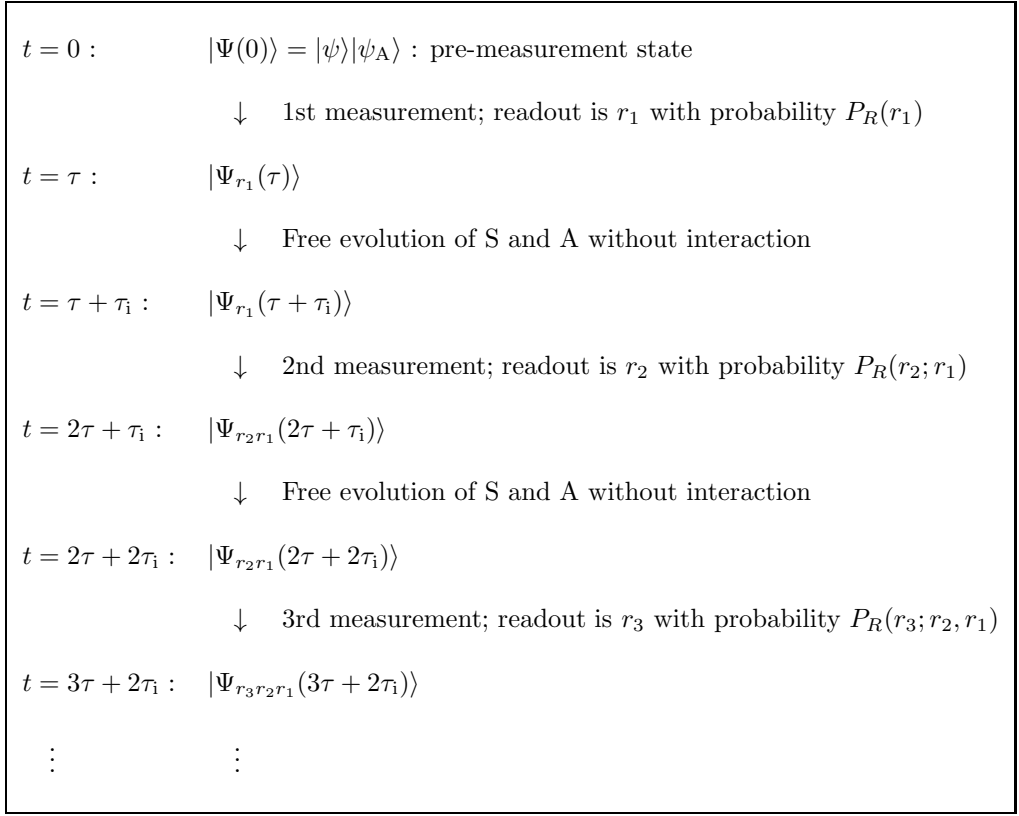

\begin{center}
\begin{tabular}[htbp]{|ll|}
\hline
& \\
$t=0$ : 
& 
$|\Psi(0) \rangle = | \psi \rangle | \psi_{\rm A} \rangle$ :
pre-measurement state 
\\ & \\
& 
\quad $\downarrow$ \quad 
1st measurement; readout is $r_1$ with probability $P_R(r_1)$
\\ & \\
$t=\tau$ : 
& 
$|\Psi_{r_1}(\tau) \rangle$ 
\\ & \\
& 
\quad $\downarrow$ \quad Free evolution of S and A without interaction
\\ & \\
$t=\tau+\tau_{\rm i}$ : 
& 
$|\Psi_{r_1}(\tau+\tau_{\rm i}) \rangle$
\\ & \\
& 
\quad $\downarrow$ \quad 
2nd measurement; readout is $r_2$ with probability $P_R(r_2; r_1)$
\\ & \\
$t=2\tau+\tau_{\rm i}$ : 
& 
$|\Psi_{r_2 r_1}(2\tau+\tau_{\rm i}) \rangle$ 
\\ & \\
& 
\quad $\downarrow$ \quad Free evolution of S and A without interaction
\\ & \\
$t=2\tau+2\tau_{\rm i}$ : 
& 
$|\Psi_{r_2 r_1}(2\tau+2\tau_{\rm i}) \rangle$
\\ & \\
& 
\quad $\downarrow$ \quad 
3rd measurement; readout is $r_3$ with probability $P_R(r_3; r_2, r_1)$
\\ & \\
$t=3\tau+2\tau_{\rm i}$ : 
& 
$|\Psi_{r_3 r_2 r_1}(3\tau+2\tau_{\rm i}) \rangle$ 
\\ & \\
\quad \vdots & \quad \vdots
\\ & \\
\hline
\end{tabular}
\end{center}
\caption{
Repeated measurements with time intervals $\tau_{\rm i}$.
}
\label{fig:rm}
\end{figure}

Assume that the Heisenberg cut is located between A and A$'$
in Fig.\ \ref{fig:vNC}.
Suppose that a measurement of Q is performed, in which 
the apparatus A interacts with S during $0 \leq t \leq \tau$
and $R$ of A is measured by A$'$ at $t = \tau$.\footnote{
Although we assume in the following equations for simplicity that 
the measurements of $R$ by $A'$ are ideal and instantaneous,
it is easy to generalize the equations to the case of 
general measurements of $R$ using, e.g., 
the operator $\hat {\cal O}_m(r)$ of the POVM measurement of
Sec.~\ref{sec:pheno}.  
}
Then, suppose that another measurement is performed, in which 
A interacts with S again during 
$\tau+\tau_{\rm i} \leq t \leq 2 \tau+\tau_{\rm i}$ 
and $R$ is again measured by A$'$ at $t = 2 \tau+\tau_{\rm i}$.
% Subsequently, the third measurement is performed during
% $2 \tau+ 2 \tau_{\rm i} \leq t \leq 3 \tau+ 2 \tau_{\rm i}$,
% and $R$ is again measured by A$'$ at $t = 3 \tau+ 2 \tau_{\rm i}$.
By repeating such sequences, one can perform {\em repeated 
measurements} of Q of S with time intervals $\tau_{\rm i}$, 
as shown in Fig.\ \ref{fig:rm}.

Repeated measurements in the limit of 
$\tau \to +0$ (while keeping $\tau_{\rm i}$ finite)
may be called {\em repeated instantaneous measurements}.\footnote{
It is sometimes called {\em pulsed measurements}.
To avoid possible confusion, however, 
we do not use this term in this article.
}
To keep $I$ of each measurement constant in this case, 
one has to increase 
the coupling constant $\xi$ of $\hat H_{\rm int}$ to infinity,
as discussed in Sec.~\ref{sec:IMasLC}.
Therefore, the repeated instantaneous measurements is a rather 
unphysical limit.

On the other hand, 
one can change $\tau_{\rm i}$ freely without changing $I$,
because $\tau_{\rm i}$ is basically independent of $I$ of each measurement.
Therefore, the limit of $\tau_{\rm i} \to +0$ is 
a physical and realistic limit, which is widely performed in real experiments.
Since the apparatus A interacts continuously with S in such 
repeated measurements, it may be called a 
{\em continuous measurement}.\footnote{
This term is widely used when 
A interacts continuously with S, 
even when 
% the measurements of $R$ by $A'$ are ideal and instantaneous,
the times and properties 
(i.e., whether ideal or general and whether instantaneous or not) 
of measurements of $R$ by $A'$ are 
not specified.
As will be shown in Sec.~\ref{sec:ua}, 
such times and properties become irrelevant
if one employs the `unitary approximation' % (Sec.~\ref{sec:ua}) 
and is interested only in $\langle X \rangle_{\rm vN}$
and/or $\langle R \rangle_{\rm vN}$,  
where $X$ is an observable of S.
% as will be shown in Sec.~\ref{sec:foarmluaCM}. 
} 

Despite the above-mentioned difference between the two limits, 
it is sometimes argued that 
repeated instantaneous measurements 
for which $(\tau_{\rm i}, \tau)=(T,0)$
is equivalent to 
continuous measurement 
for which $(\tau_{\rm i}, \tau)=(0, K T)$, where $K$
is a positive constant of order unity.  
However, this equivalence holds only for certain limited cases.
In fact, the results of Sec.~\ref{sec:rmt} show that 
they are not equivalent, sometimes much different, in general.

Note that the definition of the response time $\tau$ becomes 
ambiguous in the case of continuous measurement, because
$\hat H_{\rm int}$ is effective for all $t \geq 0$.
In this case, $\tau$ may be defined as the time scale $\tau_{\rm r}$ 
on which the probability distribution $P_R(r)$ becomes significantly 
different from the initial distribution $P_R(r) = \delta_{r, r_0}$.
Although only the order of magnitude can be determined according 
to this definition, it suffices for discussions on the Zeno effect
induced by continuous measurement.
Therefore, we will use this definition in Sec.~\ref{sec:rmt}.

%%%%%%%%%%%%%%%%%%%%%%%%%%%%%%%%%%%%%%%%%%%%%%%%%%%%%%%%%%%%%%%%%%%%%%%%%%%%%%%
% \subsubsection{Formulas for repeated measurements}
%%%%%%%%%%%%%%%%%%%%%%%%%%%%%%%%%%%%%%%%%%%%%%%%%%%%%%%%%%%%%%%%%%%%%%%%%%%%%%%
% \label{sec:foarmluaCM}
% 

Properties of repeated measurements can be calculated simply 
as a sequence of general measurements, which we have discussed so far.
In fact, by repeatedly applying 
Eqs.\ (\ref{Psi(tau)-0}), (\ref{P(r)general}) and (\ref{Psi(r)general})
(with $\tau'=\tau$), 
we obtain
\begin{equation}
P_R(r_{n}; r_{n-1}, \cdots, r_1)
=
\left\| 
\hat {\cal P}_R(r_{n}) 
e^{-i \hat H_{{\rm S}+{\rm A}} \tau} 
|\Psi_{r_{n-1} \cdots r_1}((n-1)(\tau+\tau_{\rm i})) \rangle
\right\|^2,
\label{PR-repeated}\end{equation}
\begin{equation}
|\Psi_{r_{n} \cdots r_1}(n\tau+(n-1)\tau_{\rm i}) \rangle
=
{
\hat {\cal P}_R(r_{n}) 
e^{-i \hat H_{{\rm S}+{\rm A}} \tau} 
\over
\sqrt{P_R(r_{n}; r_{n-1}, \cdots, r_1)}
}
|\Psi_{r_{n-1} \cdots r_1}((n-1)(\tau+\tau_{\rm i})) \rangle,
\label{Psi-repeated}\end{equation}
\begin{equation}
|\Psi_{r_{n} \cdots r_1}(n(\tau+\tau_{\rm i})) \rangle
=
e^{-i \hat (H_{\rm S}+H_{\rm A})\tau_{\rm i}}
|\Psi_{r_{n} \cdots r_1}(n\tau+(n-1)\tau_{\rm i}) \rangle,
\label{Psi-Psi-repeated}\end{equation}
for $n = 1, 2, \cdots$.
From these formulas, 
one can calculate everything about repeated measurements, 
including the Zeno effect.
For example, 
the expectation value 
$\langle W \rangle_{r_n \cdots r_1}$
of an observable $W$ (of S or A)
for the state after $n$ measurements, for which  
the readouts are $r_1, \cdots, r_n$, 
is given by
\begin{equation}
\langle W \rangle_{r_n \cdots r_1}
=
\langle \Psi_{r_{n} \cdots r_1}(n\tau+(n-1)\tau_{\rm i}) |
\hat W
|\Psi_{r_{n} \cdots r_1}(n\tau+(n-1)\tau_{\rm i}) \rangle.
\label{<K>rnr1}\end{equation}

\subsubsection{Unitary approximation}
%%%%%%%%%%%%%%%%%%%%%%%%%%%%%%%%%%%%%%%%%%%%%%%%%%%%%%%%%%%%%%%%%%%%%%%%%%%%%%%
\label{sec:ua}

In some cases, one is only interested in the average 
of $\langle W \rangle_{r_n \cdots r_1}$
over all possible values of the readouts.
Such an average $\langle W \rangle_{\rm vN}$ is given by
\begin{equation}
\langle W \rangle_{\rm vN}
=
\sum_{r_1, \cdots, r_n}
P_R(r_{n}; r_{n-1}, \cdots, r_1)
P_R(r_{n-1}; r_{n-2}, \cdots, r_1)
\cdots 
P_R(r_1)
\langle W \rangle_{r_n \cdots r_1}.
\label{<K>-1}\end{equation}
Using Eq.~(\ref{PR-repeated}), 
and taking $\tau_{\rm i}=0$ for simplicity,\footnote{
The corresponding formula for $\tau_{\rm i}>0$ can also be 
obtained easily.}
we can rewrite this equation as
\begin{eqnarray}
\langle W \rangle_{\rm vN}
&=&
\! \! \!
\sum_{r_1, \cdots, r_n}
P_R(r_{n-1}; r_{n-2}, \cdots, r_1)
\cdots 
P_R(r_1)
\nonumber\\
& & \qquad \times \ 
\langle \Psi_{r_{n-1} \cdots r_1}((n-1)\tau) |
e^{i \hat H_{{\rm S}+{\rm A}} \tau} \hat {\cal P}_R(r_{n}) 
\hat W
\hat {\cal P}_R(r_{n}) e^{-i \hat H_{{\rm S}+{\rm A}} \tau} 
|\Psi_{r_{n-1} \cdots r_1}((n-1)\tau) \rangle,
\nonumber\\
&=&
\! \! \!
\sum_{r_1, \cdots, r_n}
\langle \Psi(0) |
e^{i \hat H_{{\rm S}+{\rm A}} \tau} \hat {\cal P}_R(r_{1}) 
\cdots
e^{i \hat H_{{\rm S}+{\rm A}} \tau} \hat {\cal P}_R(r_{n}) 
\hat W
\hat {\cal P}_R(r_{n}) e^{-i \hat H_{{\rm S}+{\rm A}} \tau} 
\cdots
\hat {\cal P}_R(r_{1}) e^{-i \hat H_{{\rm S}+{\rm A}} \tau}
|\Psi(0) \rangle.
\nonumber\\
\label{<K>-2}
\end{eqnarray}
% In particular, when $W=X$ or $R$, where $X$ is an observable of S, 
% $\hat W$ commutes with $\hat {\cal P}_R(r_{j})$,
% and thus Eq.~(\ref{<K>-2}) reduces to 
% \begin{equation}
% \langle W \rangle_{\rm vN}
% =
% \langle \Psi(0) |
% e^{i \hat H_{{\rm S}+{\rm A}} n \tau}
% \hat W
% e^{-i \hat H_{{\rm S}+{\rm A}} n \tau} 
% |\Psi(0) \rangle
% \quad \mbox{for } \hat W = \hat R \mbox{ or } \hat X. 
% \label{<K>-3}
% \end{equation}
% Therefore, in this case 
% it is sufficient to calculate only the unitary part of the measurement process,
% and one can forget about the ideal measurement of $R$ by A$'$,
% for which we have used the projection postulate.
% That is, 
% {\em 
% if one is interested only in $\langle X \rangle_{\rm vN}$
% and/or $\langle R \rangle_{\rm vN}$,
% which are respectively the averages of $\langle X \rangle_r$
% and $\langle R \rangle_r$
% over possible values of the readout $r$, 
% one does not have to use the projection postulate at all.
% }
% This simple result will be fully utilized in Sec.~\ref{sec:rmt}.
% Note, however, that 
% if one is interested in other quantities, such 
% as the time correlation (Sec.~\ref{sec:timecorr}), 
% one needs to use the general result, 
% Eqs.~(\ref{PR-repeated})-(\ref{Psi-Psi-repeated}). 
% 
% 
% 
% 
% 
% 
% 
% 
% 
% Consider a continuous measurement 
% of an observable $W$ of S or A.
% If one is interested only in its average 
% over all possible values of the readouts, 
% such an average $\langle W \rangle_{\rm vN}$ is given by Eq.~(\ref{<K>-2}).

In practical calculations, this is often approximated by\footnote{
Unlike formula (\ref{<X>-commute})
for single measurement,
Eq.~(\ref{<K>-3}) is not a rigorous formula 
because
$\hat R$ does not commute with $\hat H_{{\rm S}+{\rm A}}$
(since if they did then $R$ would not change by the interaction, 
and thus no information would be transferred to $R$), 
except for the trivial case where
$[\hat X, \hat H_{{\rm S}+{\rm A}}] = 0$ and $\hat W = \hat X$, 
for which 
$\langle W \rangle_{\rm vN} = \langle \Psi(0) | \hat W |\Psi(0) \rangle$.
} 
\begin{eqnarray}
\langle W \rangle_{\rm vN}
\simeq
\langle \Psi(0) |
e^{i \hat H_{{\rm S}+{\rm A}} n \tau} 
\hat W
e^{-i \hat H_{{\rm S}+{\rm A}} n \tau} 
|\Psi(0) \rangle.
\label{<K>-3}
\end{eqnarray}
%% This approximation, which may be called 
%% the {\em unitary approximation}, is widely used
%% in studying the Zeno effects.
According to this approximate formula, 
one can evaluate $\langle W \rangle_{\rm vN}$ 
by simply calculating the unitary evolution,  
generated by $e^{-i \hat H_{{\rm S}+{\rm A}} t}$,
of the initial state $|\Psi(0) \rangle$ of the composite system S$+$A.
That is, one does not have to use the projection postulate at all.
We thus call 
this approximation the {\em unitary approximation}.
For {\em each} model of S$+$A, 
the validity of this approximation can % in principle 
be checked by comparing the result obtained from Eq.~(\ref{<K>-3}) with 
that obtained from Eq.~(\ref{<K>-2}).

Note, however, that {\em general} justification of the unitary approximation is 
not so simple.
In Eq.~(\ref{<K>-2}), the role of the sum of the projection operators
$\sum_{r_{j}} \hat {\cal P}_R(r_{j})$ 
($j=1, 2, \cdots, n$)
is to destroy quantum interference between states 
corresponding to different values of $r_{j}$'s.
Therefore, if environments surrounding S$+$A are taken into account, 
decoherence by the environments would induce the same effects as 
theirs \cite{dec}.
It might thus be tempting to consider that one could 
reduce Eq.~(\ref{<K>-2}) to Eq.~(\ref{<K>-3}) simple by
using this equivalence.
% replacing the projection operators with the decoherence effects
% by the environments.
However, such decoherence effects generally induce noise terms in the 
Schr\"odinger equation, which thus turns into a stochastic one.
In general, the time evolution by such a stochastic Schr\"odinger equation
cannot be described by a unitary evolution such as Eq.~(\ref{<K>-3}).
In particular, 
Eq.~(\ref{<K>-3}) is obviously wrong in the limit of $\tau \to 0$
% Furthermore, there are certain cases for which the unitary approximation is
%% obviously wrong.
%% For example, 
%% if $\tau$ in Eq.~(\ref{<K>-2}) is extremely short,
%% then Eq.~(\ref{<K>-3}) is obviously wrong 
because then the Zeno effect on $R$ should take place.
Therefore, to show the general validity of the unitary approximation 
one needs to show that 
Eq.~(\ref{<K>-2}) (or the stochastic Schr\"odinger equation)
reduces to Eq.~(\ref{<K>-3}) under certain conditions. 
%% and that the conditions are quite likely to be satisfied.
%% the unitary approximation should be used with a certain care.

When $W=X$ or $R$, where $X$ is an observable of S,
a sufficient condition for the validity of the unitary approximation 
would be that $R$ is a macroscopic variable, for the following reason.
If $R$ is a macroscopic variable, 
the quantum interference destroyed by 
$\sum_{r_{j}} \hat {\cal P}_R(r_{j})$ 
is that between macroscopically distinct states.
Such quantum interference can become significant only in
limited cases such as 
(i) a certain observable is measured which can detect 
such interference, 
(ii) the state will evolve back 
closely to the initial state,
or (iii) the Zeno effect on $R$ occurs.
Neither $X$ nor $R$ can be such an observable of case (i).
Furthermore, 
it seems unlikely that 
the `recurrent' process of case (ii) could occur 
in the time scale of a practical value of $\tau$
if $R$ is a macroscopic variable, 
because then A is a macroscopic system
which generally has many degrees of freedom and complicated dynamics.
Moreover, 
the time scale of case (iii) seems much shorter than
the time scale of the Zeno effect on $Q$, which is 
a microscopic variable.
% The time scale of the latter `recurrent' process is 
% quite long (e.g., an exponential function of the 
% degrees of freedom) for macroscopic systems
% (which have many degrees of freedom and complicated dynamics), 
% It may be worth mentioning that 
% the recurrence should be impossible in almost all situations
% even long after $\tau$
% because small perturbations from environments would destroy 
% quantum coherence which is necessary to the recurrence.

Actually, the unitary approximation is used widely 
in studying the Zeno effect without confirming its validity,
even when $R$ is taken as a microscopic variable.
However, it is generally believed (and confirmed empirically) that 
% It is known empirically that 
results obtained by this approximation 
are much better than
the results of a naive application of the projection postulate on S. 
In Sec.~\ref{sec:rmt}, 
we will employ the unitary approximation and discuss its validity
for the proposed model.

\subsection{A simple explanation of the Zeno effect
using the quantum measurement theory}
% \subsection{Mechanism of QZE using quantum measurement theory}
\label{sec:Zeno-MMT}

When a quantum system S is not measured, 
its state $| \psi \rangle$ evolves freely as
$| \psi(t) \rangle = \exp[-i \hat H_{\rm S} t] | \psi \rangle$,
and the expectation value of an observable $Q$ of S evolves as
$\langle \psi(t) |\hat Q | \psi(t) \rangle \equiv 
\langle Q(t) \rangle^{\rm free}$.
To measure $Q$, on the other hand, 
one must couple S with an apparatus A via an interaction
$\hat H_{\rm int}$ between them
in such a way that 
non-vanishing correlation is established between $Q$ and 
an readout $R$ of A,
as discussed in Sec.~\ref{sec:ev_S+A}.
As a result, S and A evolves as a coupled system as
$| \Psi(t) \rangle = 
\exp[-i (\hat H_{\rm S} + \hat H_{\rm int} + \hat H_{\rm A}) t] 
| \psi \rangle | \psi \rangle_{\rm A}$,
and 
the expectation value of $Q$ now evolves as
$\langle \Psi(t) |\hat Q | \Psi(t) \rangle \equiv 
\langle Q(t) \rangle^{\rm int}$.
Since $Q$ and $R$ undergo a coupled motion by 
$\hat H_{\rm int}$, 
$\langle Q(t) \rangle^{\rm int}$ generally evolves 
differently from 
$\langle Q(t) \rangle^{\rm free}$.
Even when A reports no signal (i.e., a negative-result measurement), 
the presence of $\hat H_{\rm int}$ does have an effect.
Obviously, the effect 
becomes larger as 
the coupling constant $\xi$ of $\hat H_{\rm int}$ is increased.

Suppose that one performs repeated measurements,
each takes $\tau$ seconds, with vanishing intervals
($\tau_{\rm i}=0$).
To make the measurements more frequent (i.e., to reduce $\tau$)
without reducing the amount of information $I$ obtained by each
measurement, one must increase $\xi$.
Therefore, as $\tau$ is decreased without reducing $I$, 
the effect of $\hat H_{\rm int}$ on S becomes larger.
If $\tau$ can thus be reduced sufficiently short 
by increasing $\xi$ enough,
the difference between 
$\langle Q(t) \rangle^{\rm int}$ and $\langle Q(t) \rangle^{\rm free}$
(for a given $t$) becomes larger, 
and
it will become possible to detect the difference by experiments.
This is the Zeno effect when $Q$ represents 
an observable that distinguishes the status (whether 
the system is decayed or not) of an unstable state.

Note that an {\em arbitrary} interaction 
with an external system does {\em not} necessarily 
affect the expectation value of $Q$.
The essence of the Zeno effect is that 
{\em the form of $\hat H_{\rm int}$ is limited 
and its strength $\xi$ is lower bounded}
by the requirement
that $\hat H_{\rm int}$ 
should create sufficient correlation between $Q$ and $R$
in order to extract non-vanishing information.
For this reason, the lifetime is {\em always} modified
in the limit of 
$\tau_{\rm i} \to 0$ and $\tau \to 0$, 
because this implies $\xi \to \infty$.
Such a universal conclusion can never be drawn for 
general interactions with (or perturbations from)
external systems.

However, as explained in Sec.~\ref{sec:IMasLC}, 
an instantaneous ideal measurement is an unrealistic limit of 
real measurements.
Therefore, the following questions need to be answered:
(i) Is the Zeno effect induced by real measurements?
(ii) Under what conditions does it occur? 
(iii) How does the decay rate of the unstable state behave
as a function of the measurement parameters, 
such as the measurement error, response time, range, and so on?
We will answer these questions in Secs.~\ref{sec:rmt} and 
\ref{sec:cqed}.

Note that for analyzing the Zeno effect it is sufficient to 
calculate the averages, over all possible values of the readout,
of expectation values of a few observables.
In fact, one is most interested in the lifetime of an unstable state,
which is the average time at which the decay occurs.
This can be expressed as the average of the expectation value
of an appropriate observable.
Therefore, as explained in Sec.~\ref{sec:ua},
{\em for analyzing the Zeno effect 
it is sufficient to calculate the unitary part of the measurement process
if one employs the unitary approximation}.\footnote{
As explained there, this approximation should be good if
$R$ is taken as a macroscopic variable. 
}
That is, unlike the conventional theories of Sec.~\ref{sec:ct}, 
one does not have to use the projection postulate. % at any points 
%in the analysis of the Zeno effect.
To {\em understand} this point, however, 
the full framework, which we have explained so far in this section,
 of the quantum measurement theory is necessary.

\subsection{Additional comments}
\label{sec:Gcmt}

We have explained all things necessary to apply 
the quantum measurement theory to the Zeno effect.
To be more complete, however, 
we will describe a few more points which will help the reader.

%%%%%%%%%%%%%%%%%%%%%%%%%%%%%%%%%%%%%%%%%%%%%%%%%%%%%%%%%%%%%%%%%%%%%%%%%%%%%%
\subsubsection{Non-triviality of the uncertainty relations}
%%%%%%%%%%%%%%%%%%%%%%%%%%%%%%%%%%%%%%%%%%%%%%%%%%%%%%%%%%%%%%%%%%%%%%%%%%%%%%
\label{sec:UCR}

The Zeno effect is a sort of backaction of measurements.
It might thus be tempting to think that 
the Zeno effect could simply be described using 
the uncertainty relations.
However, this is false.
We here explain this point, assuming 
the canonical commutation relation
$[\hat Q, \hat P]=i \hbar$ for simplicity. 

The uncertainty relation that is described in most textbooks is the 
following inequality,
\begin{equation}
\delta q \, \delta p \geq \hbar/2
\label{eq:UCR1}\end{equation}
Here, 
$(\delta q)^2 \equiv \langle \psi | (\Delta \hat Q)^2 | \psi \rangle$
and 
$(\delta p)^2 \equiv \langle \psi | (\Delta \hat P)^2 | \psi \rangle$,
where 
$\Delta \hat Q \equiv \hat Q - \langle \psi | \hat Q | \psi \rangle$
and 
$\Delta \hat P \equiv \hat P - \langle \psi | \hat P | \psi \rangle$.
This inequality is derived directly from $[\hat Q, \hat P]=i \hbar$.
On the other hand, 
in his famous gedanken experiment on the uncertainty principle,
Heisenberg claimed the following inequality,
\begin{equation}
\delta q_{\rm err} \, \delta p_{ba} \geq \hbar/2,
\label{eq:UCR2}\end{equation}
where $\delta q_{\rm err}$ and $\delta p_{ba}$ are
the measurement error in a measurement of $Q$ and
its backaction on $P$, respectively, which are quantified by 
the square roots of 
Eqs.\ (\ref{me-delta-variance}) and (\ref{ba-delta-variance}).

As stressed by Lamb \cite{Lamb69}, inequalities
(\ref{eq:UCR1}) and (\ref{eq:UCR2}) are totally different
from each other.
In the former, $\delta q$ and $\delta p$ represent 
the standard deviation of experimental data that are obtained
from {\em error-less} measurements of $Q$ and $P$, respectively,
which are performed independently of each other.
% (Here, 
% the measurement of $Q$ and that of $P$ are performed not simultaneously 
% but separately.)
Properties of the measuring apparatus are not included at all.
In this sense, inequality (\ref{eq:UCR1}) can be understood as
the uncertainty relation {\em of the pre-measurement state}.
On the other hand, $\delta q_{\rm err}$ and $\delta p_{ba}$ 
in inequality (\ref{eq:UCR2}) are
the measurement error and backaction, respectively, 
{\em of the measuring apparatus}.
They are obviously different from $\delta q$ and $\delta p$;
e.g., $\delta q_{\rm err}$ can be large even when $\delta q=0$.
Furthermore, inequality (\ref{eq:UCR1}) is never violated
by any quantum states whereas relation (\ref{eq:UCR2}) {\em can} be violated.
For example, 
suppose that we have an approximately
error-less measuring apparatus A$^{\rm errless}$ of $Q$ of a 
particle, 
and a momentum modulator M, which limits the range 
of momentum in some finite range.
% is a band-pass filter of the wavenumber 
% of the wavefunction.
We can let the pre-measurement wavefunction enter in A$^{\rm errless}$, 
and then pass through M.
Since the location of the Heisenberg cut and the 
time at which the measurement is completed can be taken arbitrary,
we can regard this composite system A$^{\rm errless}+$M as 
a single apparatus A.
For this apparatus, $\delta q_{\rm err} \simeq 0$ (by A$^{\rm errless}$)
whereas $\delta p_{ba}$ is upper limited (by M).
Therefore, 
$\delta q_{\rm err} \, \delta p_{ba} \simeq 0$, and 
inequality (\ref{eq:UCR2}) is violated.

As might be understood from this simple example, 
one can construct many different ``uncertainty products''
by combining two of 
$\delta q, \delta q_{\rm err}, \delta q_{\rm ba}$,
$\delta p, \delta p_{\rm err}$, and $\delta p_{\rm ba}$.
The lower limits, if exist, of different uncertainty products can have 
different values.
Furthermore, 
inequality (\ref{eq:UCR1}) assumes that 
the measurements of $Q$ and $P$ are performed not simultaneously 
but separately.
When $Q$ and $P$ are measured simultaneously, 
on the other hand, the uncertainty becomes larger as  
$
\delta q \, \delta p \geq \hbar
$ 
\cite{simultaneousM}.
To explore these uncertainty products, 
the quantum measurement theory is necessary.
Recently, Ozawa \cite{ozawa} have found 
certain universal relations among them, using 
the rather mathematical definition (\ref{eq:ozawa}).

It is clear from these considerations that
the Zeno effect cannot be discussed simply
using uncertainty relations.

\subsubsection{Measurement of time correlations}
\label{sec:timecorr}

Suppose that an observable $X$ is measured at $t=0$ 
using an apparatus $A_X$, 
and subsequently another observable $Y$ is measured at time $t$ ($>0$)
using another apparatus $A_Y$.
The expectation value of the product of the two readouts $R_X$ and $R_Y$
is called the {\em time correlation}, which we denote as
$\langle Y(t) X(0) \rangle$.
As a quantum-theoretical expression of this quantity, 
the following one is often employed:
\begin{equation}
\langle \psi_{\rm H} | 
\hat Y_{\rm H}(t) \hat X_{\rm H}(0) | \psi_{\rm H} \rangle,
\label{tc-naive}\end{equation}
where the subscript H denotes the corresponding quantities in the 
Heisenberg picture.
However, as stressed first by Glauber \cite{Glauber}, 
this expression is wrong except for a special case.
The correct expression is obtained, 
in a manner similar to discussions of Sec.~\ref{sec:CvsD}, 
as follows.

From Eqs.\ (\ref{Psi(tau)-0}) and (\ref{P(r)general})
(with $\tau'=\tau \equiv \tau_X$, the response time of A$_X$), 
the probability distribution of the value $r_X$ of $R_X$ is given by
\begin{equation}
P_{R_X}(r_X)
=
\left\| 
\hat {\cal P}_{R_X}(r_X) 
e^{-i \hat H_{{\rm S}+{\rm A}} \tau_X} 
| \psi \rangle | \psi_{\rm A} \rangle
\right\|^2,
\end{equation}
where A denotes the joint system of A$_X$ and A$_Y$, % $\equiv$ A$_X$+A$_Y$,
and the post-measurement state is
\begin{equation}
| \Psi_{r_X}(\tau_X) \rangle 
= 
{1 \over \sqrt{P_{R_X}(r_X)}}
\hat {\cal P}_{R_X}(r_X) 
e^{-i \hat H_{{\rm S}+{\rm A}} \tau_X} 
| \psi \rangle | \psi_{\rm A} \rangle.
\end{equation}
This state evolves into 
$
e^{-i \hat (H_{\rm S}+H_{\rm A})(t-\tau_X)} 
| \Psi_{r_X}(\tau_X) \rangle 
$
at time $t$ ($\geq \tau_X$),
which becomes the pre-measurement state of the 
measurement of $Y$. 
The probability distribution of the value $r_Y$ of $R_Y$ is 
therefore given by
\begin{equation}
P_{R_Y}(r_Y; r_X)
=
\left\| 
\hat {\cal P}_{R_Y}(r_Y) 
e^{-i \hat H_{{\rm S}+{\rm A}} \tau_Y} 
e^{-i \hat (H_{\rm S}+H_{\rm A})(t-\tau_X)} 
| \Psi_{r_X}(\tau_X) \rangle 
\right\|^2,
\end{equation}
where $\tau_Y$ is the response time of A$_Y$.
From these equations, 
the time correlation is calculated as
\begin{eqnarray}
\langle Y(t) X(0) \rangle
&=&
\sum_{r_X, r_Y} r_Y r_X P_{R_Y}(r_Y; r_X) P_{R_X}(r_X)
\nonumber\\
&=&
\sum_{r_X, r_Y} r_Y r_X 
\left\| 
\hat {\cal P}_{R_Y}(r_Y) 
e^{-i \hat H_{{\rm S}+{\rm A}} \tau_Y} 
e^{-i \hat (H_{\rm S}+H_{\rm A})(t-\tau_X)} 
\hat {\cal P}_{R_X}(r_X) 
e^{-i \hat H_{{\rm S}+{\rm A}} \tau_X} 
| \psi \rangle | \psi_{\rm A} \rangle
\right\|^2.
\label{tc-correct}\end{eqnarray}

If both measurements are ideal and instantaneous\footnote{
As noted in Sec.~\ref{sec:IMasLC}, this means that 
the coupling constant $\xi$ of $\hat H_{\rm int}$ is infinite,
and hence $\hat H_{{\rm S}+{\rm A}} \tau_X \not\to 0$
although $\tau_X \to 0$.
}, 
and if $[\hat X, \hat Y]=[\hat X, \hat H_{\rm S}]=0$, 
then from Eq.~(\ref{PRPSI-ideal}) this formula reduces to
\begin{equation}
\langle Y(t) X(0) \rangle
=
\sum_{x, y} y x
\left\| 
\hat {\cal P}_Y(y) e^{-i \hat H_{\rm S} t} \hat {\cal P}_X(x) | \psi \rangle
\right\|^2
=
\langle \psi_{\rm H} | 
\hat Y_{\rm H}(t) \hat X_{\rm H}(0) | \psi_{\rm H} \rangle.
\end{equation}
For general measurements, however, one must use
the correct formula (\ref{tc-correct}), which states
that the value of the time correlation depends on 
properties of the measuring apparatus.
In particular, the value strongly depends on the backaction of A,
because it determines the post-measurement state 
$| \Psi_{r_X}(\tau_X) \rangle$, which evolves into 
the pre-measurement state of the subsequent measurement of $Y$. 
Therefore, if one has two sets of pieces of apparatus
(A$_X$, A$_Y$) and (A$'_X$, A$'_Y$),
the value of $\langle Y(t) X(0) \rangle$ depends on 
which set is used as the measuring apparatus,
even when their measurement errors %A$_X$, A$_Y$, A$'_X$, A$'_Y$ 
are negligibly small. 
% principal origin of 
% this dependence is {\em not} the measurement errors:
% Suppose that one has two sets of pieces of apparatus
% (A$_X$, A$_Y$) and (A$'_X$, A$'_Y$).
% Even when their measurement errors %A$_X$, A$_Y$, A$'_X$, A$'_Y$ 
% are negligibly small,
% the value of $\langle Y(t) X(0) \rangle$ depends on 
% which set is used as the measuring apparatus.
Examples and experimental demonstrations of this fact are presented, e.g., 
in books on quantum optics \cite{QNoise,Mandel}.

An important implication of the discussions of this subsection is
that the Zeno effect would also depend on properties
of measuring apparatus.
This is indeed the case, as will be demonstrated in Sec.~\ref{sec:rmt}.

%%%%%%%%%%%%%%%%%%%%%%%%%%%%%%%%%%%%%%%%%%%%%%%%%%%%%%%%%%%%%%%%%%%%%%%%%%%%%%
\subsubsection{POVM measurement}
%%%%%%%%%%%%%%%%%%%%%%%%%%%%%%%%%%%%%%%%%%%%%%%%%%%%%%%%%%%%%%%%%%%%%%%%%%%%%%
\label{sec:pheno}

From Eqs.\ (\ref{Psi(tau)-0}) and (\ref{P(r)general})
(with $\tau'=\tau$), the probability distribution 
of the readout can be expressed as
\begin{eqnarray}
P_R(r)
&=& 
{\rm Tr}_{\rm S+A}  \left[
\hat {\cal P}_R(r) 
e^{-i \hat H_{\rm S+A} \tau} | \psi \rangle | \psi_{\rm A} \rangle
\langle \psi_{\rm A} | \langle \psi | e^{i \hat H_{\rm S+A} \tau} 
\hat {\cal P}_R(r) 
\right]
\nonumber\\
&=& 
{\rm Tr} \left[
\sum_m \langle r,m|
e^{-i \hat H_{\rm S+A} \tau} | \psi \rangle | \psi_{\rm A} \rangle
\langle \psi_{\rm A} | \langle \psi | e^{i \hat H_{\rm S+A} \tau} 
|r,m \rangle
\right],
\end{eqnarray}
where ${\rm Tr}_{\rm S+A}$ and ${\rm Tr}$ denotes the trace operations 
over ${\bf H}_{\rm S+A}$ and ${\bf H}_{\rm S}$, respectively.
If we define the operator $\hat {\cal O}_m(r)$ on ${\bf H}_{\rm S}$ by
\begin{equation}
\hat {\cal O}_m(r) | \psi \rangle
\equiv
\langle r,m|
e^{-i \hat H_{\rm S+A} \tau}  | \psi \rangle | \psi_{\rm A} \rangle
\quad \mbox{ for } \forall | \psi \rangle \in {\bf H}_{\rm S},
\label{def-Om}\end{equation}
then the above equation can be written as
\begin{equation}
P_R(r)
=
{\rm Tr} \left[
\sum_m \hat {\cal O}_m(r)
\hat \rho(0)
\hat {\cal O}^\dagger_m(r)
\right],
\label{P_R(r)-POVM}\end{equation}
where $\hat \rho(0) = | \psi \rangle \langle \psi |$.
Furthermore, the post-measurement state can be expressed as
\begin{eqnarray}
\hat \rho_r(\tau) 
&=&
{1 \over P_R(r)}
{\rm Tr}_{\rm A} \left[
\hat {\cal P}_R(r) 
e^{-i \hat H_{\rm S+A} \tau} | \psi \rangle | \psi_{\rm A} \rangle
\langle \psi_{\rm A} | \langle \psi | e^{i \hat H_{\rm S+A} \tau} 
\hat {\cal P}_R(r) 
\right]
\nonumber\\
&=&
{1 \over {\rm Tr} \left[
\sum_m \hat {\cal O}_m(r)
\hat \rho(0)
\hat {\cal O}^\dagger_m(r)
\right]
}
\sum_m \hat {\cal O}_m(r)
\hat \rho(0)
\hat {\cal O}^\dagger_m(r).
\end{eqnarray}
Therefore, we can calculate both $P_R(r)$ and $\hat \rho_r(\tau)$ 
from the pre-measurement state $\hat \rho(0)$ 
if the set of operators $\{ \hat {\cal O}_m(r) \}$ is given.
General properties of $\{ \hat {\cal O}_m(r) \}$ is easily obtained from
its definition (\ref{def-Om}). For example, 
\begin{equation}
\sum_r \sum_m \hat {\cal O}^\dagger_m(r) \hat {\cal O}_m(r) 
=
\hat 1
\end{equation}
because
$
\sum_r \sum_m \langle \psi_1 | \hat {\cal O}^\dagger_m(r)
\hat {\cal O}_m(r) | \psi_2 \rangle
=
\langle \psi_1 | \psi_2 \rangle
$
for arbitrary vectors $| \psi_1 \rangle$ and $| \psi_2 \rangle$.

If $\Delta$ is a set of values of $r$, then
the probability of getting the readout in $\Delta$ is given by
\begin{equation}
\sum_{r \in \Delta} P_R(r)
=
{\rm Tr} \left[
\sum_{r \in \Delta}
\sum_m \hat {\cal O}_m(r)
\hat \rho(0)
\hat {\cal O}^\dagger_m(r)
\right].
\label{P_R(r)-POVM-Delta}\end{equation}
Helstrom \cite{Helst} derived a similar 
expression in a different manner,
by considering mathematical requirements for general measurements. 
He called the association between $\Delta$ and the linear map
\begin{equation}
\hat \rho
\mapsto
\sum_{r \in \Delta} \sum_m \hat {\cal O}_m(r)
\hat \rho
\hat {\cal O}^\dagger_m(r)
\end{equation}
a {\em positive operator-valued measure} (POVM).
Therefore, a general measurement is sometimes called 
a {\em POVM measurement}.

One can in principle calculate the correct POVM using
Eq.\ (\ref{def-Om}) for each model of the measurement process.
However, 
for certain purposes, 
it is sufficient to {\em assume} some reasonable form of 
the POVM by hand.
This simplifies discussions greatly.
Such a phenomenological theory is widely used, e.g.,  in
quantum information theory \cite{Helst,NC}.
The Zeno effect can also be analyzed using 
such a phenomenological theory, although we shall not
use it in this article.

\subsubsection{Completeness of the standard laws of quantum theory}

In concluding this section, we want to stress that 
the results of this section 
show the completeness of the standard laws of quantum theory, 
which include Born's rule and the projection postulate.
One can surely obtain the correct results
by applying these laws 
if the Heisenberg cut is located at an appropriate position,
although wrong results might be obtained 
if one naively assumed the Heisenberg cut between S and A.
Furthermore, we have {\em derived} formula for POVM measurements
in Sec.~\ref{sec:pheno}, 
although some recent textbooks employed POVM measurements as 
one of the fundamental laws of quantum theory.
Therefore, 
the standard laws of quantum theory
are complete if correctly applied.

\section{Analysis of Zeno effect by quantum measurement theory}
% \verb#{sec:rmt}#
\label{sec:rmt}
%%%%%%%%%%%%%%%%%%%%%%%%%%%%%%%%%%%%%%%%%%%%%%%%%%%%%%%%%%%%%%%%%%%%%%%%%%%%%%

In the previous Section, we have reviewed 
the quantum measurement theory, 
according to which (in particular, Sec.~\ref{sec:ua}) 
one should analyze the unitary temporal evolution 
of both the target system S of measurements
and (a part of) measuring apparatus A.
Many of theoretical analyses of the Zeno effect employed this 
formalism~\cite{Kraus,Joos,PTP,FS,BB,PasNam,laser_th0,laser_th1,Sch}.
In this section, taking a photon-counting measurement 
on the decay of an excited atom as an example,
we study the Zeno effect with this formalism, 
and compare the results with those obtained in Sec.~\ref{sec:ct}.

This section is organized as follows:
In Sec.~\ref{sec:model},
a concrete Hamiltonian for the system-apparatus interaction,
as well as the physical quantities of interest, is presented.
In Sec.~\ref{sec:rff},
the effect of the system-apparatus interaction
is investigated analytically;
it is shown that the system-apparatus interaction
results in the renormalization of the form factor,
through which the decay rate of the atom is modified.
In Sec.~\ref{sec:idealm},
we consider an idealized situation 
where the detector satisfies the flat-response condition, 
Eq.~(\ref{eq:eta=tau-1});
it is observed that the conventional projection-based theory 
discussed in Sec.~\ref{sec:ct} 
is essentially reproduced under this condition.
Contrarily, in Secs.~\ref{sec:gim}, \ref{sec:eim}, \ref{sec:fm},
we consider the effects of imperfect measurements,
where the flat-response condition is not satisfied
and various phenomena beyond the conventional theory appear.

%%%%%%%%%%%%%%%%%%%%%%%%%%%%%%%%%%%%%%%%%%%%%%%%%%%%%%%%%%%%%%%%%%%%%%%%%%%%%%
\subsection{Model for the system and apparatus}
% \verb#{sec:model}#
\label{sec:model}
%%%%%%%%%%%%%%%%%%%%%%%%%%%%%%%%%%%%%%%%%%%%%%%%%%%%%%%%%%%%%%%%%%%%%%%%%%%%%%

%%%%%%%%%%%%%%%%%%%%%%%%%%%%%%%%%%%%%%%%%%%%%%%%%%%%%%%%%%%%%%%%%%%%%%%%%%%%%%
\subsubsection{Hamiltonian for atom-photon-detector system}
% \verb#{sec:Hapd}#
\label{sec:Hapd}
%%%%%%%%%%%%%%%%%%%%%%%%%%%%%%%%%%%%%%%%%%%%%%%%%%%%%%%%%%%%%%%%%%%%%%%%%%%%%%

As an example of an unstable system and a measuring apparatus 
for checking its decay, we discuss the case where
the radiative decay of an excited atom is 
%shmzshmz>
continuously monitored by counting the emitted photon. 
This is a sort of a continuous measurement (Sec.~\ref{sec:CvsD}),
where the observer judges that the atom has decayed 
if the detector (measuring apparatus) has counted a photon.
Note that this measurement is classified as 
a negative-result and indirect measurement, 
for which the curiousness of the Zeno effect is most emphasized
% in which the curiousness of the Zeno effect is emphasized
(see Secs.~\ref{sec:DvsI} and \ref{sec:PvsN}). 

%Here, 
We present again the Hamiltonian of the measured system S,
i.e., an atom and a photon field:
\beq
\hat{H}_S = \Om\sigma_+\sigma_- + 
\int \rmd\veck \ \left[(g_{\veck}\sigma_+ b_{\veck}+{\rm H.c.})
+\eps_{\veck}b_{\veck}^{\dagger}b_{\veck}\right].
% \verb#{eq:H1_ag}#
\label{eq:H1_ag}
\eeq
The unobserved decay dynamics of this system has 
already been discussed in Sec.~\ref{sec:fpus}.
We now couple a detector A to S.
In usual photodetectors,
photons are converted to elementary excitations
(typically, electron-hole pairs) in the detector.
Here, we model the detector by a spatially homogeneous absorptive medium, 
whose Hamiltonian is given by
\begin{equation}
\hat{H}_A = \sum_j \int d\veck 
\ \eps_{\veck j} c_{\veck j}^{\dagger} c_{\veck j}.
% \verb#{eq:HA1}#
\label{eq:HA1}
\end{equation}
Here, $\eps_{\veck j}$ and $c_{\veck j}$ denote 
the energy and an annihilation operator, respectively, of
the elementary excitation 
with the momentum $\veck$ 
and a set of other quantum numbers $j$
\footnote{
For example, $\veck$ is the center-of-mass momentum of
an electron-hole pair, and $j$ is 
a set of other quantum numbers for 
the electron-hole relative motion.
}.
We treat 
$c_{\veck j}$ as a bosonic operator,
which satisfies $[c_{\veck j}, c_{\veck' j'}^{\dagger}]
=\delta(\veck-\veck')\delta_{j,j'}$,
and thus the detector is here modeled by non-interacting bosons. 
Such a treatment is allowed as long as the density of excitations 
is low,\footnote{
States excited by photons are in the charge-neutral sector
of electron-hole states.
Such states can always be mapped to states of {\em interacting} 
bosons \cite{Klein,Ivanov}.
When the density of excitations is low, 
then the density of bosons in the mapped state is low, 
and thus the interactions among the bosons are negligible.
For details, see, e.g., Refs.~\cite{Klein,Ivanov}
}
which is valid in usual photodetection processes.
Usually, elementary excitations form a continuum in energy,
and the conversion from a photon to an excitation occurs irreversibly.
The interaction between photons and the elementary excitations 
may be described by adding the following photon-detector interaction term:~\cite{Sch,HBPRA}
\beq
\hat{H}_{\rm int}=\sum_j\int d\veck 
(\xi_{\veck j} b^{\dagger}_{\veck} c_{\veck j}+ \ H.c.).
% \verb#{eq:Hint1}#
\label{eq:Hint1}
\eeq
Here, the photon (of mode) $\veck$ does not couple to elementary excitations 
with a different momentum $\veck'(\neq \veck)$,
due to the translational symmetry 
inherent in spatially homogeneous systems. 

Throughout this section,
we assume that there is no excitation in the detector initially.
Then, following Sec.~\ref{sec:ff},
we can transform Eqs.~(\ref{eq:HA1}) and (\ref{eq:Hint1})
into the following form:
\bea
\hat{H}_A &=& \int \int d\veck d\om 
\ \om c_{\veck\om}^{\dagger} c_{\veck\om},
% \verb#{eq:HA}#
\label{eq:HA}
\\
\hat{H}_{\rm int}&=&\int \int d\veck d\om 
(\xi_{\veck\om} b^{\dagger}_{\veck} c_{\veck\om}+ \ H.c.),
% \verb#{eq:Hint}#
\label{eq:Hint}
\eea
where $c_{\veck\om}$ is normalized as
$[c_{\veck\om}, c_{\veck'\om'}^{\dagger}]=
\delta(\veck-\veck')\delta(\om-\om')$.
$\xi_{\veck\om}$ is the form factor for 
the photon-detector interaction, for a photon with momentum $\veck$.

The photon-detector coupling $\xi_{\veck\om}$ 
generally introduces two effects on the photonic modes:
The coupling makes the lifetimes of photons finite,
as well as it introduces slight shifts in the photonic energies. 
The latter effect appears
when $\xi_{\veck\om}$ is not a symmetric function of $\om$
about the photon energy $\epsilon_{\veck}$
(see Sec.~\ref{ssec:a_ham}).
Here, in order to neglect the energy shifts of photons,
which bring about uninteresting complexity from the viewpoint of 
the Zeno effect,
we neglect $\om$-dependence of $\xi_{\veck\om}$
and take the following form:
\beq
\xi_{\veck\om} = \sqrt{\eta_{\veck}/2\pi},
% \verb#{eq:flatband}#
\label{eq:flatband}
\eeq
which is called the {\em flat-band approximation}.
By this choice of photon-detector coupling,
the photon $\veck$ will be converted into an excitation in the detector
at a rate $\eta_{\veck}$.
The response time of the detector for the photon $\veck$ is therefore given by
\beq
\tau_{\veck}\equiv \eta_{\veck}^{-1}.
\label{eq:def_rt}
\eeq
In realistic experimental situations,
the photon-detector coupling $\eta_{\veck}$ often depends on $\veck$.
For example, 
% if the detector covers only a part of 
% the whole solid angle around the atom,
if the detector has a finite detection energy band,
$\eta_{\veck}$ is nonzero only for 
photons whose energy falls in the detection energy band.
Therefore, we retain $\veck$-dependence of $\eta_{\veck}$
in order to treat such cases.

% shmz>
In real photodetectors,
photogenerated excitations are magnified to yield macroscopic signals.
Here we neglect the magnification processes,  
regarding it as the apparatus A$'$ that performs an ideal 
measurement of the number of excitation quanta
(Sec.~\ref{sec:ev_S+A}), although actually 
it would not be ideal in general. 
% , regarding 
% the absorptive medium as the `probe system.
% (see Sec.~\ref{sec:probe}). 
% and approximately regard that an observer can know the decay 
% by generation of excitations in the detector.
Such an approximation has been successfully applied to 
many problems in quantum optics \cite{Glauber,QNoise,Mandel}.
% shmz<

%%%%%%%%%%%%%%%%%%%%%%%%%%%%%%%%%%%%%%%%%%%%%%%%%%%%%%%%%%%%%%%%%%%%%%%%%%%%%%
% shmz>
\subsubsection{Quantities of interest}
% \subsubsection{Definitions of $s(t)$, $\varepsilon(t)$ and $r(t)$}
%%%%%%%%%%%%%%%%%%%%%%%%%%%%%%%%%%%%%%%%%%%%%%%%%%%%%%%%%%%%%%%%%%%%%%%%%%%%%%

In studying the Zeno effect, 
one is interested only in 
the averages of a few observables
(such as $s(t)$, $\varepsilon(t)$ and $r(t)$, described below)
over all possible values of the readouts of the measurements. 
Furthermore, 
since this model does not corresponds to either of the 
limiting cases (i)-(iii) of Sec.~\ref{sec:ua},
\footnote{
For example, the time scale of the recurrent process
is infinite because A of this model has a continuous spectrum.
}
the unitary approximation of Sec.~\ref{sec:ua} 
is expected to be good for this model. 
Therefore, 
within the unitary approximation ,
it is sufficient to 
investigate the unitary time evolution of the joint 
quantum system S$+$A,
as explained in Sec.~\ref{sec:Zeno-MMT}.
Therefore, the projection postulate is no more necessary; 
the counteraction of measurement onto S
is naturally introduced through the interaction $\hat{H}_{\rm int}$
% (which is here the photon-detector coupling)
between the measured system and the measuring apparatus. 

In our model, the Hamiltonian of S$+$A is given by the sum of 
Eqs.~(\ref{eq:H1_ag}), (\ref{eq:HA}) and (\ref{eq:Hint});
\beq
\hat{H}=\hat{H}_S+\hat{H}_{\rm int}+\hat{H}_A.
% \verb#{eq:enlH}#
\label{eq:enlH}
\eeq
The pre-measurement state is $\sigma_+|0\rangle|0_A\rangle$,
where $|0_A\rangle$ denotes the vacuum state for $c_{\veck\om}$.
Hereafter, we denote the vacuum state for the enlarged system 
S+A, $|0\rangle|0_A\rangle$, by $|0\rangle$ for simplicity.
Since the number of total quanta,
$\hat{N}=\sigma_+ \sigma_- + \int d\veck b_{\veck}^{\dagger}b_{\veck} +
\int\int d\veck d\om c_{\veck\om}^{\dagger}c_{\veck\om}$,
is conserved (=1) in this enlarged system,
the state vector at time $t$ can be written in the following form:
\beq
|\psi (t) \rangle = \exp(-\rmi{\cal H}t)|\rmi\rangle
= f(t)\sigma_+|0\rangle
+ \int d\veck f_{\veck}(t) b_{\veck}^{\dagger}|0\rangle
+ \int \int d\veck d\om f_{\veck\om}(t) c_{\veck\om}^{\dagger}|0\rangle.
\label{eq:psi_t}
\eeq
We define the following three probabilities of physical interest:
\bea
s(t) &=& |f(t)|^2, 
% \verb#{eq:svp}#
\label{eq:svp}
\\
\varepsilon(t) &=& \int d\veck |f_{\veck}(t)|^2,
% \verb#{eq:erp}#
\label{eq:erp}
\\
r(t) &=& \int \int d\veck d\om |f_{\veck\om}(t)|^2.
% \verb#{eq:rp}#
\label{eq:rp}
\eea
$s(t)$ is the survival probability of the atom
under the photoncounting measurement. 
$\varepsilon(t)$ is the probability 
that the atom has decayed and emitted a photon
but the emitted photon has not been detected.
We therefore call $\varepsilon(t)$ the measurement error.
\footnote{
However, there are several other definitions of measurement error
(Sec.~\ref{sec:me}).
}
$r(t)$ is the probability that the atom has decayed 
and the emitted photon has been detected.
Neglecting the signal magnification process,
we can interpret $r(t)$ as the probability of getting the detector response.
One of the merits of the present analysis of measurement
is that all of these quantities of interest can be calculated.

In the following part of Sec.~\ref{sec:rmt},
we will investigate how the decay probability $1-s(t)$
is perturbed by the interaction 
with measuring apparatus, $\hat{H}_{\rm int}$.
If $1-s(t)$ is suppressed (enhanced) compared to the case of free decay
where $\hat{H}_{\rm int}=0$,
we regard that the QZE (AZE) is taking place.
One might feel strange why this judgment is not done through $r(t)$,
which is directly accessible by an observer.
This is because 
$r(t)$ does not necessarily reflect
the decay probability $1-s(t)$ faithfully
in general measurement processes:
In good measurements, $1-s(t)$ well coincides with $r(t)$
as observed in Fig.~\ref{fig:ser}(b),
but such an optimal response is expected only when
the response time of detector is much shorter than the atomic lifetime,
and when the detector is active for all relevant photons.
If the detector response is slow [see Fig.~\ref{fig:ser}(a)]
or if the detector is inactive for some photons 
[see Figs.~\ref{fig:ser2}, \ref{fig:ser3} and \ref{fig:ser4}],
$r(t)$ largely deviates from $1-s(t)$.
Note that the atomic state does have decayed 
if $1-s(t) = 1$ even when $r(t) \simeq 0$. 

%%%%%%%%%%%%%%%%%%%%%%%%%%%%%%%%%%%%%%%%%%%%%%%%%%%%%%%%%%%%%%%%%%%%%%%%%%%%%%
\subsubsection{Relation to direct measurements}
% \verb#{sec:rel_dir}#
\label{sec:rel_dir}
%%%%%%%%%%%%%%%%%%%%%%%%%%%%%%%%%%%%%%%%%%%%%%%%%%%%%%%%%%%%%%%%%%%%%%%%%%%%%%

As stated at the beginning of Sec.~\ref{sec:Hapd},
the model presented in Sec.~\ref{sec:Hapd}
% the system-apparatus interaction [Eqs.~(\ref{eq:HA}) and (\ref{eq:Hint})]
describes a case of an indirect measurement.
However, it is shown here that
a model for a {\em direct} measurement
can also be recast into the same form, 
and therefore that 
the results presented in Sec.~\ref{sec:rmt}
are applicable not % restrictedly 
only to indirect measurements but also to direct measurements.

As the unstable quantum system, 
we again employ an excited atom undergoing radiative decay, 
but we slightly change the notation:
Denoting the excited and ground states by $|a\rangle$ and $|b\rangle$,
and taking the energy of $|b\rangle$ as the origin of energy,
we rewrite the unobserved system Hamiltonian as % follows:
\beq
\hat{H}_{\rm S}^{\rm direct} = \Om |a\rangle\langle a| + 
\int \rmd\veck \ \left[(g_{\veck} b_{\veck}|a\rangle\langle b|+{\rm H.c.})
+\eps_{\veck}b_{\veck}^{\dagger}b_{\veck}\right],
% \verb#{eq:H1''}#
\label{eq:H1''}
\eeq
% Needless to say, Eq.~(\ref{eq:H1''}) is completely the same as 
which is identical to 
Eq.~(\ref{eq:H1_ag}) except for notations.
% By a direct measurement considered here,
As a model of a direct measurement, 
we assume that the atom is coupled with
an apparatus A, which has a continuum of single-electron states,
in such a way that 
as soon as the decayed electronic state $|b\rangle$ is occupied
the electron tunnels 
into the continuum % of states (for example, unbound electronic states)
with a rate $\eta$ \cite{FP_For}.
The observer knows the decay of the initial unstable state $|a\rangle$
through the population in the continuum.
Denoting an
energy eigenstate of A with energy $\mu$ 
% state vector with energy $\mu$ in the continuum 
by $|\mu\rangle$, we take the Hamiltonian of
A and the interaction with S as
% the system-apparatus interaction is given by
\beq
\hat{H}_{\rm A}^{\rm direct}+\hat{H}_{\rm int}^{\rm direct}=
\int \rmd\mu \ \mu |\mu\rangle\langle\mu| +
\int \rmd\mu \sqrt{\eta/2\pi}(|\mu\rangle\langle b|+|b\rangle\langle\mu|).
% \verb#{eq:HA''}#
\label{eq:HA''}
\eeq
Obviously, this model describes a direct measurement,
% This measurement is classified to a direct measurement,
for the measurement apparatus directly interacts with the atom 
% (more precisely, the ground state of the atom)
in order to get information on the atom.
% A similar model was employed in Ref.~\cite{FP_For}.

According to the above Hamiltonians, 
the state vector starting from the initial state $|a\rangle$
evolves in the subspace, spanned by
$\{ |a\rangle, b_{\veck}^{\dagger}|b\rangle, 
b_{\veck}^{\dagger}|\mu\rangle \}$, 
of the total Hilbert space.
This situation is similar to the model of Sec.~\ref{sec:Hapd},
for which the subspace is spanned by
$\{ \sigma_+|0\rangle, b_{\veck}^{\dagger}|0\rangle, 
c_{\veck\om}^{\dagger}|0\rangle \}$
as Eq.~(\ref{eq:psi_t}).
Actually, by regarding $|a\rangle \to \sigma_+|0\rangle$,
$b_{\veck}^{\dagger}|b\rangle \to b_{\veck}^{\dagger}|0\rangle$,
and $b_{\veck}^{\dagger}|\om-\eps_{\veck} \rangle 
\to c_{\veck\om}^{\dagger}|0\rangle$,
we can map 
the present model of a direct measurement 
exactly onto the model of Sec.~\ref{sec:Hapd}
with the photon-detector coupling constant 
$\xi_{\veck\om}=\sqrt{\eta/2\pi}$.
(Such a case, where $\xi_{\veck\om}$ has no $\veck$-dependence,
is called the case of ``flat response'' in this article
and shall be discussed extensively in Sec.~\ref{sec:idealm}.)
Thus, the model of a direct measurement, 
Eqs.~(\ref{eq:H1''}) and (\ref{eq:HA''}),
is included as a special case of the model of an indirect measurement, 
Eqs.~(\ref{eq:H1_ag}), (\ref{eq:HA}) and (\ref{eq:Hint}).

%%%%%%%%%%%%%%%%%%%%%%%%%%%%%%%%%%%%%%%%%%%%%%%%%%%%%%%%%%%%%%%%%%%%%%%%%%%%%%
\subsection{Renormalization of form factor by measurement}
% \verb#{sec:rff}#
\label{sec:rff}
%%%%%%%%%%%%%%%%%%%%%%%%%%%%%%%%%%%%%%%%%%%%%%%%%%%%%%%%%%%%%%%%%%%%%%%%%%%%%%

As an inevitable counteraction of photon-counting measurements,
the lifetimes of photons become finite.
As a result, the energies of photons are broadened, %obscured,
and the form factor %(the spectrum of atom-photon coupling, 
(see Sec.~\ref{sec:ff})
suffers modification.
Hereafter, we refer to the new form factor 
as the {\it renormalized} form factor.
In this section, we discuss how the form factor is renormalized
by the measurement, i.e.,
through the photon-detector interaction.
It should be reminded that, 
when the system is not observed ($\hat{H}_{\rm int}=0$),
the {\it original} form factor is given by
\beq
|g_{\mu}|^2 = \int d\veck |g_{\veck}|^2 \delta(\eps_{\veck}-\mu).
% \verb#{eq:g2_re}#
\label{eq:g2_re}
\eeq

In order to obtain the renormalized form factor,
we first diagonalize the photon-detector part of the Hamiltonian,
$\hat{H}_{\rm int}+\hat{H}_A$.
For this purpose, we define the coupled-mode operator $B_{\veck\mu}$~\cite{Fano,HBPRA} by
\bea
B_{\veck\mu} &=& \alp_{\veck}(\mu) b_{\veck} +
\int d\om \beta_{\veck}(\mu,\om) c_{\veck\om},
\\
\alp_{\veck}(\mu) &=& \frac{(\eta_{\veck}/2\pi)^{1/2}}
{\mu-\eps_{\veck}+\rmi \eta_{\veck}/2},
\\
\beta_{\veck}(\mu,\om) &=& \frac{\eta_{\veck}/2\pi}
{(\mu-\eps_{\veck}+\rmi \eta_{\veck}/2)(\mu-\om+\rmi\delta)}
+\delta(\mu-\om).
\eea
It can be confirmed that $B_{\veck\mu}$ is orthonormalized as
$[B_{\veck\mu}, B_{\veck'\mu'}^{\dagger}]=
\delta(\veck-\veck')\delta(\mu-\mu')$.
Inversely, the original operators, $b_{\veck}$ and $c_{\veck\om}$,
are given, in terms of $B_{\veck\mu}$, by
\bea
b_{\veck} &=& \int d\mu \alp_{\veck}^{\ast}(\mu) B_{\veck\mu},
\\
c_{\veck\om} &=& \int d\mu \beta_{\veck}^{\ast}(\om,\mu) B_{\veck\mu}.
\eea
Using the coupled-mode operators, the enlarged Hamiltonian
$\hat{H}$ is transformed into the following form:
\beq
\hat{H} = \Om \sigma_+ \sigma_- +
\int \int d\veck d\mu \ \mu B_{\veck\mu}^{\dagger} B_{\veck\mu}
+ \int \int d\veck d\mu \left[
\frac{(\eta_{\veck}/2\pi)^{1/2}g_{\veck}}
{\mu-\eps_{\veck}-\rmi \eta_{\veck}/2}\sigma_+ B_{\veck\mu} + H.c.
\right].
\eeq
Now we extract the interaction mode $\bar{B}_{\mu}$ at energy $\mu$,
employing the same method as used in Sec.~\ref{sec:ff}.
$\bar{B}_{\mu}$ is given by
\bea
\bar{B}_{\mu} &=& \bar{g}_{\mu}^{-1} \int\rmd\veck
\frac{(\eta_{\veck}/2\pi)^{1/2}g_{\veck}}
{\mu-\eps_{\veck}-\rmi\eta_{\veck}/2}B_{\veck\mu},
\\
|\bar{g}_{\mu}|^2 &=& \int\rmd\veck
|g_{\veck}|^2
\frac{\eta_{\veck}/2\pi}
{|\mu-\eps_{\veck}-\rmi\eta_{\veck}/2|^2},
% \verb#{eq:barg2}#
\label{eq:barg2}
\eea
where $|\bar{g}_{\mu}|^2$ was determined so that 
$\bar{B}_{\mu}$ is orthonormalized as 
$[\bar{B}_{\mu},\bar{B}^{\dagger}_{\mu'}]=\delta(\mu-\mu')$.
Using the interaction modes, $\hat{H}$ is further rewritten as
\beq
\hat{H} = \Om_0 \sigma_+ \sigma_- +
\int\rmd\mu \left[
\left(\bar{g}_{\mu}\sigma_+ \bar{B}_{\mu}+
{\rm H.c.}\right) + \mu \bar{B}_{\mu}^{\dagger}\bar{B}_{\mu}
\right] + \hat{H}_{\rm rest},
% \verb#{eq:renH}#
\label{eq:renH}
\eeq
where $\hat{H}_{\rm rest}$ consists of coupled modes 
which do not interact with the atom.

In the final form of the Hamiltonian, Eq.~(\ref{eq:renH}),
the atom is coupled to a one-dimensional continuum of $\bar{B}_{\mu}$
with the coupling function $\bar{g}_{\mu}$.
Thus, $|\bar{g}_{\mu}|^2$ gives the form factor 
renormalized by measurement.
It is easy to confirm that,
in the limit of $\eta_{\veck} \to 0$ for every $\veck$,
Eq.~(\ref{eq:barg2}) reduces to the original form factor, Eq.~(\ref{eq:g2_re}).

\begin{figure}%----------------------------------------------------------------
\bec
\includegraphics{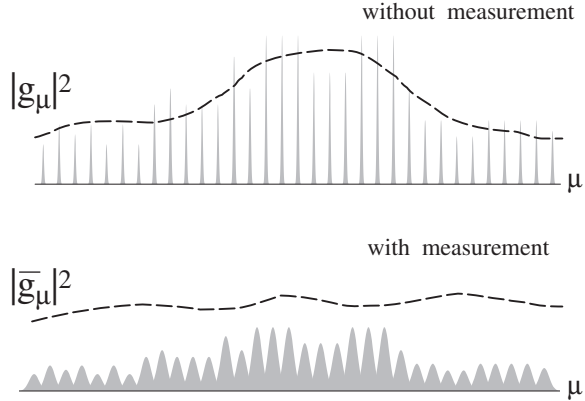}
\caption{\label{fig:illff}
Illustration of the renormalization of the form factor.
The contribution of each photon mode is given 
by a delta function when photons are not counted,
whereas it acquires finite width $\eta_{\veck}$ 
as a counteraction of measurement.
The form factor is an accumulation of 
%composed by accumulating 
contributions of all photon modes (broken line).
}% \verb#{fig:illff}#
\enc
\end{figure}%-----------------------------------------------------------------

Equations (\ref{eq:g2_re}) and (\ref{eq:barg2}) clarify 
how the form factor is renormalized
as a backaction of measurement.
When the system is not observed,
the form factor is 
an accumulation of delta functions, 
% composed by accumulating delta functions,
$|g_{\veck}|^2\delta(\mu-\eps_{\veck})$.
When one tries to measure the decay of the system by detecting an emitted photon,
the lifetime of the emitted photon becomes finite 
as inevitable counteraction of measurement.
Thus, the contribution of photon $\veck$ is energetically broadened as
\beq
|g_{\veck}|^2\delta(\mu-\eps_{\veck})
\ \ \rightarrow \ \ 
|g_{\veck}|^2 \frac{\eta_{\veck}/2\pi} 
{|\mu-\eps_{\veck}-\rmi\eta_{\veck}/2|^2},
% \verb#{eq:brdning}#
\label{eq:brdning}
\eeq
satisfying a sum rule:
\beq
\int d\mu |g_{\veck}|^2\delta(\mu-\eps_{\veck})
=\int d\mu |g_{\veck}|^2 \frac{\eta_{\veck}/2\pi} 
{|\mu-\eps_{\veck}-\rmi\eta_{\veck}/2|^2} = |g_{\veck}|^2.
% \verb#{eq:srule}#
\label{eq:srule}
\eeq
The renormalization of form factor is illustrated in Fig.~\ref{fig:illff}.

As is observed in Sec.~\ref{sec:Lff},
the decay rate is approximately given by the FGR with high accuracy.
% shmzshmz>
We may thus estimate the decay rate under measurements by the FGR.
In this case, note that when applying the FGR
 the final states 
must be defined as eigenstates of the system
in the absence of the atom-photon interaction.
Under measurements, 
the photon states ($b_{\veck}^{\dagger}|0\rangle$) 
do not satisfy this condition
because of the photon-detector interaction, 
whereas the coupled modes ($\bar{B}_{\mu}^{\dagger}|0\rangle$ 
or ${B}_{\veck\mu}^{\dagger}|0\rangle$) do.
Therefore, 
% It should be remarked that,
% shmzshmz<
in order to predict the decay rate correctly,
the FGR should be applied not to the original form factor, 
Eq.~(\ref{eq:g2_re}), 
but to the renormalized form factor, Eq.~(\ref{eq:barg2}).
Then, we obtain the decay rate under measurement as
\beq
\Gamma(\{\tau_{\veck}\}) = 2\pi |\bar{g}_{\Om}|^2
= \int\rmd\veck |g_{\veck}|^2
\frac{\eta_{\veck}}
{|\Om-\eps_{\veck}-\rmi\eta_{\veck}/2|^2},
% \verb#{eq:drum}#
\label{eq:drum}
\eeq
This quantity is the principal result on the QZE and AZE
by the quantum measurement theory:
the decay rate is modified through renormalization 
of form factor by measurement.
If $\Gamma$ is smaller (larger) than the original decay rate 
without measurement, which is given by $\Gamma = 2\pi |g_{\Om}|^2$,
we regard that the QZE (AZE) is taking place.
In Sec.~\ref{sec:idealm},
Eq.~(\ref{eq:drum}) is compared with Eq.~(\ref{eq:LGam}),
which gives the decay rate under repeated instantaneous 
measurements 
based on the projection postulate.

% shmzshmz>
%%%%%%%%%%%%%%%%%%%%%%%%%%%%%%%%%%%%%%%%%%%%%%%%%%%%%%%%%%%%%%%%%%%%%%%%%%%%%%
\subsection{Continuous measurement with flat response}
% \verb#{sec:idealm}#
\label{sec:idealm}
%%%%%%%%%%%%%%%%%%%%%%%%%%%%%%%%%%%%%%%%%%%%%%%%%%%%%%%%%%%%%%%%%%%%%%%%%%%%%%

The conventional theories of the Zeno effect (Sec.~\ref{sec:ct})
assumed that each of the repeated measurements is
instantaneous and ideal.
On the other hand, we are treating here  
a continuous measurement.
As pointed out in Sec.~\ref{sec:CvsD}, 
these measurements are quite different from each other in general.
For example, the response time $\tau_{\rm r} =0$ 
and the measurement intervals $\tau_{\rm i}>0$ in the former, 
whereas $\tau_{\rm r} > 0$ and $\tau_{\rm i}=0$ in the latter.
Regarding the Zeno effect, however, 
similarity between these different measurements 
has often been discussed~\cite{Sch}.
In this subsection, we present a case
where they indeed give similar results,
whereas in Secs.~\ref{sec:gim}, \ref{sec:eim} and \ref{sec:fm} 
we will present drastic cases where they give much different results.

To see the similarity, it is customary to consider that 
$\tau_{\rm r}$ (response time of the apparatus)
would correspond to 
$\tau_{\rm i}$ (interval of repeated instantaneous measurements)
with a possible multiplicative factor of order unity.
Furthermore, it should be noted that,
if one applies the projection postulate directly to 
the atomic states,
% by applying the projection postulate for a measurement of atomic survival,
the quantum coherences between the survived state ($\sigma_+|0\rangle$) 
and decayed states ($b_{\veck}^{\dagger}|0\rangle$)
are destroyed simultaneously, regardless of photon wavenumber $\veck$.
Therefore, 
the projection postulate implicitly assumes an idealized situation,
in which the detector is sensitive to every photon mode
with an identical response time $\tau_{\rm r}$. % \sim \tau_{\rm i}$.
In the present model, such {\it flat response} is realized by putting 
\beq
\eta_{\veck}=\tau_{\rm r}^{-1}
\ \mbox{for every $\veck$}.
\label{eq:eta=tau-1}
\eeq
In this section, 
% using the quantum measurement theory, 
using the formalism of Secs.~\ref{sec:model} and \ref{sec:rff},
we study the Zeno effect under a continuous measurement with such a detector, 
and compare the results with those obtained by the conventional theories.

%%%%%%%%%%%%%%%%%%%%%%%%%%%%%%%%%%%%%%%%%%%%%%%%%%%%%%%%%%%%%%%%%%%%%%%%%%%%%%
\subsubsection{Decay rate under flat response}
\label{sssec:drfr}
%%%%%%%%%%%%%%%%%%%%%%%%%%%%%%%%%%%%%%%%%%%%%%%%%%%%%%%%%%%%%%%%%%%%%%%%%%%%%%

\begin{figure}%----------------------------------------------------------------
\bec
\includegraphics{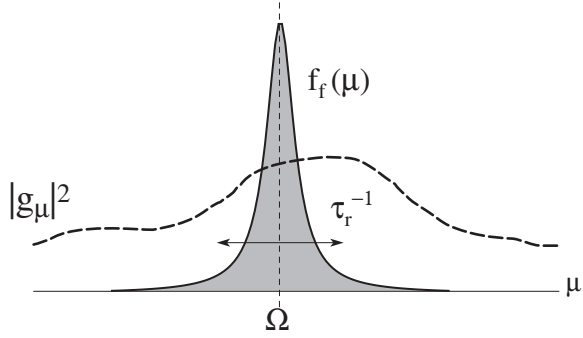}
\caption{\label{fig:drr2}
Illustration of the calculation by Eq.~(\ref{eq:Gtr}) 
of the decay rate $\Gamma(\tau_{\rm r})$
under continuous measurement with flat response.
$\Gamma(\tau_{\rm r})$ is given by integrating the form factor $|g_{\mu}|^2$
with a weight function $f_{\rm f}(\mu)$.
}
% \verb#{fig:drr2}#
\enc
\end{figure}

When the condition of flat response [Eq.~(\ref{eq:eta=tau-1})]
is satisfied,
the general expression of the measurement-modified decay rate
[Eq.~(\ref{eq:drum})] is recast into the following form,
using the definition of the form factor, Eq.~(\ref{eq:g2}):
\bea
\Gamma(\tau_{\rm r}) &=& \int d\mu |g_{\mu}|^2 \times f_{\rm f}(\mu),
\label{eq:Gtr}
\\
f_{\rm f}(\mu) &=& \frac{\tau_{\rm r}^{-1}}{|\mu-\Om-\rmi/2\tau_{\rm r}|^2},
\label{eq:ffmu}
\eea
Namely, the decay rate under a continuous measurement with flat response
is given by integrating the original form factor $|g_{\mu}|^2$
with a weight function $f_{\rm f}(\mu)$,
as illustrated in Fig.~\ref{fig:drr2}.
The weight function $f_{\rm f}(\mu)$ has the following properties:
(i) $f_{\rm f}(\mu)$ is a Lorentzian
centered at the atomic transition energy $\Om$
with a spectral width $\sim \tau_{\rm r}^{-1}$,
and (ii) $f_{\rm f}(\mu)$ is normalized as 
$\int d\mu f_{\rm f}(\mu) = 2\pi$.

%-----------------------------------------------------------------

Now a close connection 
between repeated instantaneous measurements (Sec.~\ref{sec:DRR})
and continuous measurements with flat response
has become apparent~\cite{KK2005}:
The difference between them lies merely
in the functional forms of the weight functions
$f_{\rm c}(\mu)$ of Eq.~(\ref{eq:wei1}) 
and $f_{\rm f}(\mu)$.
Thus, apart from slight quantitative discrepancy
due to the difference between $f_{\rm c}(\mu)$ 
and $f_{\rm f}(\mu)$,
the conventional theories presented in Sec.~\ref{sec:ct}
based on the projection postulate
can be essentially reproduced from the formalism 
of Secs.~\ref{sec:model} and \ref{sec:rff} 
in the special case of flat response.

Furthermore, it is of note that
the effects of these two measurements 
would coincide even at a quantitative level, 
when the measurement interval $\tau_{\rm i}$ is a stochastic variable
following a distribution function 
$P(\tau_{\rm i})=(2\tau_{\rm r})^{-1}\exp(-\tau_{\rm i}/2\tau_{\rm r})$.
In this case, the weight function 
for the repeated measurements is modified as follows:
\beq
\tilde{f}_{\rm c}(\mu)
=\int d\tau_{\rm i} P(\tau_{\rm i})\times 
\tau_{\rmi} \ {\rm sinc}^2 \left[\frac{\tau_{\rmi}(\mu-\Om)}{2}\right]
=\frac{\tau_{\rm r}^{-1}}{|\mu-\Om-\rmi/2\tau_{\rm r}|^2},
\eeq
which is identical to $f_{\rm f}(\mu)$~\cite{KK2005}.

Of course, the condition of flat response is not satisfied
in general measurement processes,
so interesting phenomena beyond the conventional theories 
are expected, which are the topics  
of Secs.~\ref{sec:gim}, \ref{sec:eim} and \ref{sec:fm}.
In the rest of Sec.~\ref{sec:idealm},
we confirm the qualitative agreement between these two formalisms
with concrete numerical examples,
based on the Lorentzian form factor [Eq.~(\ref{eq:Lff})],
for which the Zeno effect by the conventional theories
has already been revealed quantitatively 
in Sec.~\ref{sec:ct}. 

%Regarding the target unstable system,
%we assume that it has the Lorentzian form factor, Eq.~(\ref{eq:Lff}),

%%%%%%%%%%%%%%%%%%%%%%%%%%%%%%%%%%%%%%%%%%%%%%%%%%%%%%%%%%%%%%%%%%%%%%%%%%%%%%
\subsubsection{Numerical results}
%%%%%%%%%%%%%%%%%%%%%%%%%%%%%%%%%%%%%%%%%%%%%%%%%%%%%%%%%%%%%%%%%%%%%%%%%%%%%%

% shmz>
To calculate 
the three probabilities of physical interest,
$s(t)$, $\varepsilon(t)$ and $r(t)$,
we use the Green function method presented in Sec.~\ref{sec:Gfm}.
Here, an important modification is required 
due to the photon-detector interaction:
The bare photon Green function $P(\om,\veck)$,
which appears in Eqs.~(\ref{eq:slfene}) and (\ref{eq:a_to_p}),
should be replaced by the dressed photon Green function $\bar{P}(\om,\veck)$.
Following Sec.~\ref{sec:Gfm}, 
the dressed Green function is given by
\beq
\bar{P}(\om,\veck) 
% shmz
% &=& P(\om,\veck)
% +P(\om,\veck)\Sigma(\om,\veck)P(\om,\veck)
% +P(\om,\veck)\Sigma(\om,\veck)P(\om,\veck)
% \Sigma(\om,\veck)P(\om,\veck)+\cdots
% \\
=\frac{P(\om,\veck)}{1-\Sigma(\om,\veck)P(\om,\veck)},
\label{eq:drsd_p}
\eeq
where $P(\om,\veck)$ and $\Sigma(\om,\veck)$ 
are the bare Green function and the self-energy of the photon $\veck$,
respectively, which are given by
\bea
P(\om,\veck) &=& \frac{1}{\om-\eps_{\veck}+\rmi\delta},
\label{eq:barepG}
\\
\Sigma(\om,\veck) &=& \int d\mu 
\frac{|\xi_{\veck\mu}|^2}{\om-\mu+\rmi\delta}
=-\frac{\rmi\eta_{\veck}}{2}.
\label{eq:selfene_p}
\eea
Substituting Eqs.~(\ref{eq:barepG}) and (\ref{eq:selfene_p}) 
into Eq.~(\ref{eq:drsd_p}), we obtain the dressed Green function 
of the photon $\veck$ as
\beq
\bar{P}(\om,\veck) = \frac{1}{\om-\epsilon_{\veck}+\rmi\eta_{\veck}/2}.
\eeq
The change from $P(\om,\veck)$ to $\bar{P}(\om,\veck)$ represents 
the renormalization effect, which is discussed in Sec.~\ref{sec:rff}, 
in the language of the Green function method.
% First, we present the numerical results for 
% the three probabilities of physical interest,
% $s(t)$, $\varepsilon(t)$ and $r(t)$,
% which are calculated rigorously using the Green function method.
% shmz<
We can calculate $s(t)$, $\varepsilon(t)$ and $r(t)$ numerically 
using $\bar{P}(\om,\veck)$.

\begin{figure}%----------------------------------------------------------------
\bec
\includegraphics{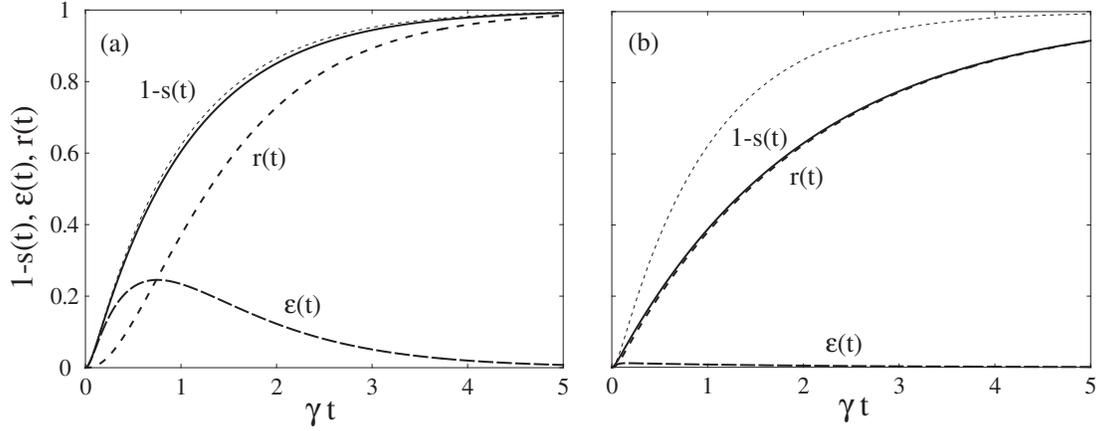}
\caption{\label{fig:ser}
Temporal evolutions of $1-s(t)$, $\varepsilon(t)$, and $r(t)$.
The parameters are chosen as 
$\Delta = 20\gamma$ (i.e., $t_{\rm j} = 0.05\gamma^{-1}$),
$|\Om - \mu_0|=0$.
The response time of the detector is chosen as
$\tau_{\rm r} = 0.5\gamma^{-1}$ in (a),
and $\tau_{\rm r} = 0.025\gamma^{-1}$ in (b).
$r(t)$ follows $1-s(t)$ with a delay time 
that is approximately given by $\tau_{\rm r}$.
Thin dotted lines show the unobserved decay probability.
}% \verb#{fig:ser}#
\enc
\end{figure}%-----------------------------------------------------------------

In Fig.~\ref{fig:ser}, the
temporal evolutions of the three probabilities are plotted.
In Fig.~\ref{fig:ser}(a),
the response of the detector is assumed to be very slow
($\tau_{\rm r} \sim \gamma^{-1}$),
in order to visualize the delay of the detector response.
As a result,
the decay dynamics $s(t)$ is almost unchanged from the unobserved case.
We can confirm that $r(t)$ follows $1-s(t)$ with a delay time 
$\sim \tau_{\rm r}$; thus, $\tau_{\rm r}$ may 
% that is approximately $\tau_{\rm r}$; thus, $\tau_{\rm r}$ may 
safely be regarded as the response time of the detector.
Recall that as noticed in Sec.~\ref{sec:rt} 
$\tau_{\rm r}$ is actually the {\em lower limit} of the response time because
additional delays in the response, 
such as delays in signal magnification processes, 
may occur in practical experiments.
In discussing fundamental physics, 
the limiting value is more significant     %important 
than practical values, which depend strongly  % vary widely 
on detailed experimental conditions.
A typical value of $\tau_{\rm r}$ for GaAs is $10^{-15}$s, which is much shorter than
the practical response times of commercial photodetectors, which 
range from $10^{-6}$ to $10^{-13}$s.
In the case of measurement of the decay of an excited atom
by semiconductor photodetectors, 
we can usually assume that 
$\tau_{\rm r} \ll \gamma^{-1}$
because $\gamma^{-1} \sim 10^{-9}$s typically. 
The results for such quick response %($\tau_{\rm r} \ll \gamma^{-1}$)
are plotted in Fig.~\ref{fig:ser}(b).
Furthermore, $r(t)$ follows $1-s(t)$ almost without decay,
and that the measurement error $\varepsilon(t)$
almost vanishes for all time.
It is observed that the decay is slowed down as compared with (a),
i.e., the QZE occurs under 
a continuous measurement with quick and flat response.
In both Figs.~\ref{fig:ser}(a) and (b),
emitted photons are all counted by the detector. 
Therefore, $r(t) \to 1$ and $\varepsilon(t) \to 0$ as $t \to \infty$.

\begin{figure}%----------------------------------------------------------------
\bec
\includegraphics{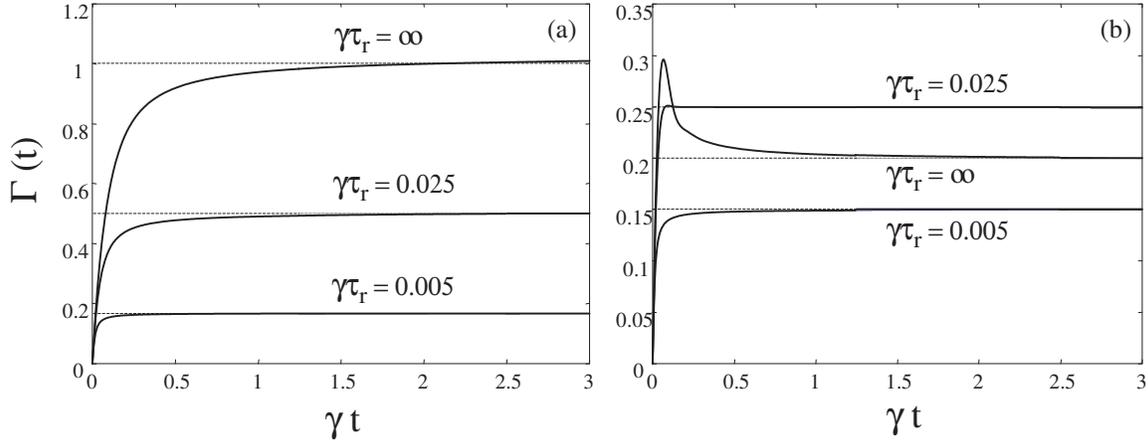}
\caption{\label{fig:decrate2}
Temporal evolution of $\Gamma(t)$, where
$\Delta = 20\gamma$ (i.e., $t_{\rm j} = 0.05\gamma^{-1}$).
$\Om-\mu_0 = 0$ in (a), and $\Om-\mu_0 = 2\Delta$ in (b).
The values of the response time %measurement intervals 
$\tau_{\rm r}$ are indicated in the figures. 
The decay rate agrees well with the FGR decay rate
(thin dotted lines), which is given by Eq.~(\ref{eq:Gam_mes}).
}% \verb#{fig:decrate2}#
\enc
\end{figure}%-----------------------------------------------------------------

In order to see more details, % visualize the decay rate, 
$\Gamma(t)=-\ln s(t)/t$ is plotted
in Fig.~\ref{fig:decrate2}
for several values of the response time $\tau_{\rm r}$ of the detector.
As for the initial behavior of $\Gamma(t)$,
it is confirmed that $\Gamma(t)$ increases linearly in time
as $\Gamma(t)=\gamma \Delta t$,
regardless of the values of
$\Om-\mu_0$ and $\tau_{\rm r}$.
This feature is completely the same as that of the free evolution 
of the atom-photon system (Fig.~\ref{fig:decrate}).
\footnote{
The initial behavior of survival probability is given by 
$s(t)=1-(\langle\hat{H}^2\rangle-\langle\hat{H}\rangle^2)t^2$,
where $\langle\cdots\rangle=\langle\rmi|\cdots|\rmi\rangle$ and
$\hat{H}$ is the enlarged Hamiltonian for S+A,
given by Eq.~(\ref{eq:enlH}).
However, $\langle\hat{H}^2\rangle-\langle\hat{H}\rangle^2=
\langle\hat{H}_S^2\rangle-\langle\hat{H}_S\rangle^2$,
i.e., the measurement terms ($\hat{H}_{\rm int}+\hat{H}_A$)
play no role in determining the initial behavior of $s(t)$.
}
On the other hand, in the later stage of decay 
($t \gtrsim t_{\rm j}$), %Delta^{-1}
it is confirmed that $\Gamma(t)$ approaches a constant value,
indicating that the decay proceeds 
exponentially with a well-defined decay rate.
It is also confirmed that the decay rate agrees well with the FGR decay rate
applied to the renormalized form factor
(thin dotted lines in Fig.~\ref{fig:decrate2}),
which will be discussed in Sec.~\ref{sssec:rfgr} in detail.
Note also that 
in Fig.~\ref{fig:decrate2}(a), where $\Om-\mu_0=0$,
the decay rate decreases monotonically 
as $\tau_{\rm r}$ is shortened.
In contrast, in Fig.~\ref{fig:decrate2}(b), where $\Om-\mu_0=2\Delta$,
an increase of the decay rate (the AZE) is observed for large $\tau_{\rm r}$,
whereas suppression of decay (the QZE) is observed for small $\tau_{\rm r}$.
% Thus, the QZE and AZE are surely described by the modern formalism 
% based on the enlarged Hamiltonian, Eq.~(\ref{eq:enlH}).

%%%%%%%%%%%%%%%%%%%%%%%%%%%%%%%%%%%%%%%%%%%%%%%%%%%%%%%%%%%%%%%%%%%%%%%%%%%%%%
\subsubsection{Renormalized FGR decay rate}
\label{sssec:rfgr}
%%%%%%%%%%%%%%%%%%%%%%%%%%%%%%%%%%%%%%%%%%%%%%%%%%%%%%%%%%%%%%%%%%%%%%%%%%%%%%

\begin{figure}%----------------------------------------------------------------
\bec
\includegraphics{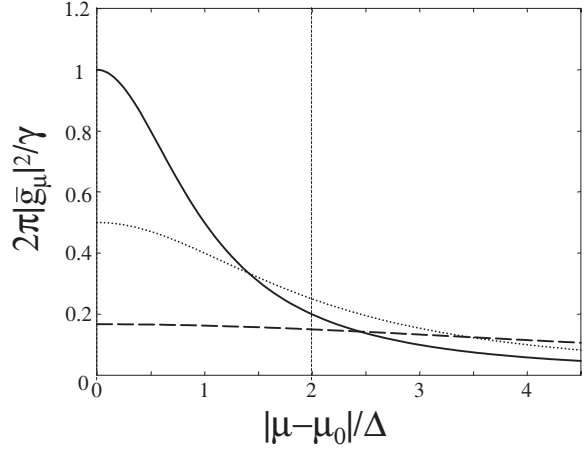}
\caption{\label{fig:rff}
Renormalization of the form factor.
The original form factor is plotted by the solid line,
and the renormalized form factors for 
$\tau_{\rm r} = 0.025\gamma^{-1}$ (slow response)
and $0.005\gamma^{-1}$ (fast response)
are plotted by dotted and dashed lines, respectively.
At $\mu=\mu_0$, $|\bar{g}_{\mu_0}|^2$ decreases monotonically 
as $\tau_{\rm r}$ becomes shorter, 
which corresponds to Fig.~\ref{fig:decrate2}(a).
In contrast, at $\mu=\mu_0+2\Delta$, 
$|\bar{g}_{\mu_0+2\Delta}|^2$ 
is increased (decreased) for slow (fast) response, 
in comparison to the unobserved case. 
This feature corresponds to Fig.~\ref{fig:decrate2}(b).
}% \verb#{fig:rff}#
\enc
\end{figure}%-----------------------------------------------------------------

We can explain the above numerical results 
in terms of the renormalization of the form factor, 
which was discussed in Sec.~\ref{sec:rff}.
Using Eqs.~(\ref{eq:g2}), (\ref{eq:Lff}), (\ref{eq:barg2}) 
and (\ref{eq:eta=tau-1}), 
we obtain the renormalized form factor as
\beq
|\bar{g}_{\mu}|^2 = \frac{\gamma}{2\pi}
\frac{\Delta(\Delta+(2\tau_{\rm r})^{-1})}{(\mu-\mu_0)^2
+(\Delta+(2\tau_{\rm r})^{-1})^2},
% \verb#{eq:ren_ff}#
\label{eq:ren_ff}
\eeq
which is, again, a Lorentzian centered at $\mu=\mu_0$.
As a result of the continuous measurement, the width of the form factor 
is broadened as $\Delta \to \Delta + (2\tau_{\rm r})^{-1}$,
satisfying the following sum rule;
\beq
\int d\mu |\bar{g}_{\mu}|^2 = \frac{\gamma \Delta}{2}.
\eeq
The renormalized form factor is plotted in Fig.~\ref{fig:rff}.
Applying the Fermi golden rule to Eq.~(\ref{eq:renH}),
the atomic decay rate is calculated as
\footnote{
One can also derive Eq.~(\ref{eq:Gam_mes}), 
combining Eq.~(\ref{eq:Lff}), (\ref{eq:Gtr}) and (\ref{eq:ffmu}).
}
\beq
\Gamma(\tau_{\rm r}) = 2\pi |\bar{g}_{\Om}|^2 = \gamma
\frac{\Delta(\Delta+(2\tau_{\rm r})^{-1})}{(\Om-\mu_0)^2
+(\Delta+(2\tau_{\rm r})^{-1})^2}.
% \verb#{eq:Gam_mes}#
\label{eq:Gam_mes}
\eeq
It is confirmed from Fig.~\ref{fig:decrate2} that
the FGR decay rate agrees well with the rigorous numerical results.

\begin{figure}%----------------------------------------------------------------
\bec
\includegraphics{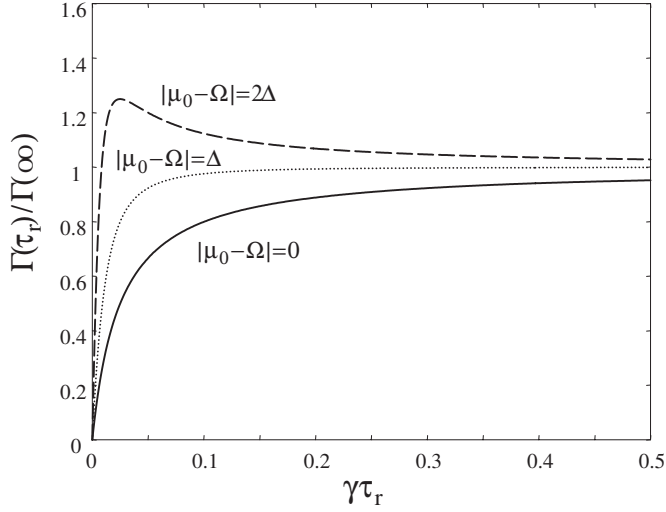}
\caption{\label{fig:norm_dr2}
Dependence of the normalized decay rate 
$\Gamma(\tau_{\rm r})/\Gamma(\infty)$
on the response time $\tau_{\rm r}$ of detector,
which is given by Eq.~(\ref{eq:norm_dr}).
The parameters are the same as Fig.~\ref{fig:norm_dr};
$\Delta=20\gamma$ ($t_{\rm j}=0.05\gamma^{-1}$)
and the values of $|\Om-\mu_0|$ are indicated in Figure.
For $|\Om-\mu_0|=2\Delta$,
the decay rate is maximized when $\tau_{\rm r}=0.025\gamma^{-1}$.
}% \verb#{fig:norm_dr2}#
\enc
\end{figure}%-----------------------------------------------------------------

In order to clarify the effect of measurement,
the decay rate under the continuous measurement is normalized by 
the unobserved decay rate $\Gamma(\infty)$ in Fig.~\ref{fig:norm_dr2}. 
It is given by
\beq
\frac{\Gamma(\tau_{\rm r})}{\Gamma(\infty)} = 
\frac{\Delta+(2\tau_{\rm r})^{-1}}{\Delta}
\frac{(\Om-\mu_0)^2+\Delta^2}{(\Om-\mu_0)^2+(\Delta+(2\tau_{\rm r})^{-1})^2}.
% \verb#{eq:norm_dr}#
\label{eq:norm_dr}
\eeq
This quantity for the case of repeated instantaneous ideal 
measurements 
has been calculated in Fig.~\ref{fig:norm_dr}.
% yet using the conventional theories.
By comparing Figs.~\ref{fig:norm_dr} and \ref{fig:norm_dr2},
we find that the results for the two cases 
agree semi-quantitatively with each other
% the two figures are qualitatively coincident,
%if we interpret the response time $\tau_{\rm r}$ 
% as the measurement interval $\tau_{\rm i}$.
if we identify 
the measurement intervals $\tau_{\rm i}$ of Fig.~\ref{fig:norm_dr}
with the response time $\tau_{\rm r}$ of Fig.~\ref{fig:norm_dr2}
as $\tau_{\rm i} \simeq 2.64 \tau_{\rm r}$.
(However, 
complete quantitative agreement is not attained:
For example, the peak values of the decay rate 
for $|\mu_0-\Om|=2\Delta$  (broken line)
are different between Figs.~\ref{fig:norm_dr} and \ref{fig:norm_dr2}.)
Furthermore, 
Eq.~(\ref{eq:norm_dr}) indicates that the Zeno effect
becomes significant when $(2\tau_{\rm r})^{-1} \gtrsim \Delta$,
i.e.,
\beq
\tau_{\rm r} \lesssim \Delta^{-1}  =  % \simeq 
t_{\rm j},
% \verb#{eq:condr}#
\label{eq:condr}
\eeq
which is certainly confirmed in Fig.~\ref{fig:decrate2}.
A similar condition, Eq.~(\ref{eq:condi}), 
has been obtained also for repeated instantaneous ideal measurements 
by the conventional theory.

The above observations demonstrate that, 
in an idealized case 
where every photon is detected with the same response time ({\it flat} response),
repeated instantaneous ideal measurements
and a continuous measurement
give similar results for the Zeno effect.
% the conventional theory based on the projection postulate
% is reproducible by unitary dynamics of the enlarged system, S+A. 
% Eq.~(\ref{eq:norm_dr}) indicates that a significant QZE or AZE
% is expected when $(2\tau_{\rm r})^{-1} \gtrsim \Delta$,
% i.e.,
% \beq
% \tau_{\rm r} \lesssim \Delta^{-1} \simeq t_{\rm j},
% % \verb#{eq:condr}#
% \label{eq:condr}
% \eeq
% which is certainly confirmed in Fig.~\ref{fig:decrate2}.
% A similar condition, Eq.~(\ref{eq:condi}), 
% has been obtained also by the conventional theory.
% This fact also supports the equivalence 
% between the measurement interval ($\tau_{\rm i}$) 
% and the response time ($\tau_{\rm r}$).

%%%%%%%%%%%%%%%%%%%%%%%%%%%%%%%%%%%%%%%%%%%%%%%%%%%%%%%%%%%%%%%%%%%%%%%%%%%%%%
\subsubsection{QZE--AZE phase diagram}
%%%%%%%%%%%%%%%%%%%%%%%%%%%%%%%%%%%%%%%%%%%%%%%%%%%%%%%%%%%%%%%%%%%%%%%%%%%%%%

\begin{figure}%----------------------------------------------------------------
\bec
\includegraphics{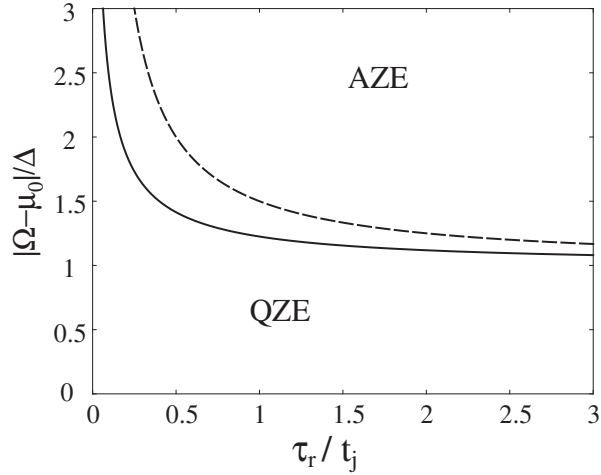}
\caption{\label{fig:phased}
The phase diagram for the QZE and the AZE, 
for a case of Lorentzian form factor and a continuous measurement
with flat response.
The solid curve divides the QZE region and the AZE region.
The dotted line shows the value of $\tau_{\rm r}$ at which 
the decay rate is maximized for each value of $|\Om-\mu_0|$.
}% \verb#{fig:phased}#
\enc
\end{figure}

By analyzing Eq.~(\ref{eq:norm_dr})
as a function of $|\Om-\mu_0|$ and $\tau_{\rm r}$,
a `phase diagram' discriminating the QZE and AZE is generated,
which is shown in Fig.~\ref{fig:phased}.
The `phase boundary' (solid curve) is given by
\beq
\tau_{\rm r}^{\rm (b)}=
\frac{\Delta}{2[(\Om-\mu_0)^2-\Delta^2]},
% \verb#{eq:pb}#
\label{eq:pb}
\eeq 
on which the decay rate is not altered 
from the free rate, i.e., $\Gamma(\tau_{\rm r}^{\rm (b)})=\Gamma(\infty)$.
The decay rate takes the maximum value,
\beq
\frac{\Gamma(\tau_{\rm r}^{\rm (m)})}{\Gamma(\infty)}
=\frac{|\Om-\mu_0|^2+\Delta^2}{2\Delta|\Om-\mu_0|}-1,
\eeq
on the dotted line, which is given by
\beq
\tau_{\rm r}^{\rm (m)}=
\frac{1}{2(|\Om-\mu_0|-\Delta)}.
\eeq
When the atomic transition energy is close to the center of the form factor
($|\Om-\mu_0| \lesssim \Delta$), only the QZE is observed.
However, in the opposite case   
($|\Om-\mu_0| \gtrsim \Delta$),
the AZE is observed dominantly except for an extremely small response time.
In this respect, one may say that 
the AZE is more widely expected than the QZE~\cite{KofNature}.
It should be remarked that
Fig.~\ref{fig:phased} agrees semi-quantitatively with Fig.~\ref{fig:pdpp},
which is the phase diagram for repeated instantaneous ideal 
measurements.

%-----------------------------------------------------------------

% shmzshmz<

%%%%%%%%%%%%%%%%%%%%%%%%%%%%%%%%%%%%%%%%%%%%%%%%%%%%%%%%%%%%%%%%%%%%%%%%%%%%%%
\subsection{Geometrically imperfect measurement}
% \verb#{sec:gim}#
\label{sec:gim}
%%%%%%%%%%%%%%%%%%%%%%%%%%%%%%%%%%%%%%%%%%%%%%%%%%%%%%%%%%%%%%%%%%%%%%%%%%%%%%

\begin{figure}%----------------------------------------------------------------
\bec
\includegraphics{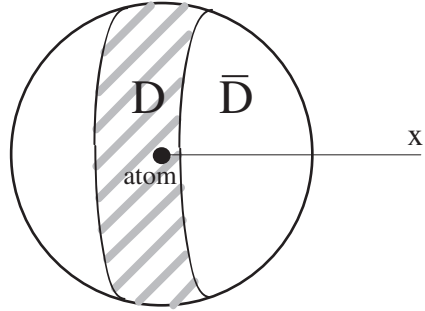}
\caption{\label{fig:gim}
Illustration of geometrically imperfect measurement.
The detector covers only a part of the whole solid angle around the atom,
and some of emitted photons are lost without being detected.
}% \verb#{fig:gim}#
\enc
\end{figure}%-----------------------------------------------------------------

\begin{figure}%----------------------------------------------------------------
\bec
\includegraphics{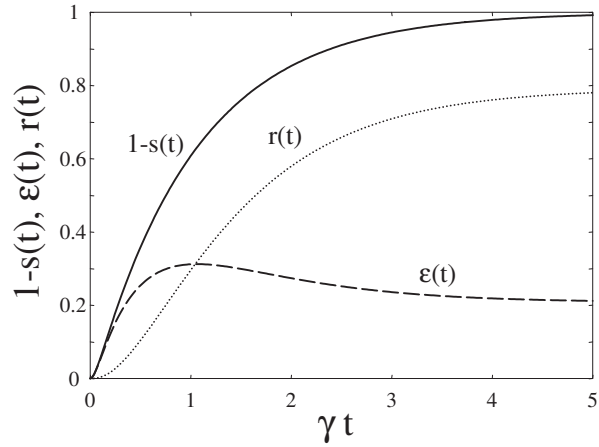}
\caption{\label{fig:ser2}
Temporal evolutions of $1-s(t)$, $\varepsilon(t)$, and $r(t)$.
The values of the parameters are the same as in Fig.~\ref{fig:ser}
(i.e., 
$t_{\rm j} = 0.05\gamma^{-1}$, $|\Om - \mu_0|=0$, $\tau_{\rm r} = 0.5\gamma^{-1}$)
except that $\ve_{\infty}=0.2$ in this Figure.
}% \verb#{fig:ser2}#
\enc
\end{figure}%-----------------------------------------------------------------

%shmzshmz>
In the previous section, we have discussed 
the Zeno effect under an continuous measurement with flat response,
in which all photons are detected with the same response time.
In the following subsections,
we discuss more realistic measurement processes,
in which the response time may be different among different 
photon modes.
%It is indispensable to employ the quantum measurement theory
%to analyze such types of measurement.

As an example of such realistic measurements,
we consider in this subsection a {\it geometrically} 
imperfect measurement~\cite{KSPRA},
% shmzshmz>
in which the detector is inactive to photons in some modes
because of a geometric condition.
For example, suppose that the photoabsorptive medium
composing the detector is sensitive 
only to the $x$-component of the electric field.
Then, the photon-detector interaction becomes proportional
to $\vece_{\veck \lambda}\cdot \vece_x$,
where $\vece_{\veck \lambda}$ is the polarization 
vector, which is perpendicular to $\veck$, 
% is a unit vector 
% oriented in the polarization direction of the photon
and $\vece_x=(1, 0, 0)$.
Therefore,
such a detector is inactive to photons 
whose wavevector $\veck$ is oriented in the $x$ direction, for example.
% because $\vece_{\veck \lambda}\cdot \vece_x$ vanishes.

To discuss essential points of the geometrically imperfect measurements,
we here consider a simplified example, in which  
the detector has an active solid angle ${\cal D}$
and an inactive solid angle $\bar{\cal D}$ around the atom,
as illustrated in Fig.~\ref{fig:gim}.
We assume that 
the detector can catch an emitted photon
with a unique response time $\tau_{\rm r}$
when the wavevector of the photon is oriented inside of ${\cal D}$;
otherwise, the detector misses the photon.
Thus, we put
\beq
\eta_{\veck}=
\begin{cases}
\tau_{\rm r}^{-1} & (\veck \in {\cal D}) \\
0 & (\veck \in \bar{\cal D}) 
\end{cases}.
% \verb#{eq:tauk}#
\label{eq:tauk}
\eeq
As for the atom-photon coupling, we assume 
the Lorentzian form factor again and take the following form:
\bea
\int_{\veck \in {\cal D}} d\veck |g_{\veck}|^2 \delta(\eps_{\veck}-\mu)
&=& \frac{(1-\ve_{\infty})\gamma}{2\pi}\frac{\Delta^2}{\Delta^2+(\mu-\mu_0)^2},
% \verb#{eq:Lff1}#
\label{eq:Lff1}
\\
\int_{\veck \in \bar{\cal D}} d\veck |g_{\veck}|^2 \delta(\eps_{\veck}-\mu)
&=& \frac{\ve_{\infty}\gamma}{2\pi}\frac{\Delta^2}{\Delta^2+(\mu-\mu_0)^2}.
% \verb#{eq:Lff2}#
\label{eq:Lff2}
\eea
The newly introduced parameter $\ve_{\infty}$
represents the probability that the emitted photon 
is lost without being detected.
For example, 
if % $\bar{\cal D} = \emptyset$ and 
spontaneous emission occurs spherically symmetrically
(i.e., $g_{\veck}$ is independent of the direction of $\veck$),
and if 
the detector is sensitive only to the $x$-component of the electric field,
then $\ve_{\infty}=1/3$.
In general cases such as the case of the dipole radiation,
$\ve_{\infty}$ also depends on the direction of 
the transition dipole of the atom.

The temporal behaviors of $s(t)$, $\ve(t)$ and $r(t)$ 
are plotted in Fig.~\ref{fig:ser2}.
Contrarily to the case of flat response,
which is plotted in Fig.~\ref{fig:ser},
$\ve(t) \to \ve_{\infty}$ $(\neq 0)$ and 
$r(t) \to 1-\ve_{\infty}$ $(\neq 1)$
even in the limit of $t \to \infty$.
Using Eqs.~(\ref{eq:drum}), (\ref{eq:barg2}), (\ref{eq:Lff1}), 
(\ref{eq:Lff2}) and (\ref{eq:tauk}), 
we can calculate the decay rate as
\bea
\Gamma(\tau_{\rm r}, \ve_{\infty}) &=& 
(1-\ve_{\infty})\gamma
\frac{\Delta(\Delta+(2\tau_{\rm r})^{-1})}{(\Om-\mu_0)^2+(\Delta+(2\tau_{\rm r})^{-1})^2}
+
\ve_{\infty}\gamma
\frac{\Delta^2}{(\Om-\mu_0)^2+\Delta^2}
\\
&=& \ve_{\infty}\Gamma(\infty) + (1-\ve_{\infty})\Gamma(\tau_{\rm r}).
\eea
Here, $\Gamma(\infty)$ is the free decay rate
and $\Gamma(\tau_{\rm r})$ is the decay rate under 
continuous measurement with flat response,
Eq.~(\ref{eq:Gam_mes}). 
Thus, the decay rate is simply given by these mixture
under geometrically imperfect measurement.
% of 
% the unobserved decay rate and the ideally observed rate.
It is therefore clear that 
if $\ve_{\infty} \sim 1$ then 
the decay rate in this case differs much from 
that under repeated instantaneous ideal measurements.
% which agrees semi-quantitatively with $\Gamma(\tau_{\rm r})$. 
Although this result may sound rather trivial, 
more surprising examples
will be presented 
in the following subsections.
% This is the simplest, but maybe rather trivial, example of 
% the discrepancy between the results for general measurements
% and those of ideal measurements.
% In the following subsections, we will present more drastic examples.
% shmzshmz<

%%%%%%%%%%%%%%%%%%%%%%%%%%%%%%%%%%%%%%%%%%%%%%%%%%%%%%%%%%%%%%%%%%%%%%%%%%%%%%
\subsection{Quantum Zeno effect by energetically imperfect measurement}
% \verb#{sec:eim}#
\label{sec:eim}
%%%%%%%%%%%%%%%%%%%%%%%%%%%%%%%%%%%%%%%%%%%%%%%%%%%%%%%%%%%%%%%%%%%%%%%%%%%%%%

\begin{figure}%----------------------------------------------------------------
\bec
\includegraphics{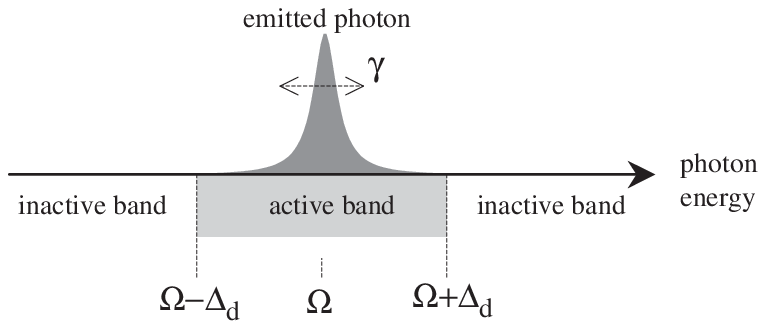}
\caption{\label{fig:eim}
Illustration of an energetically imperfect measurement.
The gray spike represents the energy spectrum of an emitted photon,
which is defined by 
$I(\om)=\lim_{t \to \infty}\int d\veck \delta(\eps_{\veck}-\om)
|\langle 0|b_{\veck} 
e^{-\rmi H_S t} \sigma_+|0 \rangle|^2$.
The detector is sensitive only to photons within the active detection band,
which spans $(\Om-\Delta_{\rm d}, \Om+\Delta_{\rm d})$.
In the model of Secs.~\ref{sec:eim} and \ref{sec:fm},
where Eq.~(\ref{eq:flatff}) is assumed,
$I(\om)$ %=|g_{\om}|^2|\bar{A}(\om)|^2 \propto |\bar{A}(\om)|^2$,
becomes proportional to the dressed atomic Green function $\bar{A}(\om)$,
which is given by Eq.~(\ref{eq:expansion}). 
}% \verb#{fig:eim}#
\enc
\end{figure}%-----------------------------------------------------------------

Actual materials composing photodetectors are sensitive 
only to photons within a restricted energy range,
which is the source of another kind of imperfectness of measurement.
Thus, we are led to consider {\it energetically} imperfect measurement~\cite{KSPRL},
% shmzshmz>
where, as illustrated in Fig.~\ref{fig:eim},
the range of the measurement of photons by the 
photodetector does not cover all the energy range of a photon.
In other words, 
the photodetector has a finite detection band.
% where the photodetector has a finite detection band,
% as illustrated in Fig.~\ref{fig:eim}.
% In other words, the range of the measurement of photons by the 
% photodetector does not cover all the energy range of a photon.
To simplify the discussion,  
% As the simplest situation, 
we consider the case where 
the detector responds with an identical response time $\tau_{\rm r}$ 
to a photon if its energy $\eps_{\veck}$ falls 
in the detection band as $|\eps_{\veck}-\Om|<\Delta_{\rm d}$,
whereas it does not respond to photons outside of this detection band.
% shmzshmz<
Thus, we take the following form for the photon-detector coupling:
\beq
\eta_{\veck}=\eta_{\eps_{\veck}}=
\begin{cases}
\tau_{\rm r}^{-1} & (|\eps_{\veck}-\Om|<\Delta_{\rm d}), \\
0 & ({\rm otherwise}). 
\end{cases}
% \verb#{eq:tauk2}#
\label{eq:tauk2}
\eeq

It is of note that when $\eta_{\veck}$ depends only on the photonic energy,
namely, $\eta_{\veck}=\eta_{\eps_{\veck}}$,
the formula Eq.~(\ref{eq:barg2}) for the renormalized form factor 
is recast into the following simplified form:
\beq
|\bar{g}_{\mu}|^2 = \int\rmd\om
|g_{\om}|^2
\frac{\eta_{\om}/2\pi}
{|\mu-\om-\rmi\eta_{\om}/2|^2}.
\label{eq:barg2_sim}
\eeq

%%%%%%%%%%%%%%%%%%%%%%%%%%%%%%%%%%%%%%%%%%%%%%%%%%%%%%%%%%%%%%%%%%%%%%%%%%%%%%
\subsubsection{A model for an exact exponential decay}
\label{sssec:exact_exp}
%%%%%%%%%%%%%%%%%%%%%%%%%%%%%%%%%%%%%%%%%%%%%%%%%%%%%%%%%%%%%%%%%%%%%%%%%%%%%%

As for the atom-photon coupling,
we treat a special case where the unobserved form factor 
is given by a constant function:
\beq
|g_{\mu}|^2 = \int d\veck |g_{\veck}|^2 \delta(\mu-\eps_{\veck})
=\frac{\gamma}{2\pi},
% \verb#{eq:flatff}#
\label{eq:flatff}
\eeq
%shmzshmz>
which is the $\Delta \to \infty$ limit of a Lorentzian form factor.
One reason why we have employed this form factor
is that  
% Although this form factor is employed to simplify the presentation,
we expect that the qualitative results will not be much different 
for other cases if the unobserved form factor has a finite $\Delta$.
Another reason, which is more important, 
is that the above model extracts most clearly 
a drastic feature
of the Zeno effect under energetically imperfect measurement.
To see this, 
we note the following points peculiar to the above form factor:
% should be noted:
(i) The survival probability exactly follows 
the exponential decay law as $s(t)=\exp(-\gamma t)$,
i.e., the jump time $t_{\rm j}$ $(=\Delta^{-1})$ is zero
(see Sec.~\ref{sec:Lff}).
(ii) The conventional theory therefore predicts that the system 
undergoes neither the QZE nor the AZE
% under repeated instantaneous ideal measurements
(see Sec.~\ref{sec:QZE}).
(iii) Neither effect can be induced in this system 
by a continuous measurement with flat response,
% where the detector responds to every photon
% with an identical response time (see Sec.~\ref{sec:idealm}).
which is proven to yield equivalent results to the conventional theory
(see Sec.~\ref{sssec:drfr}).
However, we will show in the following part of this subsection that
the QZE {\em can} be induced when the measurement is energetically imperfect.
%shmzshmz<

%%%%%%%%%%%%%%%%%%%%%%%%%%%%%%%%%%%%%%%%%%%%%%%%%%%%%%%%%%%%%%%%%%%%%%%%%%%%%%
\subsubsection{Numerical results}
%%%%%%%%%%%%%%%%%%%%%%%%%%%%%%%%%%%%%%%%%%%%%%%%%%%%%%%%%%%%%%%%%%%%%%%%%%%%%%

\begin{figure}%---------------------------------------------------------------
\bec
\includegraphics{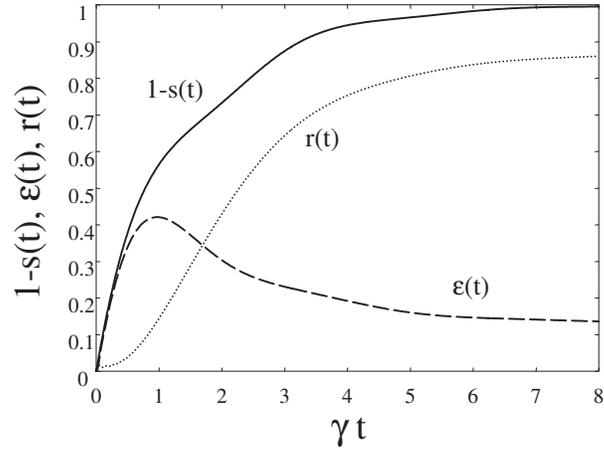}
\caption{\label{fig:ser3}
Temporal evolutions of $1-s(t)$, $\varepsilon(t)$, and $r(t)$.
The parameters are chosen as follows:
$\Delta_{\rm d} = 2\gamma$ and $\gamma \tau_{\rm r} = 0.5$.
}% \verb#{fig:ser3}#
\enc
\end{figure}%-----------------------------------------------------------------
\begin{figure}%---------------------------------------------------------------
\bec
\includegraphics{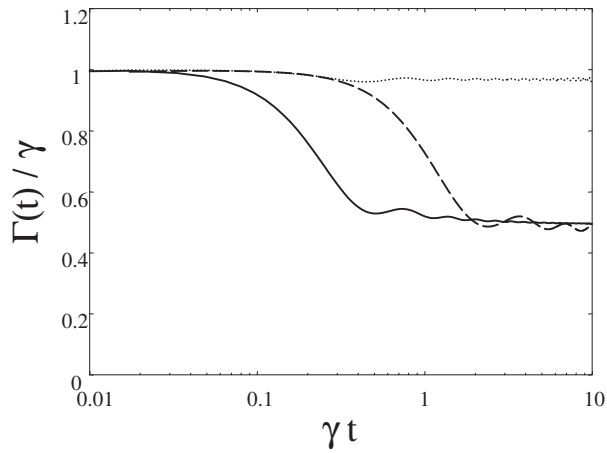}
\caption{\label{fig:logs3}
Temporal evolution of $\Gamma(t)$.
$\{\Delta_{\rm d}, \tau_{\rm r}\}$ are chosen at
$\{10\gamma, 0.05\gamma^{-1} \}$ (solid line),
$\{ 2\gamma, 0.25\gamma^{-1} \}$ (broken line) and
$\{10\gamma, \gamma^{-1}    \}$  (dotted line).
}% \verb#{fig:logs3}#
\enc
\end{figure}%-----------------------------------------------------------------

First we present the numerical results based on the Green function method.
Using the fact that the lineshape of emitted photons is 
an exact Lorentzian with width $\gamma/2$, 
the probability of obtaining the detector response
is naively expected to be 
$r(\infty)=(2/\pi)\arctan (2\Delta_{\rm d}/\gamma)$.
Therefore, significant measurement error will result
when the detection bandwidth $\Delta_{\rm d}$ is small as
$\Delta_{\rm d} \lesssim \gamma$.
In Fig.~\ref{fig:ser3}, temporal behaviors of $1-s(t)$, $\ve(t)$ and $r(t)$
are plotted, for a case of narrow detection band
($\Delta_{\rm d} = 2\gamma$).
The probability of photodetection is 
in good agreement with naive estimation,
$r(\infty)=(2/\pi)\arctan (2\Delta_{\rm d}/\gamma) = 0.84$.
By looking at the decay probability $1-s(t)$,
it is observed that the decay is slightly suppressed for 
$t \gtrsim \gamma^{-1}$.

The change of the decay rate is more emphasized in Fig.~\ref{fig:logs3},
where temporal evolution of $\Gamma(t)=-\ln s(t)/t$ is plotted 
for three different values of $\Delta_{\rm d}$ and $\tau_{\rm r}$.
It should be recalled that 
$\Gamma(t)$ reduces to a constant function ($=\gamma$)
when the atom is not measured.
This feature is contrary to the models with a finite jump time,
where $\Gamma$ always approaches zero as $t \to 0$
% where $\Gamma(t\to 0)$ becomes always zero
as a result of quadratic decrease of $s(t)$
(see, e.g., Fig.~\ref{fig:decrate}).
We find the following two-stage behavior of $\Gamma(t)$ 
in Fig.~\ref{fig:logs3}:
Initially, the decay rate is identical to the unobserved rate $\gamma$,
whereas the decay proceeds with a suppressed rate in the later stage.
For example, when $\Delta_{\rm d}= 10\gamma$ and $\tau_{\rm r} = 0.05\gamma^{-1}$
(solid line in Fig.~\ref{fig:logs3}),
the decay rate changes from $\gamma$ to $0.5\gamma$
at $t \sim 0.1\gamma^{-1}$.
Since the atom is kept almost undecayed 
at the crossover time 
$[s(t\sim 0.1\gamma^{-1})\simeq 0.9]$,
significant decay occurs in the second stage with a suppressed rate.
Thus, the QZE is surely taking place for the exponentially decaying 
system.

%%%%%%%%%%%%%%%%%%%%%%%%%%%%%%%%%%%%%%%%%%%%%%%%%%%%%%%%%%%%%%%%%%%%%%%%%%%%%%
\subsubsection{Conditions for QZE}
% \verb#{sec:cqze}#
\label{sec:cqze}
%%%%%%%%%%%%%%%%%%%%%%%%%%%%%%%%%%%%%%%%%%%%%%%%%%%%%%%%%%%%%%%%%%%%%%%%%%%%%%

\begin{figure}%----------------------------------------------------------------
\bec
\includegraphics{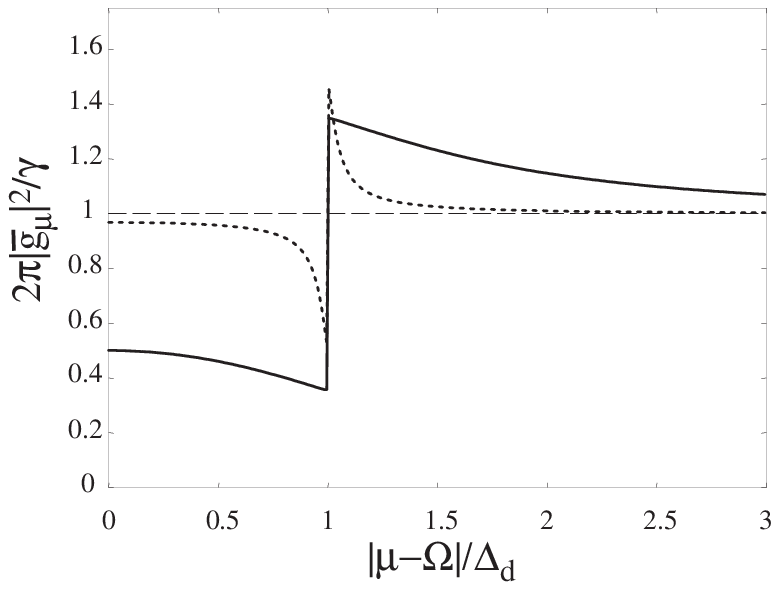}
\caption{\label{fig:ff3}
Renormalization of form factor.
The original form factor without measurement,
which corresponds to $\tau_{\rm r} \to \infty$ limit, 
is a constant function (thin broken line).
The renormalized form factors are plotted for 
$\tau_{\rm r}\Delta_{\rm d} = 0.5$ (solid line)
and $\tau_{\rm r}\Delta_{\rm d} = 10$ (broken line). 
}% \verb#{fig:ff3}#
\enc
\end{figure}%-----------------------------------------------------------------

Here, we explore the underlying mechanisms of the 
two-stage behavior of decay rate,
and clarify the condition for inducing the QZE.
Using Eqs.~(\ref{eq:tauk2}), (\ref{eq:barg2_sim}) and (\ref{eq:flatff}),
the renormalized form factor is calculated as 
\bea
|\bar{g}_{\mu}|^2 &=& \frac{\gamma}{4\pi^2}\int d\om 
\frac{\eta_{\om}}{|\mu-\om-\rmi\eta_{\om}/2|^2}
\\
&=& \frac{\gamma}{2\pi^2}\left[
\pi\theta(|\mu-\Om|-\Delta_{\rm d}) + \arctan (2\tau_{\rm r}(\mu-\Om+\Delta_{\rm d})) 
-\arctan (2\tau_{\rm r}(\mu-\Om-\Delta_{\rm d}))
\right]
\eea
where $\theta(x)$ is a step function.
The renormalized form factor is plotted in Fig.~\ref{fig:ff3}
for three different values of $\tau_{\rm r}\Delta_{\rm d}$.
The form factor is modified locally around the band edge 
in case of large $\tau_{\rm r}\Delta_{\rm d}$,
whereas global modification 
occurs for small $\tau_{\rm r}\Delta_{\rm d}$.
$|\bar{g}_{\mu}|^2$ approaches the unobserved value $\gamma/2\pi$
in the limit of $|\mu-\Om| \to \infty$,
regardless of $\tau_{\rm r}\Delta_{\rm d}$.
Considering that the value of the form factor 
at $\mu=\Om$ is given by 
$|\bar{g}_{\Om}|^2 = (\gamma/\pi^2)\arctan(2\tau_{\rm r}\Delta_{\rm d})$,
we obtain the condition for significant decrease of the form factor
at $\mu=\Om$ as
\beq
\tau_{\rm r}\Delta_{\rm d} \lesssim 1.
% \verb#{eq:cond1}#
\label{eq:cond1}
\eeq

The two-stage behavior can be understood 
with a help of the perturbation theory in $g$.
Applying the lowest-order perturbation 
to the renormalized Hamiltonian Eq.~(\ref{eq:renH}), 
we obtain the decay probability as
\beq
1-s(t)=\int\rmd\mu\ |\bar{g}_{\mu}|^2
\frac{\sin^2[(\mu-\Om)t/2]}{[(\mu-\Om)/2]^2}.
\eeq
Taking into account that the main contribution in the integral 
comes from the region of $\mu$ satisfying 
$|\mu-\Om| \lesssim 2\pi t^{-1}$,
we evaluate the right-hand side in two limiting cases:
In the case of $t \ll \Delta_{\rm d}^{-1}$,
$|\bar{g}_{\mu}|^2$ can be approximated by 
$|\bar{g}_{\infty}|^2=\gamma/2\pi$,
which coincides with the free decay rate $1-s(t)=\gamma t$;
whereas in the opposite case of $t \gg \Delta_{\rm d}^{-1}$,
$|\bar{g}_{\mu}|^2$ can be approximated by $|\bar{g}_{\Om}|^2$,
which gives the suppressed decay rate $1-s(t)=2\pi|\bar{g}_{\Om}|^2t$.
Thus, the decay rate changes from the free rate  
to the suppressed rate at $t\sim\Delta_{\rm d}^{-1}$.
We can confirm that this statement agrees 
with the results in Fig.~\ref{fig:logs3}.

Now the conditions for inducing the QZE 
in exponentially decaying systems are clarified:
(i) The decay rate in the second stage 
should be significantly suppressed from the free decay rate.
This condition is expressed by inequality (\ref{eq:cond1}).
(ii) The transition from the first to the second stage 
should occur before the atom decays.
Since the survival probability 
at the crossover time ($t\sim\Delta_{\rm d}^{-1}$)
is roughly given by 
$s(\Delta_{\rm d}^{-1})\simeq\exp(-\gamma/\Delta_{\rm d})$,
this condition is expressed as 
$\exp(-\gamma/\Delta_{\rm d})\simeq 1$, i.e.,
\beq
\gamma \ll \Delta_{\rm d},
% \verb#{eq:cond2}#
\label{eq:cond2}
\eeq
which means that the detection band should completely cover 
the radiative linewidth of the atom.
Note that 
if the detection bandwidth is not so large ($\Delta_{\rm d} \sim \gamma$)
the {\it partial quantum Zeno effect} takes place,
where suppression of decay starts during the decay 
(at $t \sim \gamma^{-1}$).
The behavior of $s(t)$ in Fig.~\ref{fig:ser3}
serves as an example of the partial QZE.

To summarize this subsection,
when the detector has a finite detection bandwidth
the QZE can be induced even in a system which 
exactly follows the exponential decay law.
The conditions for inducing the QZE on the response time and the bandwidth
are given by inequalities (\ref{eq:cond1}) and (\ref{eq:cond2}),
respectively.
One might immediately notice that these results seemingly contradict
with the well-known wisdom on the QZE,
%shmzshmz<
which states that neither the QZE nor the AZE takes place 
in exactly exponentially decaying systems,
as has been shown in Sec.~\ref{sec:QZE}.
This point will be discussed in Sec.~\ref{sec:rel_ct}.

%%%%%%%%%%%%%%%%%%%%%%%%%%%%%%%%%%%%%%%%%%%%%%%%%%%%%%%%%%%%%%%%%%%%%%%%%%%%%%
\subsection{Quantum anti-Zeno effect by false measurement}
% \verb#{sec:fm}#
\label{sec:fm}
%%%%%%%%%%%%%%%%%%%%%%%%%%%%%%%%%%%%%%%%%%%%%%%%%%%%%%%%%%%%%%%%%%%%%%%%%%%%%%

\begin{figure}%----------------------------------------------------------------
\bec
\includegraphics{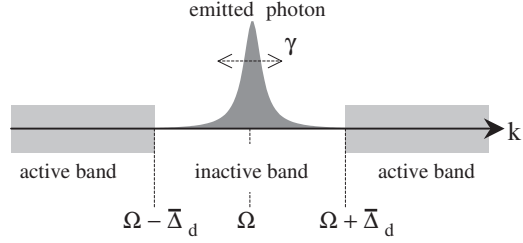}
\caption{\label{fig:fm}
Illustration of the false measurement.
The detector is insensitive to photons within the inactive detection band,
which spans $(\Om-\bar{\Delta}_{\rm d}, \Om+\bar{\Delta}_{\rm d})$.
}% \verb#{fig:fm}#
\enc
\end{figure}%-----------------------------------------------------------------

In Sec.~\ref{sec:eim},
we have observed that the QZE can be induced even in systems
which exactly follows the exponential decay law,
when the measurement is energetically imperfect.
There, the detector was assumed to be active only for photons 
close to the atomic transition energy,
as shown in Fig.~\ref{fig:eim}.
In this section, we consider the opposite situation,
where the active band of the detector does not match 
the energy of a photon emitted from the atom,
as illustrated in Fig.~\ref{fig:fm}~\cite{KKPRL}.
To study this case, 
we assume the following form for $\eta_{\veck}$:
\beq
\eta_{\veck}=\eta_{\eps_{\veck}}=
\begin{cases}
\tau_{\rm r}^{-1} & (|\eps_{\veck}-\Om|>\bar{\Delta}_{\rm d}) \\
0 & ({\rm otherwise}) 
\end{cases}.
% \verb#{eq:tauk3}#
\label{eq:tauk3}
\eeq
A photon which has the atomic transition energy $\Omega$ cannot 
be detected by such an detector.
We here refer to such measurements as {\it false measurements}.
In most  of previous discussions on the Zeno effects,
it was assumed that measuring apparatus 
can detect the decay with a high efficiency,
because the Zeno effect is supposed to appear only weakly
if measurements on the target system are ineffective,
as has been confirmed in Sec.~\ref{sec:gim}
for the case of continuous measurement with flat response.
However, we will show that the Zeno effect can take place 
even under false measurements.

\subsubsection{Natural linewidth}

As for the atom-photon coupling,
we again employ Eq.~(\ref{eq:flatff}),
by which the atomic decay follows an exact exponential decay law.
The spectrum of the emitted photon therefore becomes
a Lorentzian centered around the atomic transition energy $\Om$
with width $\gamma$,
as illustrated in Fig.~\ref{fig:fm}.
% Actually, the spectrum of the emitted photon
% has a finite broadening (the natural linewidth).

If $\gamma$ were larger than $\bar{\Delta}_{\rm d}$, 
then the probability that an emitting photon is detected 
would become large. 
However, we consider the opposite case where 
$\gamma \ll \bar{\Delta}_{\rm d}$,
for which the detection efficiency would be expected to be very small.
When $\bar{\Delta}_{\rm d}=10\gamma$, for example, 
% the measuring apparatus seems to have almost no interaction
% with the target system,
% when the inactive bandwidth $\bar{\Delta}_{\rm d}$ is much larger 
% than the natural linewidth $\gamma$ of emitted photon.
%
we can estimate, 
noting that the lineshape is an exact Lorentzian,
the fraction of photons emitted in the active band 
as 
$1-(2/\pi)\arctan(2\bar{\Delta}_{\rm d}/\gamma)
\simeq 3.2\%$.
Therefore, it is naively expected that 
almost no photons would be counted by the detector 
and that such a false measurement would not affect 
the decay dynamics of the atom significantly.
We will show that this naive expectation is wrong, 
by numerically solving 
the Sch\"{o}dinger equation in the next subsection.
The 
% physical interpretation of this surprising result, 
% as well as the 
conditions for inducing the Zeno effect
will be described in 
Sec.~\ref{sec:cond-fm}.

%%%%%%%%%%%%%%%%%%%%%%%%%%%%%%%%%%%%%%%%%%%%%%%%%%%%%%%%%%%%%%%%%%%%%%%%%%%%%%
\subsubsection{Numerical results}
% \verb#{sec:AZEnum}#
\label{sec:AZEnum}
%%%%%%%%%%%%%%%%%%%%%%%%%%%%%%%%%%%%%%%%%%%%%%%%%%%%%%%%%%%%%%%%%%%%%%%%%%%%%%

\begin{figure}%----------------------------------------------------------------
\bec
\includegraphics{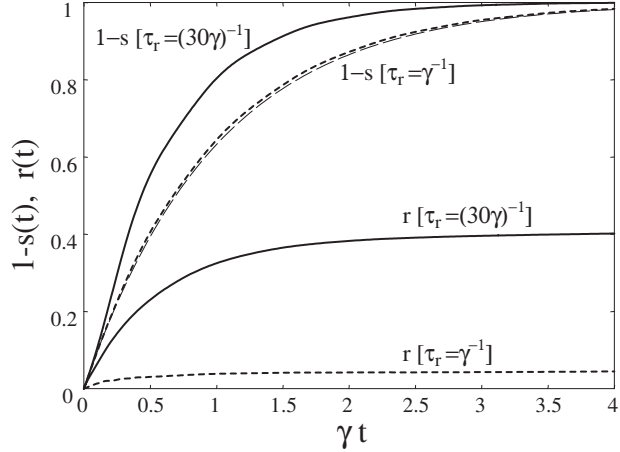}
\caption{\label{fig:ser4}
Temporal evolutions of $1-s(t)$ and $r(t)$.
The dotted and solid lines show the results for
$\tau_{\rm r}=\gamma^{-1}$ (slow detector response)
and $\tau_{\rm r}=(30\gamma)^{-1}$ (fast detector response).
The inactive bandwidth $\bar{\Delta}_{\rm d}$ is $10\gamma$.
The thin broken line shows the decay probability
for unobserved case, where $1-s(t)=1-e^{-\gamma t}$.
}% \verb#{fig:ser4}#
\enc
\end{figure}%-----------------------------------------------------------------
\begin{figure}%----------------------------------------------------------------
\bec
\includegraphics{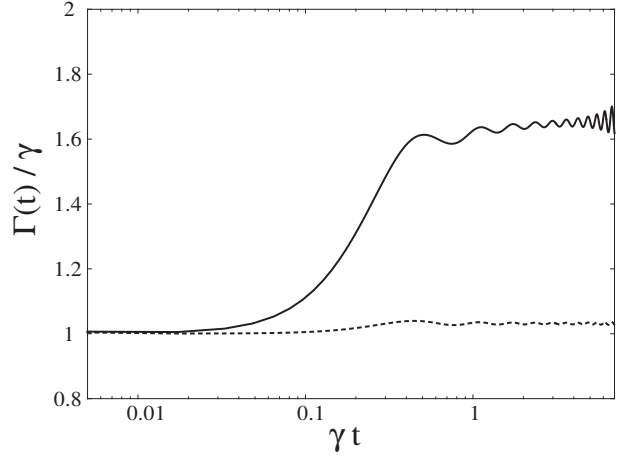}
\caption{\label{fig:logs4}
Temporal evolution of $\Gamma(t)$,
where $\bar{\Delta}_{\rm d}=10\gamma$, 
and $\tau_{\rm r}=\gamma^{-1}$ (dotted line) and $(30\gamma)^{-1}$ (solid line).
The system decays with the unobserved decay rate $\gamma$
for $t \lesssim \bar{\Delta}_{\rm d}^{-1}$,
and with the enhanced decay rate $2\pi |\bar{g}_{\Om}|^2$
for $t \gtrsim \bar{\Delta}_{\rm d}^{-1}$.
}% \verb#{fig:logs4}#
\enc
\end{figure}%-----------------------------------------------------------------

The temporal behaviors of $1-s(t)$ and $r(t)$ are drawn in Fig.~\ref{fig:ser4},
where the inactive bandwidth $\bar{\Delta}_{\rm d} \ (=10\gamma)$ 
is much larger than $\gamma$,
and the {\it false} measurement is realized.
When the detector response is slow 
($\tau_{\rm r}=\gamma^{-1}$, dotted lines in Fig.~\ref{fig:ser4}),
the decay probability is almost unchanged from that of the unobserved case,
i.e., $1-s(t) \simeq 1-e^{-\gamma t}$.
In this case, although a photon is emitted upon decay, 
detection of the emitted photon is almost unsuccessful,
i.e., $r(t) \simeq 0$.
Such behaviors of $1-s(t)$ and $r(t)$ 
agree with the above naive expectation on false measurements.

However, when the detector response is fast
($\tau_{\rm r}=(30\gamma)^{-1}$, solid lines in Fig.~\ref{fig:ser4}),
we find that the detection probability 
of the emitted photon becomes surprisingly large ($\sim 40\%$).
Furthermore, the decay is significantly promoted, 
which is nothing but the AZE.
In Fig.~\ref{fig:logs4},
$\Gamma(t)=-\ln s(t)/t$ is plotted to visualize the decay rate.
The figure clarifies that 
the decay rate changes from the unobserved rate $\gamma$
to the enhanced rate $\bar{\gamma}$
($\simeq 1.6 \gamma$) at $t \sim 0.1\gamma^{-1}$.
Because the atom is kept almost excited at that moment,
it decays with the enhanced rate.
This result would be quite unexpected, considering that
the energy of the emitted photon lies almost completely 
in the inactive band of the detector
and therefore that the detector seemingly 
cannot touch the target system.

%%%%%%%%%%%%%%%%%%%%%%%%%%%%%%%%%%%%%%%%%%%%%%%%%%%%%%%%%%%%%%%%%%%%%%%%%%%%%%
\subsubsection{Conditions for AZE}
\label{sec:cond-fm}
%%%%%%%%%%%%%%%%%%%%%%%%%%%%%%%%%%%%%%%%%%%%%%%%%%%%%%%%%%%%%%%%%%%%%%%%%%%%%%

\begin{figure}%----------------------------------------------------------------
\bec
\includegraphics{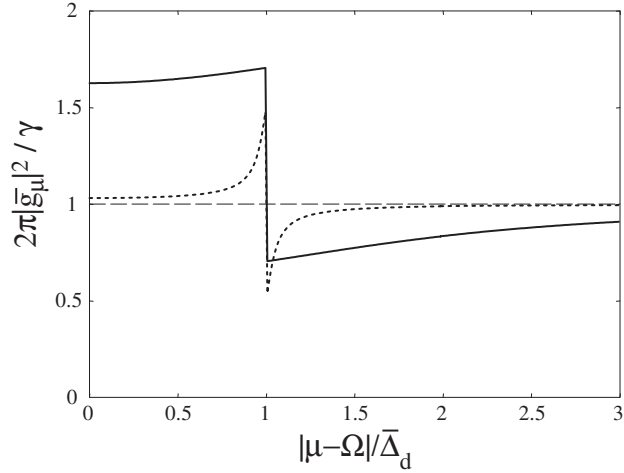}
\caption{\label{fig:ff4}
Plot of the renormalized form factor $|\bar{g}_{\mu}|^2$ 
under the measurement, for $\tau_{\rm r}\bar{\Delta}_{\rm d} = \infty$ (thin broken line),
$10$ (dotted line), and $1/3$ (solid line).
}% \verb#{fig:ff4}#
\enc
\end{figure}%-----------------------------------------------------------------

The unexpected results shown in Sec.~\ref{sec:AZEnum} can be understood
in terms of the renormalized form factor.
It is given by
\beq
|\bar{g}_{\mu}|^2 = \frac{\gamma}{2\pi^2}\left[
\pi+\pi\theta(\bar{\Delta}_{\rm d}-|\mu-\Om|)
+\arctan(2\tau_{\rm r}(\mu-\Om-\bar{\Delta}_{\rm d}))
-\arctan(2\tau_{\rm r}(\mu-\Om+\bar{\Delta}_{\rm d}))
\right],
\eeq
which is plotted in Fig.~\ref{fig:ff4}.
Contrary to Fig.~\ref{fig:ff3},
the form factor is increased at the atomic transition energy,
$\mu=\Om$.
In case of false measurements,
the form factor is always increased inside the inactive band,
which implies that 
false measurements always result in the enhancement of decay
(the AZE).
\footnote{
When $\eta_{\veck}$ depends only on the photonic energy 
as $\eta_{\veck}=\eta_{\eps_{\veck}}$
and the detector has an inactive band $I$,
the decay rate Eq.~(\ref{eq:drum}) is recast into the following form:
\beq
\Gamma = 2\pi |\bar{g}_{\Om}|^2
= \int_{\om \notin I}\rmd\om |g_{\om}|^2
\frac{\eta_{\om}}{|\Om-\om-\rmi\eta_{\om}/2|^2}
+
2\pi \int_{\om \in I}\rmd\om |g_{\om}|^2 \delta(\Om-\om),
\eeq
where $|g_{\om}|^2$ is the original form factor
[not restricted to the flat ones, Eq.~(\ref{eq:flatff})].
In case of false measurements where $\Om \in I$,
the second term gives the unobserved decay rate, $2\pi |g_{\Om}|^2$.
Because the first term is positive,
$\Gamma \geq 2\pi |g_{\Om}|^2$ in general.
}

Following Sec.~\ref{sec:cqze},
the conditions for inducing the AZE 
is summarized as follows:
(i) The decay rate in the second stage $\bar{\gamma}$
should be significantly enhanced from the free decay rate.
This condition is expressed by the following inequality:
\beq
\tau_{\rm r} \bar{\Delta}_{\rm d} \lesssim 1,
\eeq
because $\bar{\gamma}$ is given by
\beq
\bar{\gamma}=2\pi|\bar{g}_{\Om}|^2 
= \gamma [2-2\pi^{-1}\arctan(2\tau_{\rm r} \bar{\Delta}_{\rm d})].
\label{eq:decrate}
\eeq
(ii) The transition from the first to the second stage 
should occur before the atom decays.
This condition is expressed as 
\beq
\gamma \ll \bar{\Delta}_{\rm d},
\label{ineq:cond4}
\eeq
which means that % the {\it false} measurement is realized.
the probability of detecting a photon is naively 
expected to be very {\em small}.

The former condition can be understood intuitively as follows:
The lifetime of a virtually emitted photon in the active band
is estimated by $\delta t \sim (\delta E)^{-1} \sim \bar{\Delta}_{\rm d}^{-1}$,
using the uncertainty principle.
The anti-Zeno effect takes place if the detector response 
$\tau_{\rm r}$ is quick enough to fix a virtual photon,
which is accomplished by $\tau_{\rm r} \lesssim \bar{\Delta}_{\rm d}^{-1}$. 

% shmzshmz>
\subsection{Discussions}
\label{sec:D5}

%%%%%%%%%%%%%%%%%%%%%%%%%%%%%%%%%%%%%%%%%%%%%%%%%%%%%%%%%%%%%%%%%%%%%%%%%%%%%%
\subsubsection{Relation to conventional theories}
% \verb#{sec:rel_ct}#
\label{sec:rel_ct}
%%%%%%%%%%%%%%%%%%%%%%%%%%%%%%%%%%%%%%%%%%%%%%%%%%%%%%%%%%%%%%%%%%%%%%%%%%%%%%

We have observed that the QZE or AZE can be induced 
even in exactly exponentially decaying systems, for which 
$t_{\rm j} = 0$, 
if the measurement is energetically imperfect.
This fact seemingly contradicts with the conventional theories,
which state that neither the QZE nor the AZE can be induced 
in such systems.
However, it should be stressed that this conventional wisdom
was proved only for repeated instantaneous ideal measurements.
Therefore, the relation of the present theory to the 
conventional theories can be seen by taking 
the limits of flat response,
\footnote{
The flat response is one of necessary conditions for reducing 
to repeated instantaneous ideal measurements.
Therefore, it is sufficient for the present purpose to 
show that the Zeno effect disappears for the flat response.
}
as we have done in Sec.~\ref{sec:idealm}, as follows.
 
% is included as a special case of the theory presented here:
% As we have mentioned in Sec.~\ref{sec:idealm}, 
% the conventional theory is reproducible from the present formalism
% in the limit of ideal measurement.
%
Regarding the case of Sec.~\ref{sec:eim},
the limit of flat response
is obtained by taking $\Delta_{\rm d} \to \infty$.
Then, inequality (\ref{eq:cond1}) 
can never be satisfied by a finite $\tau_{\rm r}$
for exponentially decaying systems,
and the QZE does not take place. 
Similarly, regarding the case of Sec.~\ref{sec:fm},
the flat response is obtained by putting $\bar{\Delta}_{\rm d} \to 0$.
Then, inequality (\ref{ineq:cond4}) 
can never be satisfied
for exponentially decaying systems,
and the AZE does not take place. 
% , 
% which leads us to conclude that 
% the AZE never takes place for exponentially systems
% by ideal measurement.
%
We have thus obtained the conventional wisdom
from the present formalism by taking 
the limit of flat response.
It is therefore seen that 
the present theory serves as an extension 
of the conventional theories to realistic situations,
where the detection range of the detector
is always finite.

\subsubsection{Physical interpretation}
\label{sec:po}

% It has been observed in Sec.~\ref{sec:eim}
% that measurement on photons which are close to 
% (far from) the atomic transition energy $\Om$
% tends to suppress (accelerate) the decay of the atom.
In Secs.~\ref{sec:eim} and \ref{sec:fm},
the following two opposite effects of measurement
has been revealed:
(a) Measurement on photons which are close to 
the atomic transition energy $\Om$
tends to suppress the decay of the atom.
(b) Measurement on photons which are far from $\Om$
tends to accelerate the decay of the atom.
When the detector is active for all photons,
these two opposite effects appear simultaneously
and weaken each other.
This is the reason why
a `worse' detector which possesses a finite inactive band
is more advantageous in inducing the QZE or AZE
than a `better' detector which is sensitive to all photons. 
Particularly, when the unobserved form factor is a constant function
(in other words, the system decays with an exact exponential decay law),
the two opposite effects cancels out completely;
the form factor suffers no modification at all,  
no matter how short $\tau_{\rm r}$ would be.
This cancellation mechanism can be understood 
with a help of Fig.~\ref{fig:illff};
although every photonic mode is energetically broadened 
as a counteraction of measurement,
such individual broadenings are perfectly smeared out 
by the $\veck$-integration in Eq.~(\ref{eq:barg2}),
and are not reflected in the renormalized form factor.

In general systems with a nonzero jump time, %finite $\Delta$, 
however, effects (a) and (b) are 
not completely canceled out even in the case of the flat response.
For example, 
in case of a Lorentzian form factor with finite $\Delta$,
the system suffers the QZE or AZE by 
a continuous measurement with flat response,
as discussed in Sec.~\ref{sec:idealm}.
There, it was observed that the QZE (AZE) dominantly takes place
when the atomic transition energy $\Om$ is close to (far from)
the central energy $\mu_0$ of the form factor.
This fact can also be understood in terms of the competition
between effects (a) and (b):
When $\Om$ and $\mu_0$ is close,
effect (a) dominates effect (b),
resulting in the QZE; 
when $\Om$ and $\mu_0$ is far apart,
effect (b) dominates effect (a),
resulting in the AZE.

\subsubsection{Discussions and remarks on the model}
\label{sec:disc5}

In this section, we have analyzed the Zeno effect using a specific model, 
Eqs.~(\ref{eq:H1_ag})-(\ref{eq:Hint1}).
%
% and assuming the measurement intervals $\tau_{\rm i} =0$
% (i.e., continuous measurement).
%
% For example, the model can also describe the case of 
% finite intervals $\tau_{\rm i} >0$.
% The time evolution in such a case is obtained by simply inserting 
% the free evolutions during the intervals, into the equations of this
% section.
% It is then obvious that the Zeno effect is weakened compared with 
% the case of $\tau_{\rm i} =0$.
%
% Furthermore, mathematically, 
% the case of direct measurement can be described by the same Hamiltonian.
% In fact, by rewriting Eqs.~(\ref{eq:H1_ag})-(\ref{eq:Hint1})
% in terms of ``interaction modes'' $B_\mu$ of S (Sec.~\ref{sec:imff}), 
% we find that 
% A interacts {\em directly} with the interaction modes that involves 
% the atomic degrees of freedom.
%
% Although % the results of this section would describe 
We expect that the results based on this model 
would cover most of essential elements of the Zeno effect.
For example, although the model assumes an indirect measurement
we have shown in Sec.~\ref{sec:rel_dir} 
that direct measurements are included as a special case of this model.
However, the following points are worth mentioning about the model.

Firstly, the model is linear, i.e., 
Eqs.~(\ref{eq:H1_ag})-(\ref{eq:Hint1}) are bilinear in 
the creation and annihilation operators.
Although this seems to be a good approximation to an effective
Hamiltonian for photon-counting measurements by standard photodetectors, 
the model cannot describe other experimental setups, of course.
For other experimental setups, the quantitative results of this section 
would become much different, although we think that 
the qualitative results would be similar. %the same.

Secondly, 
we have computed 
the response time and the measurement error %, and the range of measurement, 
as relevant parameters characterizing measurements.
As discussed in Sec.~\ref{sec:rt}, 
they are actually the lower limits of the response time and
measurement error, 
respectively, 
within the model of 
Eqs.~(\ref{eq:H1_ag})-(\ref{eq:Hint1}),
because additional delay and/or measurement error can take place
in subsequent processes such as the signal magnification process
in a photodetector. 
Although the performance of actual measuring devices would be worse, 
the limiting values are 
most important 
% more significant
% than practical values, which depend strongly  
% on detailed experimental conditions, 
in discussing fundamental physics,
as emphasized in Sec.~\ref{sec:rt}.

Thirdly, we have assumed that the detector signal is obtained 
from the average population,  
which is given by Eq.~(\ref{eq:rp}),
of the elementary excitations.
This may be detected by subsequent magnifying processes
through, say, avalanche processes.
Note however that % the model does not 
Eqs.~(\ref{eq:H1_ag})-(\ref{eq:Hint1}) do not 
exclude the possibility of other methods of getting 
the detector signal.
For example, the signal may be obtained from 
the off-diagonal elements
$f_{\veck\om}^*(t) f_{\veck'\om'}(t)$,
where $f_{\veck\om}(t)$ is defined in Eq.~(\ref{eq:psi_t}).
What method is used is not uniquely determined by 
Eqs.~(\ref{eq:H1_ag})-(\ref{eq:Hint1}), which describe 
the dynamics of the systems inside the Heisenberg cut
(Sec.~\ref{sec:HC}).
To determine the full experimental setup, 
we must also specify the systems outside 
the Heisenberg cut.
Among many methods of getting the signal using
Eqs.~(\ref{eq:H1_ag})-(\ref{eq:Hint1}),
we have assumed that the detection of  
the average population would be the most efficient method, 
thus giving the fastest response.
Since we are interested in the lower limit of the response time
as discussed in Sec.~\ref{sec:rt}, 
the calculation of the average population suffices for our purpose.

% However, it is also possible, at least in principle, to get a signal from 
% virtual excitation of the elementary excitations.
% For example, the virtual excitation would induce 
% a change of a refractive index (optical nonlinearity),
% which may be detected by a probe optical beam \cite{NO}.
% Or, if the optical material does not have the spatial inversion 
% symmetry, the virtual excitation would generates an electrical pulse
% (electro-optic effect), which may be detected by an external 
% circuit \cite{NO,SY}.
% from real population of an elementary excitation that is 
% described by $c_{\veck\om}$.
% However, it is also possible, at least in principle, to get a signal from 
% virtual excitation of the elementary excitations.
% For example, the virtual excitation would induce 
% a change of a refractive index (optical nonlinearity),
% which may be detected by a probe optical beam \cite{NO}.
% Or, if the optical material does not have the spatial inversion 
% symmetry, the virtual excitation would generates an electrical pulse
% (electro-optic effect), which may be detected by an external 
% circuit \cite{NO,SY}.

Finally, our model assumes a homogeneous photoabsorptive media for the 
photodetector.
The case where the photodetector is separated spatially from 
the atom will be discussed in Sec.~\ref{sec:atom-in-inhomo}.

\section{Relation to cavity quantum electrodynamics} 
\label{sec:cqed}
%%%%%%%%%%%%%%%%%%%%%%%%%%%%%%%%%%%%%%%%%%%%%%%%%%%%%%%%%%%%%%%%%%%%%
%%%%%%%%%%%%%%%%%%%%%%%%%%%%%%%%%%%%%%%%%%%%%%%%%%%%%%%%%%%%%%%%%%%%%

\subsection{Relation between the Zeno effect and other phenomena}
\label{sec:otherphenomena}

If one is interested in the Zeno effect in the broad
sense (Sec.~\ref{sec:info}), 
only the decay rate will be relevant, whereas 
quantities characterizing measurements, 
such as the measurement error and 
the amount of information obtained by measurement, 
would be irrelevant.
In such a case, 
any change of the decay rate of the target system, 
which is induced by interaction with external systems, %an 
could be called a Zeno effect 
even when no information can be obtained by the interaction process.
Therefore, the Zeno effect in the broad sense 
is not necessarily connected with measurements 
(as discussed in Sec.~\ref{sec:info}),
and, consequently, is closely related to 
various phenomena %studied 
in many different fields of physics~\cite{eiQZE1,eiQZE2,eiQZE3}.

Such phenomena include, for example, 
(i) the motional narrowing \cite{KTH,AW,Mil},
in which the width of an excitation in a solid is reduced by  
perturbations from external noises or  environments, 
(ii) Raman scattering processes \cite{NO,Raman},
in which the transition rate between atomic levels
is modulated by external fields of photons or phonons,
and (iii) the cavity quantum electrodynamics 
(abbreviated as the {\it cavity QED})~\cite{c_and_r2,cQED,Mabu}, 
in which electrodynamics is modified by the presence of 
optical cavities.
These phenomena
can be considered as examples of 
the Zeno effect in the broad sense, and vice versa.

Regarding the models of continuous measurements which are
employed in Sec.~\ref{sec:rmt}, 
in particular, the physical configurations are
quite similar to those of the cavity QED.
In fact, in Sec.~\ref{sec:rmt}
we have studied  
effects of photon-counting measurements
on the decay dynamics of an excited atom,
assuming that 
the photon-counting measurement is accomplished by the interaction 
between photons and photoabsorptive media.
Therefore, if we focus only on the decay dynamics of the atom %rate, 
it can be simply said that 
we have studied effects of the 
photoabsorptive media surrounding the atom on the decay rate.
This is a subject of the cavity QED.
Because of this similarity, 
we discuss in this section
the relation between the cavity QED 
and the results of Sec.~\ref{sec:rmt}.

%%%%%%%%%%%%%%%%%%%%%%%%%%%%%%%%%%%%%%%%%%%%%%%%%%%%%%%%%%%%%%%%%%%%%
%%%%%%%%%%%%%%%%%%%%%%%%%%%%%%%%%%%%%%%%%%%%%%%%%%%%%%%%%%%%%%%%%%%%%
\subsection{Modification of form factor in cavity QED}
%%%%%%%%%%%%%%%%%%%%%%%%%%%%%%%%%%%%%%%%%%%%%%%%%%%%%%%%%%%%%%%%%%%%%
%%%%%%%%%%%%%%%%%%%%%%%%%%%%%%%%%%%%%%%%%%%%%%%%%%%%%%%%%%%%%%%%%%%%%

In discussions of the cavity QED, 
the optical media surrounding an atom are 
usually treated as passive media.
That is, the dynamics of (elementary excitations in) the 
optical media is usually disregarded.
This should be contrasted with the discussions 
in Sec.~\ref{sec:rmt}, where 
microscopic dynamics of the photoabsorptive media
plays an important role as the readout of the measuring apparatus.
If we focus only on the dynamics of the atom, however, 
both the cavity QED and the Zeno effect
can be understood from a unified viewpoint,
using the form factor of the atom-photon interaction.
To see this, note that  
the form factor is sensitive to the optical environment.
For example, the form factor has a continuous spectrum
when the atom is in the free space (see Sec.~\ref{sssec:freespace}).
However, when the atom is surrounded by mirrors, 
the eigenmodes are discretized and 
the form factor becomes a line spectrum (see Sec.~\ref{sssec:cc}).
The decay dynamics is affected by the optical environment
through the modification of the form factor.
This explains the cavity QED, as well as the Zeno effect
in the broad sense.

In the discussions of the cavity QED, 
the optical environment is treated as a passive media,
which is characterized by the dielectric constant $\varepsilon(\vecr)$.
It may depend on the space coordinate $\vecr$ because
the optical environment is spatially inhomogeneous in general.
Since $\varepsilon(\vecr)$ takes complex values, 
its effects on the form factor may be classified into two:
One is the effect of the real part, 
which results in reconstruction of optical eigenmodes, 
and the other is the effect of the imaginary part, 
which induces 
energetic broadening of optical eigenmodes.
We will explain them in the following subsections.

%%%%%%%%%%%%%%%%%%%%%%%%%%%%%%%%%%%%%%%%%%%%%%%%%%%%%%%%%%%%%%%%%%%%%
\subsubsection{Reconstruction of eigenmodes}
% \verb#{sssec:recon}#
\label{sssec:recon}
%%%%%%%%%%%%%%%%%%%%%%%%%%%%%%%%%%%%%%%%%%%%%%%%%%%%%%%%%%%%%%%%%%%%%

Suppose a situation where non-absorptive optical media 
is distributed in space~\cite{rf:Jona}.
Non-absorptive optical media are characterized 
by real dielectric constants.
The Maxwell equation for the electric field is given by
\beq
\nabla \times  \nabla \times \vecE(\vecr, t)=
-\ve(\vecr) \frac{\partial^2}{\partial t^2}\vecE(\vecr, t),
% \verb#{eq:Maxwell}#
\label{eq:Maxwell}
\eeq
where $\varepsilon (\vecr)$ is a real function,
representing the spatial distribution of dielectric constant.
Then, a set of eigenmode functions $\vecf_j(\vecr)$
and eigenfrequencies $\eps_j$,
where $j$ is an index of eigenmodes,
are obtained as the stationary solutions of the Maxwell equations.
When the atom is placed at $\vecr$,
the atom-photon coupling constant is given by Eq.~(\ref{eq:apc1}).
Thus, the eigenenergies $\eps_j$ of the photon modes 
are dependent on the spatial distribution $\varepsilon(\vecr)$ 
of the dielectric constant;
the atom-photon coupling constant $g_j$ is dependent
also on the atomic position $\vecr$,
in addition to $\varepsilon(\vecr)$.
It should be remarked that,
when the optical media is spatially homogeneous,
the eigenmode functions are given by plane waves. 
Then, $\vecr$-dependence of $g_j$ appears only in the phase of $g_j$
and the magnitude $|g_j|^2$ is independent of $\vecr$,
so the atomic decay takes place independently of its position.
However, in general, $|g_j|^2$ depends on the atomic position 
when optical media are distributed inhomogeneously in space;
as a result, the dynamics of the atom becomes
strongly sensitive to the atomic position $\vecr$.

\begin{figure}%----------------------------------------------------------------
\bec
\includegraphics{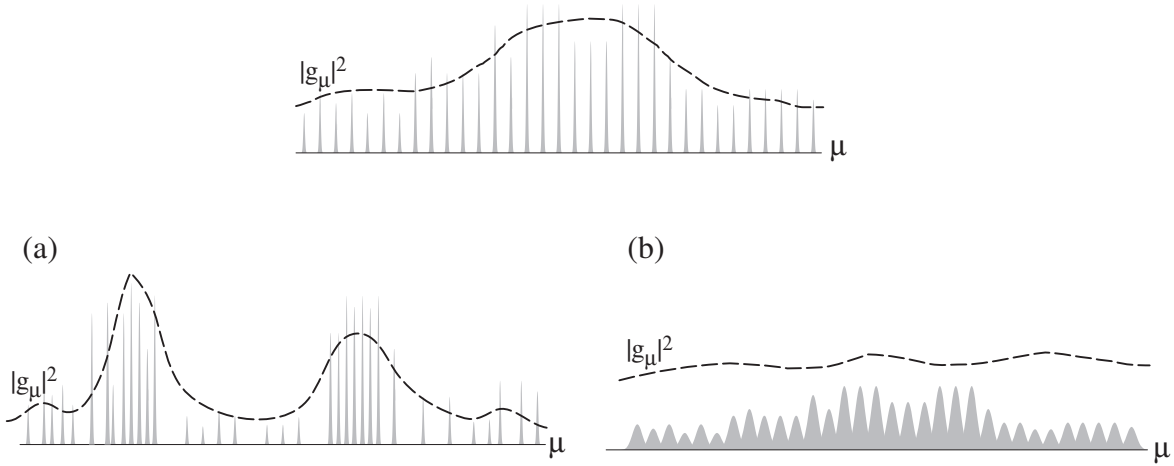}
\caption{\label{fig:mod_ff}
Two mechanisms for modification of a form factor 
in the cavity-QED:
(a) reconstruction of eigenmodes, and 
(b) broadenings of eigenmodes.
}% \verb#{fig:mod_ff}#
\enc
\end{figure}

The form factor, which is given by
\beq
|g_{\mu}|^2 = \sum_j |g_j|^2 \delta (\eps_j-\mu),
\eeq
is modified through $\eps_j$ and $g_j$.
Modification of the form factor by reconstruction of eigenmodes
is illustrated in Fig.~\ref{fig:mod_ff}(a).
The contribution of each photonic mode 
remains a delta function in this case,
but the strength ($|g_j|^2$) and position ($\eps_j$)
of a delta function is modified.

%-----------------------------------------------------------------

Two representative examples of this type of modification
of the form factor are as follows:
(i) Only the atomic position $\vecr$ is changed,
keeping the spatial distribution $\varepsilon(\vecr)$ unchanged.
Then, the positions of the delta functions ($\eps_j$) are unchanged,
but the strength ($|g_j|^2$) is changed,
depending on the eigenmode function $\vecf_j(\vecr)$.
(ii) When the size of a perfect cavity is increased,
the eigenenergies are red-shifted.
$|g_j|^2$ shows a complicated behavior, 
depending on the relative position of the atom 
to the cavity as explained in (i). 
Generally, $|g_j|^2$ is decreased proportionally 
to the inverse of the cavity volume.

%%%%%%%%%%%%%%%%%%%%%%%%%%%%%%%%%%%%%%%%%%%%%%%%%%%%%%%%%%%%%%%%%%%%%
\subsubsection{Broadening of eigenenergies}
% \verb#{sssec:brd}#
\label{sssec:brd}
%%%%%%%%%%%%%%%%%%%%%%%%%%%%%%%%%%%%%%%%%%%%%%%%%%%%%%%%%%%%%%%%%%%%%

We now consider the case where absorption of photons by
optical media is significant.
In photoabsorptive media,
photons are converted within finite lifetimes
to elementary excitations in the media,
such as electron-hole pairs, excitons, {\it etc}.
Therefore, 
in the presence of photoabsorptive media,
eigenmodes of photons do not exist in a strict sense;
they should be regarded as quasi-eigenmodes
with finite lifetimes.
The situation is phenomenologically described by
a complex eigenenergy of the mode,
the imaginary part of which is inversely proportional 
to the lifetime of the mode.

We have already observed an example of 
broadening of eigenenergies in Sec.~\ref{sec:rmt}
in the context of the Zeno effect.
The schematic view is given again in Fig.~\ref{fig:mod_ff}(b).
When there is no absorption of photons,
the form factor is composed of delta functions, 
each of which represents the contribution of each eigenmode.
In the presence of a detector,
which absorbs photons with finite lifetimes,
a delta function is broadened to be a Lorentzian
as in Eq.~(\ref{eq:brdning}),
satisfying a sum rule, Eq.~(\ref{eq:srule}).

It should be remarked that the broadening 
of a cavity mode was observed in Sec.~\ref{ssec:LC},
through the coupling between the cavity mode and external photon modes.
The difference between the systems considered 
in Secs. \ref{ssec:LC} and \ref{sec:rmt} lies in the point that
both the cavity mode ($a$) and external modes ($b_{\om}$) 
are of photonic origin in Sec.~\ref{ssec:LC},
whereas a photon mode ($b_{\veck}$) is coupled 
to non-photon modes (elementary excitations 
in the detector, $c_{\veck\om}$) in Sec.~\ref{sec:rmt}.
In the system of Sec.~\ref{ssec:LC},
a diagonalized mode ($B_{\om}$) still 
represents a photonic eigenmode,
which extends over the whole space,
both inside and outside of the cavity.
Actually, by solving the Maxwell equation
treating the mirrors as an optical medium 
with real dielectric constants,
one can obtain eigenmode functions for $B_{\om}$. 
Thus, the broadening of a cavity mode
observed in Sec.~\ref{ssec:LC}
should be classified as an example of reconstruction of eigenmodes,
discussed in Sec.~\ref{sssec:recon}.

%%%%%%%%%%%%%%%%%%%%%%%%%%%%%%%%%%%%%%%%%%%%%%%%%%%%%%%%%%%%%%%%%%%%%
%%%%%%%%%%%%%%%%%%%%%%%%%%%%%%%%%%%%%%%%%%%%%%%%%%%%%%%%%%%%%%%%%%%%%
\subsection{Atom in a homogeneous absorptive medium}
% \verb#{ssec:a_ham}#
\label{ssec:a_ham}
%%%%%%%%%%%%%%%%%%%%%%%%%%%%%%%%%%%%%%%%%%%%%%%%%%%%%%%%%%%%%%%%%%%%%
%%%%%%%%%%%%%%%%%%%%%%%%%%%%%%%%%%%%%%%%%%%%%%%%%%%%%%%%%%%%%%%%%%%%%

In the previous subsection,
two mechanisms for modification of the form factor 
in general cavity-QED systems are described.
In most photoabsorptive materials,
these two mechanisms usually appear simultaneously.
In order to see this point,
we revisit the Hamiltonian for an atom 
embedded in a homogeneous photoabsorptive media,
which was used in Sec.~\ref{sec:rmt}
for discussion of the Zeno effect.
Here, we consider a case without the flat-band 
assumption, Eq.~(\ref{eq:flatband}).
Following the same mathematics as Sec.~\ref{sec:rff}, 
one can confirm that the contribution of 
the photon $\veck$ to the form factor
is modified as follows:
\beq
|g_{\veck}|^2 \delta (\mu - \eps_{\veck})
\rightarrow
\frac{|g_{\veck}|^2}{2\pi}
\left( \frac{\displaystyle i}{\displaystyle
\mu - \eps_{\veck} + 
\int d\om \frac{|\xi_{\veck\om}|^2}{\om-\mu-i\delta}}
+{\rm c.c.}\right).
% \verb#{eq:gen_rff}#
\label{eq:gen_rff}
\eeq
The integral in the denominator of the RHS of Eq.~(\ref{eq:gen_rff})
becomes significant when $|\mu - \eps_{\veck}|$ is small.
We may therefore approximate this quantity roughly by
\bea
\int d\om \frac{|\xi_{\veck\om}|^2}{\om-\mu-i\delta}
&\simeq&
\int d\om \frac{|\xi_{\veck\om}|^2}{\om-\eps_{\veck}-i\delta}
% \verb#{eq:app1a}#
\label{eq:app1a}
\\
&=& 
P\int d\om \frac{|\xi_{\veck\om}|^2}{\om-\eps_{\veck}}
+i\pi |\xi_{\veck, \eps_{\veck}}|^2
% \verb#{eq:app1b}#
\label{eq:app1b}
\\
& \equiv &
-\Delta \eps_{\veck}+i\eta_{\veck}/2
% \verb#{eq:app1c}#
\label{eq:app1c}
\eea
Using the above quantity, Eq.~(\ref{eq:gen_rff}) is transformed into 
a more transparent form:
\beq
|g_{\veck}|^2 \delta (\mu - \eps_{\veck})
\rightarrow
|g_{\veck}|^2 \frac{\eta_{\veck}/2\pi}
{|\mu-(\eps_{\veck}+\Delta \eps_{\veck})+i\eta_{\veck}/2|^2}.
% \verb#{eq:gen_rrf2}#
\label{eq:gen_rrf2}
\eeq
Eqs.(\ref{eq:app1b}) and (\ref{eq:app1c}) reveal that 
asymmetry in the photon-medium coupling 
$|\xi_{\veck\om}|^2$ about the photon energy $\eps_{\veck}$
results in a shift of the eigenfrequency, $\Delta \eps_{\veck}$.
Generally, the photon-medium coupling simultaneously 
induces both broadening and energy shift.

In the discussion of the Zeno effect in Sec.~\ref{sec:rmt},
shifts of the eigenenergy are neglected 
by the flat-band assumption, Eq.~(\ref{eq:flatband}),
and only the broadening effect is picked up.
Under this approximation,
fairly good agreement with the % shmz
conventional theories of the Zeno effect has been achieved,
as has been shown in Sec.~\ref{sec:idealm}.
It would be important to point out that
only the broadening effect is assumed
in the conventional theory of the Zeno effect:
by using the projection postulate to the atomic state,
coherences between the undecayed ($|x \rangle$) 
and decayed ($|g,\veck\rangle$) states
are lost without shifting the energy of photons.
In actual photon counting processes,
eigenenergies of photons would be slightly shifted 
through the interaction with a detector,
which also affect the atomic decay rate.

%%%%%%%%%%%%%%%%%%%%%%%%%%%%%%%%%%%%%%%%%%%%%%%%%%%%%%%%%%%%%%%%%%%%%
%%%%%%%%%%%%%%%%%%%%%%%%%%%%%%%%%%%%%%%%%%%%%%%%%%%%%%%%%%%%%%%%%%%%%
\subsection{Atom in an inhomogeneous absorptive medium}
\label{sec:atom-in-inhomo}
%%%%%%%%%%%%%%%%%%%%%%%%%%%%%%%%%%%%%%%%%%%%%%%%%%%%%%%%%%%%%%%%%%%%%
%%%%%%%%%%%%%%%%%%%%%%%%%%%%%%%%%%%%%%%%%%%%%%%%%%%%%%%%%%%%%%%%%%%%%

In discussion of the Zeno effect in Sec.~\ref{sec:rmt},
we have considered a situation in which an atom is embedded
in a homogeneous absorptive material.
However, in actual situations of photon counting,
photodetectors composed of photoabsorptive material
are spatially separated from the target atom,
as illustrated in Fig.~\ref{fig:ill_realm}.
In the following two subsections, we discuss
what would be expected for the decay rate % shmz
when the detectors are spatially separated from the atom.
We shall describe results of the cavity QED in 
subsection \ref{sec:atom-in-inhomo-CQED},
and apply them to the Zeno effect in subsection \ref{sec:separated}.

\subsubsection{Cavity QED}
\label{sec:atom-in-inhomo-CQED}

For simplicity, we consider the following situation here:
The space is assumed to be one-dimensional,
extended in the $x$-direction,
and the photon field is treated as a scalar field.
The optical materials are modeled by 
bulk absorptive dielectric materials 
occupying $|x|>l/2$, which has a complex dielectric constant
$\eps(\om)=[\eta(\om)+i\kappa(\om)]^2$.

\begin{figure}%----------------------------------------------------------------
\bec
\includegraphics{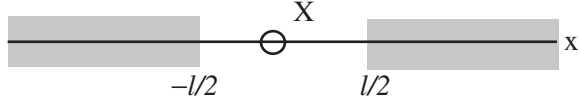}
\caption{\label{fig:ill_realm}
Arrangement of the target atom and absorptive media
composing detectors.
The atom is located at $x=X$, whereas the absorptive media
are placed in the regions $x \leq -l/2$ and $x \geq l/2$.
}% \verb#{fig:ill_realm}#
\enc
\end{figure}%-----------------------------------------------------------------

\begin{figure}%----------------------------------------------------------------
\bec
\includegraphics{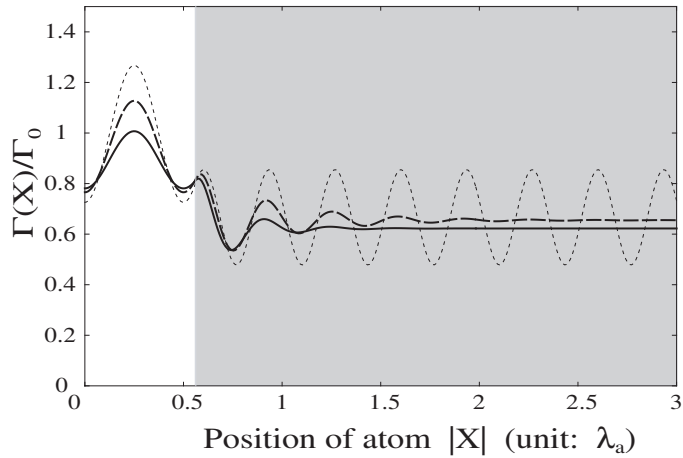}
\caption{\label{fig:pddr}
Position dependence of the decay rate 
(normalized by the decay rate in free space).
Only the $X>0$ region is plotted 
because $\Gamma(X)$ is an even function.
The length of the vacuum region is chosen at $l=(9/8) \lambda_a$,
where $\lambda_a=2\pi c/\Om$ is 
the wavelength of the emitted photon.
$\eta =1.5$ and $\kappa=0$ (thin dotted line),
0.2 (broken line), and 0.4 (solid line).
}% \verb#{fig:pddr}#
\enc
\end{figure}

First, we preliminarily consider a case
where the medium is not absorptive
and its dielectric constant is given by $\eps(\om)=\eta^2$
in the relevant frequency region under consideration. %shmz
In this case, as has been discussed in Sec.~\ref{sssec:recon},
photonic eigenmodes % at frequency $\om$ 
are reconstructed according to the following equation:
\beq
-\frac{\partial^2}{\partial x^2}E =
-\ve(x) \frac{\partial^2}{\partial t^2}E
\eeq
which is a scalar field version of Eq.~(\ref{eq:Maxwell}).
Reviving the light velocity $c$ here,
the eigenmodes at frequency $\om$ are given by
\bea
f_{\om, +}(x)&=& % \sqrt{\frac{2}{L}}\times
\begin{cases}
\displaystyle 
\frac{C_1}{\sqrt{C_1^2+C_2^2}}\cos \eta kx +
\frac{C_2}{\sqrt{C_1^2+C_2^2}}\sin \eta k|x| &
(|x|\geq l/2)
\\
\displaystyle 
\frac{1}{\sqrt{C_1^2+C_2^2}}\cos kx &
(|x|\leq l/2)
\end{cases}
\\
f_{\om, -}(x)&=& % \sqrt{\frac{2}{L}}\times
\begin{cases}
\displaystyle 
\frac{x}{|x|}\frac{D_1}{\sqrt{D_1^2+D_2^2}}\cos \eta kx +
\frac{D_2}{\sqrt{D_1^2+D_2^2}}\sin \eta kx &
(|x|\geq l/2)
\\
\displaystyle 
-\frac{1}{\sqrt{D_1^2+D_2^2}}\sin kx &
(|x|\leq l/2)
\end{cases}
\eea
where %$\eta(\om)$ is abbreviated as $\eta$, 
$k=\om/c$,
and the index $\pm$ represents the parity of the eigenmode.
The coefficients, $C_1$, $C_2$, $D_1$ and $D_2$,
are given by
\bea
\left(\begin{matrix}
C_1 \\ C_2
\end{matrix}\right)
&=&
\left(\begin{matrix}
\cos(\eta kl/2) & -\sin(\eta kl/2) \\
\sin(\eta kl/2) & \cos(\eta kl/2) 
\end{matrix}\right)
\left(\begin{matrix}
\cos(kl/2) \\ -\eta^{-1}\sin(kl/2)
\end{matrix}\right)
\\
\left(\begin{matrix}
D_1 \\ D_2
\end{matrix}\right)
&=&
\left(\begin{matrix}
\cos(\eta kl/2) & -\sin(\eta kl/2) \\
\sin(\eta kl/2) & \cos(\eta kl/2) 
\end{matrix}\right)
\left(\begin{matrix}
\sin(kl/2) \\ \eta^{-1}\cos(kl/2)
\end{matrix}\right)
\eea
As has been clarified in Secs.~\ref{sec:fpus} and \ref{sec:rmt}, % shmz
the atomic decay rate can be well evaluated by
the Fermi golden rule.
The decay rate (normalized by 
the decay rate $\Gamma_0$ in a free space)
is given by
\beq
\Gamma(X)/\Gamma_0 = \eta^{-1}\sum_{\sigma=+,-}
|f_{\Om, \sigma}(X)|^2,
\eeq
which is plotted in Fig.~\ref{fig:pddr} by a thin dotted line
as a function of the atomic position $X$.
It is observed that the atomic decay rate becomes strongly 
sensitive to the atomic position $X$,
reflecting the spatial form of the eigenmodes 
at the atomic transition energy $\Om$.
In other words, the cavity effect appears strongly
both in the vacuum region ($|X| \leq l/2$)
and the medium region ($|X| \geq l/2$).

%-----------------------------------------------------------------

Next, we proceed to discuss a case
where the medium is absorptive ($\kappa \neq 0$)
and plays the role of a photodetector.
In order to handle this case,
the absorptive material is modeled by an assembly of
harmonic oscillators,
and the canonical quantization is performed 
for both the photon field and the material~\cite{KS1996}.
As the absorptivity $\kappa$ is increased,
the effect of broadening of eigenmodes (Sec.~\ref{sssec:brd}) 
becomes more significant.
The position dependence of the decay rate 
is plotted in Fig.~\ref{fig:pddr},
for weakly absorptive ($\kappa=0.2$, broken line)
and strongly absorptive ($\kappa=0.4$, solid line) cases.
It is observed that,
when the atom is placed deeply inside of the material
($|x|-l/2 \gg \lambda_a/\kappa$, 
where $\lambda_a=2\pi c/\Om$ is the wavelength of emitted photon),
the cavity effect is completely smeared out
and the decay rate becomes identical to 
that of an atom embedded in an homogeneous medium.
In this case, the decay is solely determined 
by the nature of the material,
as in Sec.~\ref{ssec:a_ham}.
Contrarily, when the atom is placed in the vacuum region,
the position dependence remains even 
when the material is strongly absorptive.
Fig.~\ref{fig:pddr} indicated that the decay rate in this case
is basically determined by the eigenmodes 
for the non-absorptive case ($\kappa = 0$).

To summarize the results derived from the cavity QED, 
the decay rate would surely be affected by the presence of 
the optical materials 
even when they are placed separately from the atom (Fig.~\ref{fig:ill_realm}).
In this case, the physical origin of modification of the decay rate 
should mainly be attributed to reconstruction of eigenmodes,
not to broadening of eigenmodes.
In other words, the boundary effect would dominate. % the QZE.
Although a one-dimensional case is considered %observed 
here, the same qualitatively prediction would follow 
in the three-dimensional case also,
if the optical materials well surround the atom to form 
an optical cavity.

%%%%%%%%%%%%%%%%%%%%%%%%%%%%%%%%%%%%%%%%%%%%%%%%%%%%%%%%%%%%%%%%%%%%%
%%%%%%%%%%%%%%%%%%%%%%%%%%%%%%%%%%%%%%%%%%%%%%%%%%%%%%%%%%%%%%%%%%%%%
\subsubsection{
Zeno effect by detectors spatially
separated from the target atom}
\label{sec:separated}
%%%%%%%%%%%%%%%%%%%%%%%%%%%%%%%%%%%%%%%%%%%%%%%%%%%%%%%%%%%%%%%%%%%%%
%%%%%%%%%%%%%%%%%%%%%%%%%%%%%%%%%%%%%%%%%%%%%%%%%%%%%%%%%%%%%%%%%%%%%

It is sometimes argued that 
the term `Zeno effect' should be restricted to experiments 
where pieces of measuring apparatus exert a nonlocal negative-result
effect on a microscopic system \cite{HW}.
We can discuss the Zeno effect in this restricted sense as follows, 
using the results of the cavity QED presented above and 
consideration about the causality. 

First of all, % it should be noted that 
we must distinguish the following three cases.
Let the distance between the atom and the detectors be $l/2$, 
as shown in Fig.~\ref{fig:ill_realm}.
Suppose that the detectors are placed
at $t=t_{\rm d}$,\footnote{
Or, one can change the absorption spectrum of the detector materials, 
which was previously placed at $t < t_{\rm d}$, 
through, e.g.,  
the electro-optical effect \cite{NO} by applying an electric field
at $t=t_{\rm d}$.
}  
and the atom is excited to an unstable state at $t=0$.
The three cases to be distinguished are:
\begin{itemize}
\item[a)] 
The detectors are placed long before the atom is excited, i.e., 
$t_{\rm d} < 0$ and $c |t_{\rm d}| \gg l$, where 
$c$ is the light velocity.

\item[b)] 
The detectors are suddenly placed after the atom is excited, i.e., 
$t_{\rm d} > 0$.

\item[c)] 
The detectors are placed long before the atom is excited, but
are suddenly removed at $t=t'_{\rm d}$ after the atom is excited, i.e., 
$t_{\rm d} < 0 < t'_{\rm d}$ and $c |t_{\rm d}| \gg l$.
\end{itemize}

In case a), the calculations in Sec.~\ref{sec:atom-in-inhomo-CQED} 
hold, except for a very short time scale ($< 1/\Omega$) 
for which the rotating-wave
approximation employed in Eq.~(\ref{eq:H1_ag}) may be wrong.
Therefore, the results of Sec.~\ref{sec:atom-in-inhomo-CQED}
are valid at least when $t_{\rm j} > 1/\Omega$.
We thus conclude that the Zeno effect can take place in case a).
Physically, this may be understood as follows.
The materials composing the detectors modify the vacuum state of 
photons \cite{cQED,AAA}.
Although the modification requires a finite time to complete, 
it is completed before the atom is excited.
Therefore, the dynamics of the atom is affected by the modified 
photon vacuum \cite{cQED,AAA}. i.e., by the presence of the detectors.

In case b), on the other hand, 
the modification of the photon vacuum is not completed
when the atom is excited.
It is obvious from the causality that the dynamics of the 
atom is not affected during $t < l/2c$.
Therefore, if the free decay rate $\gg c/l$,
then the decay rate is not modified, 
i.e., the Zeno effect does not occur.
If, on the other hand, the free decay rate $\lesssim c/l$,
then the Zeno effect can take place.
To analyze case b) in detail, one must take account of the 
dynamics of the photon field, as in Refs.~\cite{OS,SOK}.
Note that, as shown there, 
photons will be emitted from the `false vacuum'
during the modification process of the photon vacuum.

Case c) is in some sense a combination of cases a) and b).
The dynamics of the atom is affected by the 
detector materials during $t <  t'_{\rm d} + l/2c$.
Therefore, the Zeno can take place 
if the modified decay rate $\gtrsim 1/(t'_{\rm d} + l/2c)$.
If, on the other hand, 
the modified decay rate $\ll 1/(t'_{\rm d} + l/2c)$,
then the observed decay rate will become the free
decay rate.

Although we believe these conclusions at the time of writing, 
more elaborate works may be necessary to draw more definite
conclusions. A related discussion is given in Ref.~\cite{HM}.

%%%%%%%%%%%%%%%%%%%%%%%%%%%%%%%%%%%%%%%%%%%%%%%%%%%%%%%%%%%%%%%%%%%%%%%%%%%%%%
\section{Experimental studies on the Zeno effect}
% \verb#{sec:exper}#
\label{sec:exper}
%%%%%%%%%%%%%%%%%%%%%%%%%%%%%%%%%%%%%%%%%%%%%%%%%%%%%%%%%%%%%%%%%%%%%%%%%%%%%%

In the preceding sections, we have discussed the Zeno effect, 
assuming 
the photodetection measurement on a radiatively decaying excited atom 
as a typical of quantum unstable states and measurements on them.
However, it is difficult to confirm experimentally 
the Zeno effect in such a system for the following reasons:
(i) The response time $\tau_{\rm r}$ of the detector
cannot be controlled easily,
because it is determined by the material parameters,
i.e., the interaction between photons and constituting materials of the detector.
(ii) The jump time $t_{\rm j}$ is small and therefore 
the decay dynamics is not easily perturbed by measurement.
If we simply estimate,
taking the positiveness of photon energies into account,
the bandwidth of the form factor 
by $\Delta \sim \Om$ (transition frequency),
$t_{\rm j}$ is estimated at $t_{\rm j} \sim \Om^{-1} \sim 10^{-15}$s.

It is generally believed that the jump time 
is very short in most of truly decaying states,
whose form factors are energetically broad.
Therefore, in most attempts at experimental confirmation of the 
Zeno effect,
oscillating systems are selected as the target 
system~\cite{PasNam,Kwiat,Kita1,Kita2,neutron,Luis}.
In such systems, the initial quadratic behavior lasts for a much longer time
as compared with truly decaying states,
and the Zeno effect is expected to be induced more easily.
For this reason, 
experiments on the Zeno effect have been performed
mainly for oscillating systems
such as Rabi-oscillating atoms~\cite{Cook,Itano,Ball,Itano2,CB}.
However, significance of the Zeno effect is obscured 
in oscillating systems,
particularly when one considers long-time behaviors 
of the survival probability.

Recently, there are several attempts to observe the Zeno effect 
in truly decaying states.
Here we introduce two of them,
both of which have long duration of deviation from the exponential law
(i.e., large jump time, $t_{\rm j}$)
and well devised measuring methods
with high controllability of measurement intervals.
One is the first observation of both effects
in atomic tunneling phenomena~\cite{exp_t2,FGMR}.
The other one is theoretical indication 
in the parametric down conversion processes~\cite{pdc1,pdc2,pdc3,pdc4}.
In addition to presenting them, 
we will also discuss in this section how to 
{\em avoid} the Zeno effect in general measurements, 
which are designed not to detect the Zeno effect but
to measure the free decay rate accurately.

%%%%%%%%%%%%%%%%%%%%%%%%%%%%%%%%%%%%%%%%%%%%%%%%%%%%%%%%%%%%%%%%%%%%%%%%%%%%%%
% \subsection{Experiments on unstable systems}
\subsection{Atomic tunneling phenomenon}
%%%%%%%%%%%%%%%%%%%%%%%%%%%%%%%%%%%%%%%%%%%%%%%%%%%%%%%%%%%%%%%%%%%%%%%%%%%%%%
\label{sec:tunnel}

\begin{figure}%----------------------------------------------------------------
\bec
\includegraphics{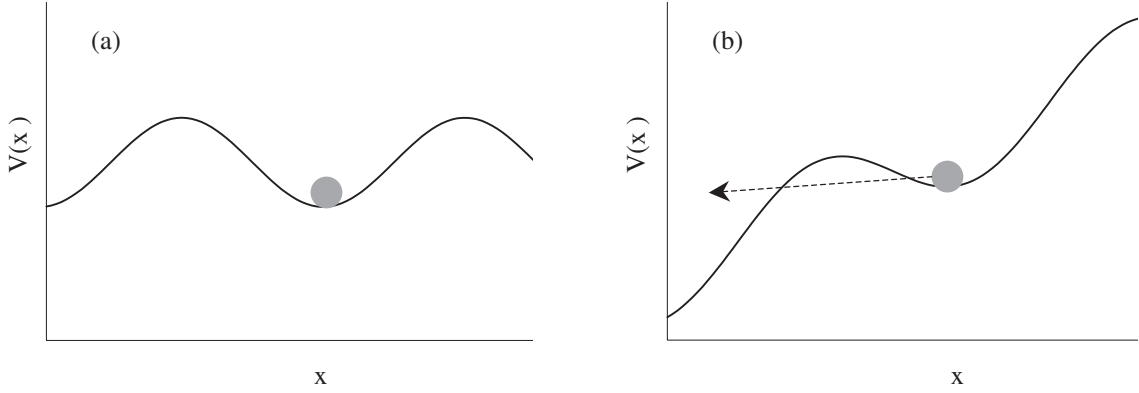}
\caption{\label{fig:wash}
The shape of the washboard potential Eq.~(\ref{eq:wash}),
under (a) small acceleration and (b) large acceleration.
The bound state is almost isolated in (a),
whereas, in (b), it may decay to unbounded modes 
extending in the negative $x'$ region.
}% \verb#{fig:wash}#
\enc
\end{figure}%-----------------------------------------------------------------

The experimental observation of the Zeno effect  
in truly unstable states is a difficult problem. 
Although the initial deviation from the exponential decay law 
is predicted in theory,
the duration of this period is supposed to be extremely short 
and the deviation is undetectable in most unstable states.
Recently, however, this deviation was successfully observed 
in tunneling phenomena of trapped atoms 
in an optical potential~\cite{exp_t2,FGMR}.

In their experiment, ultracold sodium atoms are trapped 
in standing wave of light,
which serves as an optical potential.
The optical potential is controllable in time
and may be accelerated as follows:
\beq
V(x,t)=V_0 \cos (2 k_L x - k_L a t^2).
\eeq
where $k_L$ is the wavenumber of light and $a$ is the acceleration.
The trapped atoms are driven by this potential.
In the moving frame fixed to the potential ($x'=x-at^2/2$),
the atoms suffer inertial force and 
the effective potential for the atoms becomes
\beq
V(x')=V_0 \cos (2 k_L x')+max',
% \verb#{eq:wash}#
\label{eq:wash}
\eeq
where $m$ is the atomic mass.
Thus, a tilted washboard potential can be obtained,
which contains a controllable parameter $a$.

The forms of the potential $V(x')$ is drawn in Fig.~\ref{fig:wash}.
When the acceleration is small ($a=a_{\rm trans}$),
high potential barriers are formed,
as shown in Fig.~\ref{fig:wash}(a).
By adequate choice of $V_0$ and $a_{\rm trans}$,
Fischer {\it et. al.} have realized a situation where
only the lowest state is bounded in each potential minimum.
This bound state is almost isolated from other modes 
by high potential barriers around the minimum.
Contrarily, under a large acceleration ($a=a_{\rm tunnel}$),
the left (right) potential is lowered (heightened),
as shown in Fig.~\ref{fig:wash}(b).
In this case, the bound state can couple 
through the left barrier to external modes,
which extend semi-infinitely in the negative $x'$ region
and therefore constitute a continuum in energy.
The bound state is no more stable
and decays to unbound states with a finite lifetime.
Thus, one may freely switch these two sorts of situations
(isolated bound state and unstable bound state)
by controlling the parameter $a$.

The measurement of the survival probability is devised as follows.
(i) As the initial state, the system is prepared in the potential 
of Fig.~\ref{fig:wash}(a)
by taking $a=a_{\rm trans}$ for $t<0$.
(ii) For $0<t<t_{\rm tunnel}$, 
the atoms are accelerated strongly with $a=a_{\rm tunnel}$
and the effective potential is changed to that of Fig.~\ref{fig:wash}(b).
In this period, tunneling to unbound states may take place.
(iii) For $t_{\rm tunnel}<t<t_{\rm tunnel}+t_{\rm interr}$,
$a$ is set to $a_{\rm trans}$ again
and bounded atoms are isolated.
In this period,
the bounded atoms are accelerated by the potential 
whereas the escaped atoms are not,
resulting in separating the survived (bounded) atoms 
and decayed (unbounded) atoms in the velocity space.
(iv) Finally, the optical potential is removed and 
the velocity distribution is measured,
from which one can infer the survival probability $s(t_{\rm tunnel})$.
If $t_{\rm interr}$ is long enough 
to guarantee complete separation in the velocity space,
this process serves as an ideal measurement of survival of the atoms.

In this manner, the survival probability under free temporal evolution
was successfully measured,
and deviations from the exponential decay law 
was confirmed experimentally for the first time.
The measured jump time is very long ($t_{\rm j} \sim 10\mu$s) in this system.
Furthermore,
by inserting several periods of small acceleration 
(in which the survival and decayed atoms are separated)
into the tunneling period,
they measured the survival probability under repeated measurements,
and succeeded in confirming both the QZE and the AZE.

However, it is worth mentioning that 
the measurement process through (iii) to (iv) is a sort of
direct measurements, 
the Zeno effect by which is often criticized 
as not being the genuine Zeno effect,
as discussed in Sec.~\ref{sec:DvsI}.
In fact, the potential for the atoms is altered when 
the measurement is performed.
Furthermore, 
the horizontal axes of Figs.~3-5 of the excellent experiment of
Ref.~\cite{FGMR} are
not the real time $t$ but $t-$(measuring times).
More elaborate experiments, which are free from these points, 
are therefore desired to observe the Zeno effect more clearly.

%%%%%%%%%%%%%%%%%%%%%%%%%%%%%%%%%%%%%%%%%%%%%%%%%%%%%%%%%%%%%%%%%%%%%%%%%%%%%%
\subsection{Parametric down conversion process}
%%%%%%%%%%%%%%%%%%%%%%%%%%%%%%%%%%%%%%%%%%%%%%%%%%%%%%%%%%%%%%%%%%%%%%%%%%%%%%
\label{sec:pdc}

\begin{figure}%----------------------------------------------------------------
\bec
\includegraphics{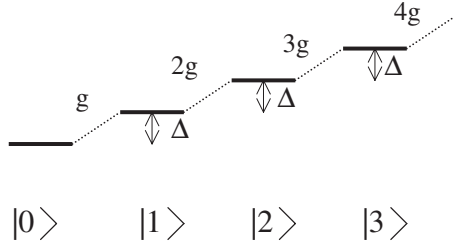}
\caption{\label{fig:tbm}
The energy diagram of Hamiltonian (\ref{eq:Hpdc}).
$|n\rangle = (n!)^{-1}(a_{\rm s}^{\dagger}a_{\rm i}^{\dagger})^n|0\rangle$
denotes the $n$ photon pair state.
}% \verb#{fig:tbm}#
\enc
\end{figure}%-----------------------------------------------------------------

In this subsection, we introduce a theoretical indication
of the possibility of observing the Zeno effect 
in parametric down conversion process~\cite{pdc1,pdc2,pdc3,pdc4}.
Although actual experimental observation of the effects
has not been reported in this system yet,
this system is an interesting candidate for actual experiments,
because the procedures for both continuous and discrete measurements
are proposed in this system.

In this process, a pump photon (frequency $\om_{\rm p}$)
spontaneously decays 
in a second-order nonlinear material
into a pair of signal and idler photons (frequencies $\om_{\rm s}$ and $\om_{\rm i}$)
satisfying the energy conservation law, $\om_{\rm p}=\om_{\rm s}+\om_{\rm i}$.
Using the semiclassical approximation for the pump field
and switching to the interaction representation,
the effective Hamiltonian for the signal and idler photons
are given by
\beq
{\cal H}=\frac{\Delta}{2}(a_{\rm s}^{\dagger}a_{\rm s}+a_{\rm i}^{\dagger}a_{\rm i})
+g(a_{\rm s}^{\dagger}a_{\rm i}^{\dagger}+a_{\rm s}a_{\rm i}),
% \verb#{eq:Hpdc}#
\label{eq:Hpdc}
\eeq
where $a_{\rm s}$ and $a_{\rm i}$ are the annihilation operators
for the signal and idler photons.
$\Delta$ is determined by the phase mismatch,
and $g$ depends on the intensity of the pump beam
and the second-order susceptibility of the material.
Here, the space coordinate in the propagating direction 
is regarded as the time coordinate.
Initially, there are no photons in the signal and idler modes,
the quantum state of which is denoted by $|0\rangle$.

The instability of the vacuum state is understood 
by regarding Eq.~(\ref{eq:Hpdc}) as a tight-binding Hamiltonian
in the photon number space.
Fig.~\ref{fig:tbm} shows the energy diagram of the Hamiltonian.
$|n\rangle = (n!)^{-1}(a_{\rm s}^{\dagger}a_{\rm i}^{\dagger})^n|0\rangle$
denotes a state with $n$ photon pairs.
The hopping energy between $|n\rangle$ and $|n+1\rangle$
is given by $(n+1)g$,
whereas there is an energy mismatch of $\Delta$ 
between neighboring sites.
The energy diagram becomes similar to Fig.~\ref{fig:enedi}
after diagonalizing $n\ge 1$ states.
When $g\gg \Delta$, the eigenstates of the Hamiltonian 
are delocalized in the number space,
and the initial state can decay to larger number states.
Contrarily, when $g\ll\Delta$,
the eigenstates are almost localized in each site.
In such a case, the initial state cannot decay completely.
Indeed, the survival probability is analytically given by
\beq
s(t)=\left[ \cosh^2(\sqrt{g^2-\Delta^2/4}\ t) +
\frac{\Delta^2}{4g^2-\Delta^2}
\sinh^2(\sqrt{g^2-\Delta^2/4}\ t) \right]^{-1}.
\eeq
Therefore, the system becomes a truly decaying one
when $|\Delta|$ is small enough to satisfy
$|\Delta|<2g$.
The jump time in this system is given, for $\Delta=0$,
by
\beq
t_{\rm j} \sim g^{-1}.
\eeq
This means that the jump time may be controllable 
in terms of the pump intensity,
which is a suitable nature for experimental observation of the 
Zeno effect.

The decay products in this experiment are the down-converted photons.
As for the measurement of them,
two schemes, discrete and continuous, are proposed.
First, we introduce the discrete type.
To this end, one divides the nonlinear crystal into several pieces,
and insert mirrors after each piece in order to remove the idler photon.
By detecting the removed photon by a photodetector,
one can perform a discrete photon number measurement
for each piece of nonlinear crystals.
The measurement intervals  % of measurements
are determined by the length of each piece,
so the control of the measurement intervals is flexibly done in this scheme.
Or, if one is only interested in the Zeno effect 
in the broad sense (Sec.~\ref{sec:info}), 
the photodetector for the removed photon is not necessary because
the removed photon will soon become entangled 
with environmental mechanical degrees of freedom,
which 
results in the decoherence 
between the states with no photon pair (survived state)
and one photon pair (decayed state).
This suffices for the modulation of the decay rate, as discussed 
generally in Secs.~\ref{sec:info} and \ref{sec:otherphenomena}.

Next, we introduce a continuous type of measurement.
In order to measure the photon number continuously,
one let the idler mode interact with the meter mode $b$,
which propagates in parallel with the idler mode,
through the third-order nonlinearity.
The Hamiltonian reads
\beq
{\cal H}={\cal H}+\kappa a_{\rm i}^{\dagger}a_{\rm i} b^{\dagger}b.
\eeq
The inference of the decayed moment is carried out as follows:
If the meter mode travels with an idler photon for time $t$,
the amplitude of the meter mode acquires a phase shift by
$\phi=\kappa t$.
By measuring this phase shift, 
one may infer the decayed moment.
As for the response time of this detecting device,
the uncertainty in the inferred time of decay
should be interpreted as the response time.
Noticing that the uncertainty $\Delta\phi$ in the phase measurement
is evaluated by $\langle b^{\dagger}b \rangle^{-1}$,
the response time $\tau_{\rm r}$ is approximately given by
\beq
\tau_{\rm r} = \Delta t \sim \frac{1}{\kappa \langle b^{\dagger}b \rangle}.
\eeq
Thus, the response time is revealed to be determined 
by the intensity of the meter mode,
which is easily controllable. 

%%%%%%%%%%%%%%%%%%%%%%%%%%%%%%%%%%%%%%%%%%%%%%%%%%%%%%%%%%%%%%%%%%%%%
\subsection{Evasion of the Zeno effect in general experiments}
\label{sec:evasion}
%%%%%%%%%%%%%%%%%%%%%%%%%%%%%%%%%%%%%%%%%%%%%%%%%%%%%%%%%%%%%%%%%%%%%

So far, we have discussed how to observe the Zeno effect.
On the other hand, in general experiments,
one usually wants to {\em avoid} the Zeno effect 
in order to get correct results~\cite{KSPRA,evasion1}.
Considering recent rapid progress of experimental techniques
and diversification of experimental objects, 
we expect that the QZE or AZE would % be significant
slip in results of advanced experiments not designed to detect it.
To avoid the Zeno effect, 
one must design the experimental setup to break at least 
one of the conditions for observing these effects \cite{KSPRA}.

For example, 
when performing an experiment with a high 
time resolution such that $\tau_{\rm r} \lesssim t_{\rm j}$, 
then the results of Sec.~\ref{sec:gim}
suggest that $\varepsilon_{\infty}$ should be {\em increased}.
If $\varepsilon_{\infty}$ cannot be increased 
to keep the sensitivity of such a high-speed measurement,  
then one should adjust parameters in such a way that
$\tau_{\rm r}$ lies on the boundary between the QZE and AZE, 
i.e., on the solid line in Fig.~\ref{fig:phased}.
Or, alternatively, 
one should calibrate the observed value using
our results, such as Eq.~(\ref{eq:norm_dr}),  
to obtain the free decay rate.
Such consideration would become important to future experiments.

%%%%%%%%%%%%%%%%%%%%%%%%%%%%%%%%%%%%%%%%%%%%%%%%%%%%%%%%%%%%%%%%%%%%%%%%%%%%%%
\section{Summary}
\label{sec:summary}
%%%%%%%%%%%%%%%%%%%%%%%%%%%%%%%%%%%%%%%%%%%%%%%%%%%%%%%%%%%%%%%%%%%%%%%%%%%%%%

The quantum Zeno effect (QZE) and the quantum anti-Zeno effect (AZE),
which are simply called the Zeno effect, 
% on decay of unstable states 
have been clearly understood for the case of 
repeated instantaneous ideal measurements,
for which one can take account of 
influence of the measurements simply 
by application of the projection postulate to the target system.
% {\bf KKin}
% taking the effect of measurement by application of the projection postulate.
%% Here, the term `instantaneous' means that 
%% the response time $\tau_{\rm r}$ of the measuring apparatus 
%% is much shorter than the measurement intervals $\tau_{\rm i}$
%% and the jump time $t_{\rm j}$ of the unstable state.
%% On the other hand, the term `ideal' means that 
%% the post-measurement state is given exactly by the projection postulate, 
%% which implies many conditions such as 
%% the measurement error is zero.
% {\bf KK out}
However, % these conditions are not strictly satisfied in 
real physical measurements are not instantaneous and ideal.
Theories of the Zeno effect induced by such general measurements
have been developed only recently.
In this article, 
after reviewing the results of conventional theories,
we have presented 
the results of such general theories of the Zeno effect.
We have also reviewed briefly 
the quantum measurement theory,
on which these general theories are based,
as well as experimental studies on the Zeno effect on 
monotonically 
%truly 
decaying states.

In Sec.~\ref{sec:fpus}, we have reviewed 
the quantum dynamics of an unstable state of an isolated system.
The dynamics is completely determined by the form factor $g_\mu$, 
defined by Eq.~(\ref{eq:g2}).
It is shown that the survival probability $s(t)$ of the unstable state 
decreases quadratically with time at the beginning of decay,
and later follows the exponential decay law, % with the decay rate
the rate of which is given by Eq.~(\ref{eq:FGR-0}).
Such an initial deviation from the exponential decay law
plays a vital role in the Zeno effect.

In Sec.~\ref{sec:ct}, combining the projection postulate
with the result for $s(t)$ obtained in Sec.~\ref{sec:fpus},
we have reproduced the results of the conventional theories
for the Zeno effect by repeated instantaneous ideal measurements.
The decay rate under the measurement is plotted 
as a function of the measurement intervals $\tau_{\rm i}$ 
in Fig.~\ref{fig:norm_dr}.
This figure demonstrates that repeated measurements 
does not necessarily result in suppression of decay:  
Depending on $\tau_{\rm i}$ 
and the form factor $g_\mu$, % of original unstable system,
the opposite effect, 
i.e., acceleration of decay, may take place, 
which is called the AZE.

These conclusions have been drawn 
under the assumptions that 
measurements are instantaneous and ideal.
However, as discussed in Sec.~\ref{sec:IMasLC},
real measurements are not strictly  %physical 
instantaneous and ideal.
To explore the Zeno effect by
realistic measurement processes,
we should apply the quantum measurement theory, 
which is briefly summarized in 
Sec.~\ref{sec:mt}.
Its key observation is that 
not only the system S to be observed but also 
the measuring apparatus A should obey the laws of quantum theory.
Therefore, one must analyze 
the time evolution of the joint quantum system S$+$A using 
the laws of quantum theory, as schematically shown in 
Figs.\ \ref{fig:S+A} and \ref{fig:ev}.
The information about the observable $Q$ to be measured is 
transferred to the readout observable $R$ of A,
through an interaction $\hat H_{\rm int}$ between S and A.
The measurement of $Q$ is thus reduced to a measurement of $R$
by another apparatus (or observer) A$'$.
Actually, 
A$'$ may be measured by a third apparatus A$''$, 
and A$''$ by A$'''$, and so on.
Such a sequence, as shown in Fig.\ \ref{fig:vNC}, 
is called the von Neumann chain, 
the basic notions of which are summarized in Sec.~\ref{sec:vNC}.
We have summarized the prescription for analyzing general measurements 
in Sec.~\ref{sec:prescription}.
As explained in Sec.~\ref{sec:prop-gm}, 
properties of general measurements are characterized by
the response time, measurement error, 
range of measurement, the amount $I$ of 
information obtained by measurement,
backaction of measurement, and so on.
An interaction process between S and A can be called 
a measurement process only when $I$ is large enough,  at least 
$I \gtrsim 1$.
Although this point is crucial when 
discussing many problems about measurement, 
it is disregarded in discussions of the Zeno effect 
in the broad sense 
(Sec.~\ref{sec:info}).
To discuss the Zeno effect, 
we have summarized in Sec.~\ref{sec:additional}
more characterizations of measurements, including  
% repeated, continuous, 
direct, indirect, positive-result, and negative-result.
It is sometimes argued that the terms 
QZE and AZE should be restricted to effects induced by 
indirect and negative-result measurements.
We have also explained repeated measurements,
including repeated instantaneous measurements
and continuous measurement. 
% The unitary approximation, which is often made
% in order to simplify the calculation, 
% is explained in Sec.~\ref{sec:ua}.
% Simplification and approximations, which are often made
% in order to make the calculation tractable, 
% are described in Sec.~\ref{sec:simple}.
% The approximations of Secs.~\ref{sec:probe} and \ref{sec:ua}
% are used in Sec.~\ref{sec:rmt}.
A simple explanation of the Zeno effect
using the quantum measurement theory is given in 
Sec.~\ref{sec:Zeno-MMT}.
It is also shown that
within the unitary approximation, which is explained in Sec.~\ref{sec:ua},
one does not have to use the projection postulate % at any points 
in the analysis of the Zeno effect. 
A few more points, which will help the reader,
concerning the quantum measurement theory are described in 
Sec.~\ref{sec:Gcmt}.

In Sec.~\ref{sec:rmt},
we have analyzed the Zeno effect using the quantum measurement theory.
We employ a model which describes a continuous indirect negative-result
measurement of an unstable state (Sec.~\ref{sec:model}).
A typical example described by this model is the continuous 
measurement of the decay of an excited atom using % state of an
a photodetector, which detects a photon emitted by the atom upon decay.
The measuring apparatus (photodetector)
is modeled by bosonic continua coupled to photons,
and the quantum dynamics is investigated 
for the enlarged system composed of the atom, photon and photodetector.
In Sec.~\ref{sec:rff}, it is shown  that
the form factor is renormalized 
as a counteraction of the measurement,
as illustrated in Fig.~\ref{fig:illff}.
The renormalized form factor $\bar{g}_{\mu}$ determines whether 
the decay is suppressed (the QZE) or enhanced (the AZE).
In Sec.~\ref{sec:idealm},
we have applied this formalism to the case 
where the response of the detector is flat, 
i.e., 
the photodetector responds to every mode of photons 
with an identical response time.
The decay rate is plotted as a function of the response time 
$\tau_{\rm r}$ in Fig.~\ref{fig:norm_dr2}.
The results almost coincide with those
of the conventional theories of Sec.~\ref{sec:ct}, which assume 
repeated instantaneous ideal measurements,
if we identify 
(apart from a multiplicative factor of order unity)
the measurement intervals $\tau_{\rm i}$ of Fig.~\ref{fig:norm_dr}
with the response time $\tau_{\rm r}$ of Fig.~\ref{fig:norm_dr2}.
In contrast, we show in Secs.~\ref{sec:gim}, \ref{sec:eim} and \ref{sec:fm}
that drastic differences emerge if the response is not flat.
The non-flat response may be interpreted as 
imperfectness in the measurement, such as  
geometrical imperfectness and energetic imperfectness.
In Sec.~\ref{sec:gim}, a geometrically imperfect measurement is discussed,
where the photodetector does not cover the full solid angle around the atom.
In this case, the Zeno effect takes place partially,
the amount of which is proportional to $1-\varepsilon_\infty$, 
the asymptotic photodetection probability. % in a long time scale.
In Sec.~\ref{sec:eim}, an energetically imperfect measurement is discussed,
where the photodetector responds to photons within 
a limited energy range $\Delta_{\rm d}$, 
called the active band.
Surprisingly, 
when the detector is energetically imperfect,
the Zeno effect can take place even for systems 
which follow the exact exponential decay law,
which was believed never to undergo the Zeno effect
according to the conventional theories.
This fact serves as an counterexample to an intuition
that imperfect measurements are disadvantageous for inducing the Zeno effect.
In Sec.~\ref{sec:fm}, a false measurement is discussed, 
where the detector cannot detect a photon 
whose energy is close to the atomic transition energy $\Omega$,
because the active band of the detector does not cover this energy. 
Interestingly, 
the AZE takes place even by such %Zeno effect
a false measurement if the detector response 
$\tau_{\rm r}$ is quick enough.
Relation between these results and
the conventional theories are discussed in
Sec.~\ref{sec:rel_ct}, and the physical interpretation
in Sec.~\ref{sec:po}.
Discussions and remarks on our model
are described in Sec.~\ref{sec:disc5}.

In Sec.~\ref{sec:cqed}, 
relation between the Zeno effects and other phenomena, 
such as the motional narrowing,
is discussed.
In particular, 
we discuss the close relationship between 
the cavity quantum electrodynamics (QED)
and the Zeno effect by a continuous 
indirect measurement.
Using the results of the cavity QED, 
we also discuss in Sec.~\ref{sec:atom-in-inhomo}
the Zeno effect in the case where the detectors are spatially 
separated from the target atom. 

Finally, in Sec.~\ref{sec:exper}, 
we discuss experimental studies on the Zeno effect.
In most attempts at experimental confirmation of the QZE,
oscillating states were selected as the unstable states.
However, significance of the Zeno effect is obscured 
in oscillating systems.
We have introduced two attempts to observe the Zeno effect
in truly decaying systems (Secs.~\ref{sec:tunnel} and \ref{sec:pdc}),
both of which have long duration (i.e., long $t_{\rm j}$) 
of deviation from the exponential decay law
and well devised measuring methods
with high controllability of measurement intervals.
We have also discussed
in Sec.~\ref{sec:evasion} 
how to {\em avoid} the Zeno effect in general measurements, 
which are designed not to detect the Zeno effect but
to measure the free decay rate accurately.

In conclusion, the quantum measurement theory has revealed
many interesting and surprising facts about the Zeno effect 
by general measurements.
We hope that future works, both experimental and theoretical, 
will confirm and/or extend these results, and thereby explore
more deeply into the Zeno effect.

%%%%%%%%%%%%%%%%%%%%%%%%%%%%%%%%%%%%%%%%%%%%%%%%%%%%%%%%%%%%%%%%%%%%%%%%%%%%%%

%%%%%%%%%%%%%%%%%%%%%%%%%%%%%%%%%%%%%%%%%%%%%%%%%%%%%%%%%%%%%%%%%%%%%%%%%%%%%%

\end{document}